# Compositional Cyber-Physical Systems Theory

GEORGIOS BAKIRTZIS

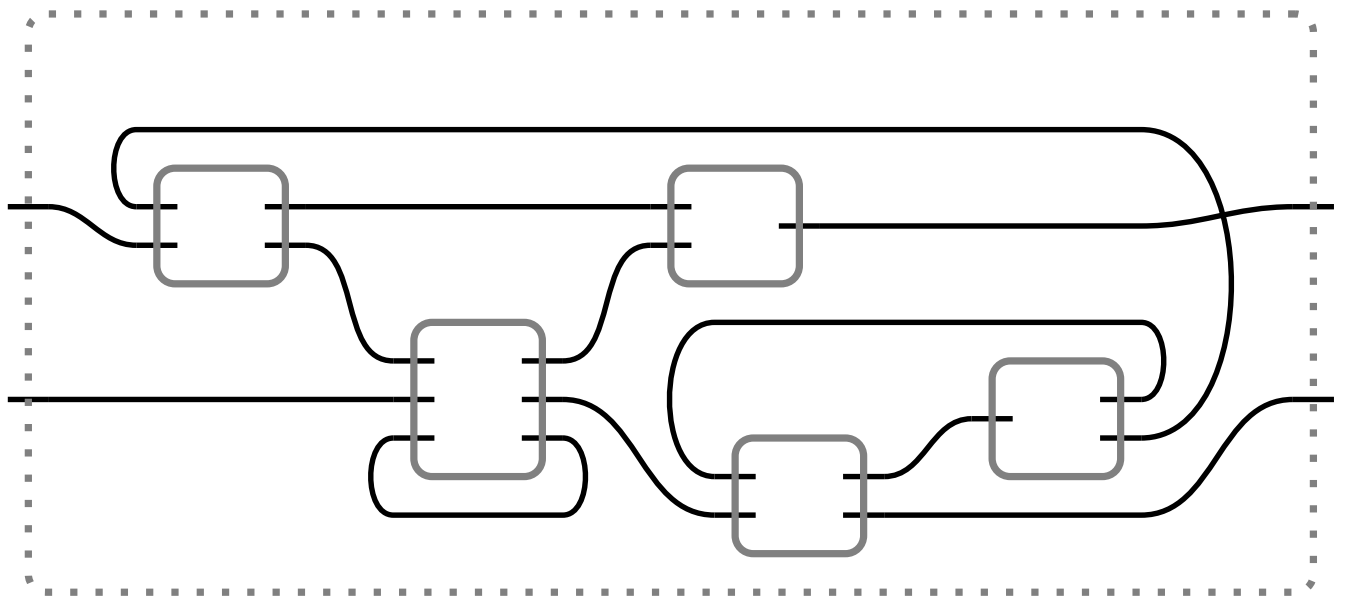

*A dissertation submitted in partial fulfillment of the requirements for the degree of Doctor of Philosophy at the University of Virginia.*

To Charon for the one obol

# Précis

A major impedance to engineering safe and secure cyber-physical systems is the lack of formal relationships between different types of models necessary for design. These various models are necessary because of the coupled physical and computational dynamics present in cyber-physical systems as well as the different properties system designers want to assure about a system. Each of the individual models has a set of rules describing what operations are allowed and which are not, including how to compose elements together in a way that is correct. These can mathematically be seen as algebras. However, the algebras in the engineering of correct and complete *requirements*, the specification and validation of dynamical *behavior*, and the identification of software and hardware *architectures* to carry out the necessary functions are distinct and can potentially lead to designing-in hazardous behavior in safety critical cyber-physical systems.

This dissertation builds a *compositional* cyber-physical systems theory to develop concrete semantics relating the above diverse views necessary for safety and security assurance. In this sense, composition can take two forms. The first is composing larger models from smaller ones within each individual formalism of requirements, behaviors, and architectures which can be thought of as *horizontal* composition—a problem which is largely solved. The second and main contribution of this theory is *vertical* composition, meaning relating or otherwise providing verified composition across requirement, behavioral, and architecture models and their associated algebras. In this dissertation, we show that one possible solu-tion to vertical composition is to use tools from category theory. Category theory is a natural candidate for making both horizontal and vertical composition formally explicit because it can relate, compare, and/or unify different algebras.

Ultimately, category theory reframes the problem of abstraction, either in the management of mathematical structures or system models by positioning a problem in its most natural domain. Category theory does not model the internal structure of the objects it acts upon. Instead, a categorical formalism perceives an object through its relationships with other objects and not by what the object is individually. Indeed, in this context we focus on abstraction, which we see as determining only what is essential in each layer of a given model. This allows us to talk about how things are *related* instead of focusing on how things *are*. This mindset as applied to systems theory gives rise to a circumspection of the system where we do not examine a system by its individual elements but by looking at the compositional structure of the system, which includes both the individual constituents and their interconnections. This is all to say that through compositional cyber-physical systems theory we can give concrete meaning to abstraction and refinement in cyber-physical system models, which can assist with the specification (and eventual validation) of increasingly complex systems.

Using this relational understanding of modeling we formalize categorically behavior and architecture using the systems as algebras framework, where boxes are subsystems and wires are connections between subsystems. This is a two step process. First we define the interface of each box as well as the way in which the boxes ought to be interconnected to compose the total system (chapter 2). Second, we assume a behavioral formalism for each box that is congruent with the behavior of other boxes based on the way they are interfaced and connected (chapter 3). We apply this framework to safety through the means of contracts (chapter 4) and to security through the means of tests and actions (chapter 5). Finally, we show how these different algebras and categorical structures can be used to mathematically implement verified composition (chapter 6).

Ithaka gave you the marvelous journey.
Without her you wouldn't have set out.
She has nothing left to give you now.

— C. P. Cavafy

# Acknowledgments

Some say that a doctorate is a Herculean task. They are unequivocally wrong. It is an Odyssean journey. A postmortem calculation reveals that there were many challenges, similar to the Scylla and Charybdis and the Sirens. Instead of some demigod power I possess, it was my company that helped me overcome those, often by sacrificing something of themselves.

I cannot but start with my grandfather and physics professor, Anastasios. While I was struggling with visualizing mathematics in my formative years, he adjusted his teaching style to the laboratory, which was the only way that I could learn at the time. He understood that mathematics is not the only way and made significant effort to fix my intuition with wires, tubes, pendulums, and other such physical devices we could make and discuss in his basement. He spent an enormous amount of time indulging an inept physical investigator, never once assuming I should have known something or that I was too far behind to understand some complicated physical reality.

I have to equally start with my other grandfather, Giorgos. I spent a lot of time working alongside him at our family business. It was from him that I built my work ethic, which has supported me through thick and thin almost like muscle memory. It was there that I learned how to actually *speak* to people and felt the sense of community the most.

*They are both the reason I am a scholar.*

My mother, Olga, gave me the opportunity to study in a different country far away from home and equipped me with the belief that I could achieve my goals as long as I truly wanted them. My father, Dimitris, followed my academic journey closely and has often been the reminder that a person ought to have other interests and when those interests are genuine, they ought to give some time from work to explore them. My sister, Evi, has been a constant emotional support, a grounding presence, and the person that pulled me out and showed me the diverse reality of experience outside of academia when I most needed it. My brother, Anastasios, took the role of the practitioner and challenged me to ground my theoretical results to the needs of my field. My aunt, Vennie, and uncle, Giorgos, gave me a home in the New World and took me in as their own child. My aunt and godmother, Eleni, taught me real freedom from within. My grandmothers, Voula and Rita, have always been there for support, interesting discussion, and lots of food.

Besides family, I have been greatly assisted by a plethora of friends. They are the ones with which I have raised my wine glass and celebrated in plenty of good times. Dimitris acts as my complaint box. Garrett is my pessimistic echo chamber. Tara is my precious gemstone. Beck takes me to nature in earnest. Kayla shares my love for wine. Emma was present for a while. Angelica tells it to me how it is.

In academia, I have to thank Carl Elks, who took a risk with me as I started graduate school. Then, peacefully giving the baton to Cody Fleming, who allowed me all the freedom in the world to drink from the unending tank of knowledge by removing all noise. David Evans discussed extensively my work and gave me direction by asking all the right questions. This dissertation wouldn't have been possible without my collaborators Fabrizio Genovese and Christina Vasilakopoulou. David Spivak was always available to explain complicated mathematics when I was lost. Edward Lee gave me advice when I needed it the most and he is part of the reason I can continue being a scholar after this dissertation.

Most of all this journey was sustained by the influence of the *Poet*. Those who give their soul for the rest of us to breathe freely within life's continuous struggle.

# Contents

# Nomenclature

| | |
|---|---|
| $\mathscr{A}, \mathscr{B}, \mathscr{C}, \cdots$ | matrices |
| $\mathbf{C}$ | a generic category |
| $A, B, C, \cdots$ | objects in a category |
| $\mathrm{id}$ | identity morphism |
| $f$ or $\xrightarrow{f}$ | morphism in a category |
| $g \circ f$ | composition (right to left) |
| $\mathrm{Hom}_{\mathbf{C}}[-,-]$ | homomorphism within the category $\mathbf{C}$ |
| $X \otimes Y$ | monoidal product (left to right) |
| $\begin{array}{ccc} A & \xrightarrow{f} & B \\ h\downarrow & & \downarrow g \\ C & \xrightarrow{k} & D \end{array}$ | commutative diagram standing for equation $g \circ f = k \circ h$ |
| $\cong$ | isomorphism |
| $(\mathbf{V}, \otimes, I)$ | a generic monoidal category |
| $F$ or $\xrightarrow{F}$ | functor |
| $F(A)$ or $FA$ | functor application on objects |
| $F(f)$ or $Ff$ | functor application on morphisms |
| $\mathrm{Nat}\,[-,-]$ | natural transformation |
| $\mathbf{C}/C$ | slice category over object $C$ |
| **Set** | the category of sets and functions |
| **Lin** | the category of linear spaces and linear maps |
| $\times$ | cartesian product of sets (or linear spaces) |
| $\Delta : X \to X \times X$ | duplication function |
| **Cat** | the category of categories and functors |
| **W** | the category of wiring diagrams (with types in **Set**) |
| $\mathbf{W}_{\mathbf{Lin}}$ | the category of wiring diagrams (with types in **Lin**) |
| $\mathcal{M}$ | the algebra of Moore machines; a functor $\mathbf{W} \to \mathbf{Cat}$ |
| $\mathcal{L}$ | the algebra of LTIS; a functor $\mathbf{W}_{\mathbf{Lin}} \to \mathbf{Cat}$ |
| $\mathcal{C}$ | the algebra of (static) contracts |
| $\mathcal{B}$ | a generic behavior algebra; could be $\mathcal{M}$ or $\mathcal{L}$ (among others) |
| $(X, S)$ | system with $X \in \mathbf{W}$ and $S \in \mathcal{B}(X)$ |
| $K_{\mathcal{B}X}$ | Knowledge database of systems of type $\mathcal{B}(X)$ |
| $\mathcal{B}(X) \xrightarrow{\Theta} \mathbf{Set}$ | test on $\mathcal{B}, X$ |

"I like the cover," he said. "Don't Panic. It's the first helpful or intelligible thing anybody's said to me all day."

— Douglas Adams

# 0 DON'T PANIC

Software kills [104].

From 1985 to 1987, the cancer treatment machine Therac-25 produced lethal doses of radiation, unbeknownst to those operating it [101, 103, 105]. The difference from previous models? Its safety mechanisms were, for the first time, fully controlled by software.

In 1991, a floating-point error drifted the internal clock of the U.S. Army's Patriot missile defense system by one third of a second, causing it to fail to intercept an incoming Scud missile in Saudi Arabia [38]. Twenty-eight U.S. soldiers were killed and roughly 100 others were injured.

In 1996, about 40 seconds into flight the Ariane-5 heavy-life space launch vehicle veered off course, graphically dismantled in air, and exploded [100]. Part of the software shipped with Ariane-5's inertial reference system was reused from Ariane-4 and were found to be responsible for the accident [131]. Scientists lost a huge number of instrumentation that was going to be used for experiments that took years to orchestrate.

In 2005, the Athens affair [140] showed the first glimpses of how hacking computing resources can lead to unexpected at the time significant losses in the "real world." Network switches were infiltrated to redirect and duplicate phone call signals to the perpetrators shadow phones. Among the tapped phones included politicians, embassy workers, the head of ministry of defense etc. This was the first attack on specialized equipment that not only we became aware of but was publicized openly partially to indicate that a changing world was coming.

In 2010, Stuxnet caused the destruction of centrifuges for separating nuclear material, crippling the Iranian nuclear program [92]. A potential future where security could significantly violate safety was now in sight.

In 2015, CrashOverride caused a blackout in Ukraine [151]. In Kiev—the capital of Ukraine—700,000 people had no electricity in freezing temperatures.

In 2017, Triton was found to have taken control of the safety instrumentation systems of a petrochemical plant in Saudi Arabia [63]. A first! Malware with the sole purpose of causing accidents.

In 2018 and 2019, two Boeing 737 MAX planes crashed due to a combination of sensor, software, and design problems that caused the plane to repeatedly push its nose downward in response to manual input from the pilot [164]. More than 300 people died.

Unfortunately, these accidents are not isolated and they are bound to increase as we rely on software to control physical systems. The main takeaway from the above examples is that introducing software means also introducing assumptions that could be erroneous. If we draw an analogy, software does not fail because it is in some sense the blueprint rather than the building. It is our specification and mental models that program in hazardous behavior. This dissertation asserts that part of improving the status quo in specification and validation can be assisted with the use of models equipped with a formal theory of composition.

To design safe and secure systems engineers use expertise, intuition, and inertia to decide how different analyses factor into higher-level, harder questions about safety, security, and resiliency. However, the deciding factors that bring those different analyses together is ad-hoc and in practice informal and that current practice of, for example, using requirements documents to design safety critical cyber-physical systems is insufficient and can lead to disastrous results.

It is recognized today that to manage and formalize that the various forms of expertise required to engineer cyber physical systems, it is imperative to use models in addition to streamlining and tracing results from varied analyses methods. To make this trace explicit we can formalize (some of) that expertise by reconstructing existing workflows and best practices in compositional terms.

This agrees with the intuition of how we put systems together. Kalman [168] broke it down to two fundamental elements.

1. Getting the physics right.
2. The rest is mathematics.[1]

This can be seen as the behavioral approach to systems theory, which was later formalized by Willems [168]. However, the behavioral approach by itself is incomplete to develop a general theory of cyber-physical systems modeling. The motivation and main contribution of this dissertation is a compositional cyber-physical system theory that, in addition to the behavioral approach to systems theory, is capable of decomposing to candidate architectures of software and hardware and an ability to restrict behavior in both behavior and architecture by formally representing requirements as contracts. Importantly, it addresses the above issues by producing traceable models, which also makes them interoperable.

It is very unlikely we will figure out how to make systems 100 percent safe and secure 100 percent of the time, but modeling practices are among the most effective tools we have. Developing further modeling capabilities through the use of compositional methods, we can unlock software's vast possibilities with greater confidence that our systems will not cause harm.

[1] The introduction of software does not really require an augmentation to these principles because software can be represented as a logical system and specified using graph-theoretic terms.

The sciences do not try to explain,
they hardly even try to interpret,
they mainly make models.

— John von Neumann

# 1 Prolegomena

Cyber-physical systems theory is not yet a science. The relationship between fundamental concepts has not been solidified, particularly across model domains. This is not surprising, because engineers work differently than scientists: while scientists build models for things, engineers use models to build things [94, 97] (figure 1.1). Nevertheless, concrete theory in engineering requires both modes of thinking. This criticism is not aimed at the individual fields manifesting concurrently in cyber-physical systems, such as control, systems theory, or software engineering. But even currently, when we have recognized this interplay between disparate fields in cyber-physical systems, the individual research programs seem to be conducted in their own silos. For this reason, the foremost problem facing cyber-physical systems assurance today is one of unification—of the challenges, for example between the codependence of security and safety in these systems, and solutions, such as the lack of formal traceability between requirements, behaviors, and architecture. Unification towards a compositional cyber-physical systems theory is the subject of this dissertation.

## 1.1 Cyber-physical systems pose new challenges

Cyber-physical systems are composed of computing platforms, control systems, sensors, and communication networks. These systems provide critical service capabilities in a number of engineering domains, including transportation, medical devices, and power, to name a few. The design of cyber-physical systems poses new challenges because of the intertwined nature of digital control with physical processes and the environment. As autonomy and coordination between a multitude of such systems becomes commonplace, there is an increasing need to formally assure not only their individual behavior but also the emergent properties that arise from the behavior of the composite system. One such emergent property is safety, which could be further imperiled by insecurity in cyber-physical systems. When cyber-physical systems exhibit unwanted behaviors, they can transition to hazardous states and then lead to accidents. To avoid such undesirable outcomes it is necessary to provide evidence of correct behavior before deployment, during the design phase of the systems lifecycle. Design changes in later stages of the system lifecycle cost more and are less effective [150]. It is only possible to produce this evidence early—when a realized system is not yet available—through the management and use of various models [22].

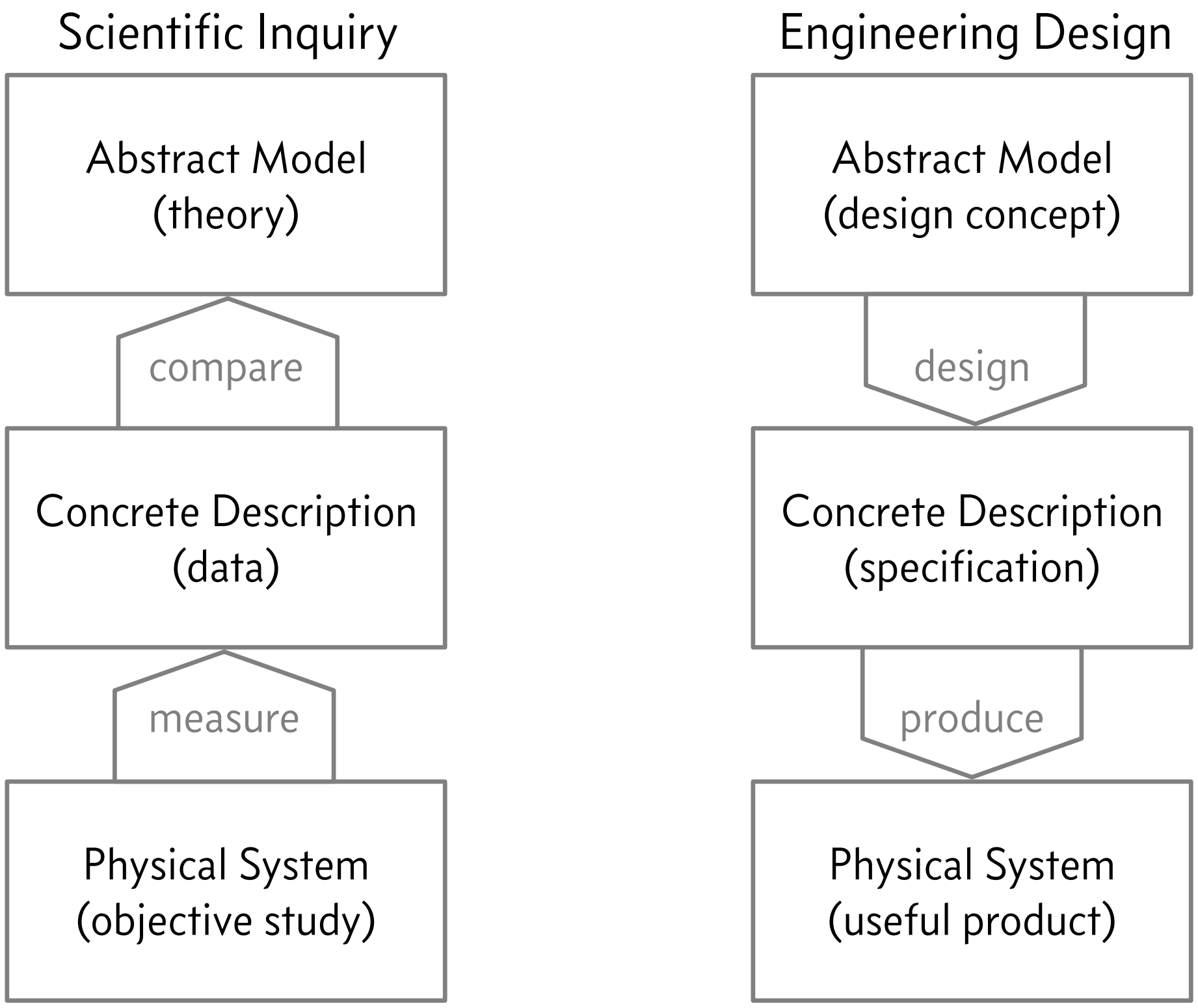

Figure 1.1: In scientific inquiry, the value of the model is how well it matches the system. In engineering design, the value of a system is how well it matches the model (adapted from Dexler [51]).

Undesirable behavior stemming from emergent properties has led to serious accidents [127]. For instance, a recent case, involves the first accidental death of a pedestrian by an indecision made by an autonomous vehicle, in which the vehicle chose *not* to stop in the face of an obstacle [106, 114, 124, 125]. A human supervisor was present, but this consists of one of the ironies of automation: humans are not good at keeping focus during such situations, but it is the automation itself that forces the humans into those very situations [14, 160].

Even more troubling are instances where unsafe and uncontrolled behavior—that at times has led to casualties—is a direct result of an exploit [73, 120]. This is supported by a series of notable and dangerous cyber-physical attacks including Stuxnet [92], Havex [143], BlackEnergy [79], and CrashOverride [151]. Significantly, Triton is the first known example of malware with the sole intention of disabling safety systems to instigate accidents [63].[1]

As designers strive for unprecedented levels of autonomy in every day life, the more engineering design problems become safety critical.[2] To assure safe and secure behavior we cannot rely only on a small number of experts that attempt to mitigate security issues after the design solution has been finalized. Instead, security and safety have to become part of the engineering design process itself—assessed and applied often and throughout but, significantly, starting from the early phases of the system's lifecycle (figure 1.2).

The design methodology of cyber-physical systems requires the study and appli-cation of several different areas (figure 1.3). These areas, such as control, have been studied at length at the individual level. However, to date there is still a lack of formal relation outside of each individual branch. The formal unification or otherwise formal composition between the branches that are used to engineer cyber-physical systems can lead to improvements in specification, managing the inherent complexity of different models and methodologies, and ultimately to better verification and validation of the systems we design and deploy. Additionally, the formal compositional modeling and analysis of such systems will allow for modularity and interoperability of system models. For example, an existing embedded systems model that is explicitly decomposed from some simulated dynamical behavior can then be further augmented with further design choices by a mechanical engineer to include more precise definitions and mathematical models of an "airframe" model entity.

[1] In any given successful attack it is common to discuss the innovative portion of the exploit chain, especially if the eventual target is part of the operational technology (OT) layer. But, the bread and butter of every successful attack is still basic malware, utilities, and intrusion tactics, techniques, and procedures that permeate in the information technology (IT) layer [155].

[2] Incidentally, a large number of sectors apply automation so rapidly to the point where it should be considered unchecked and careless in addition to unprecedented.

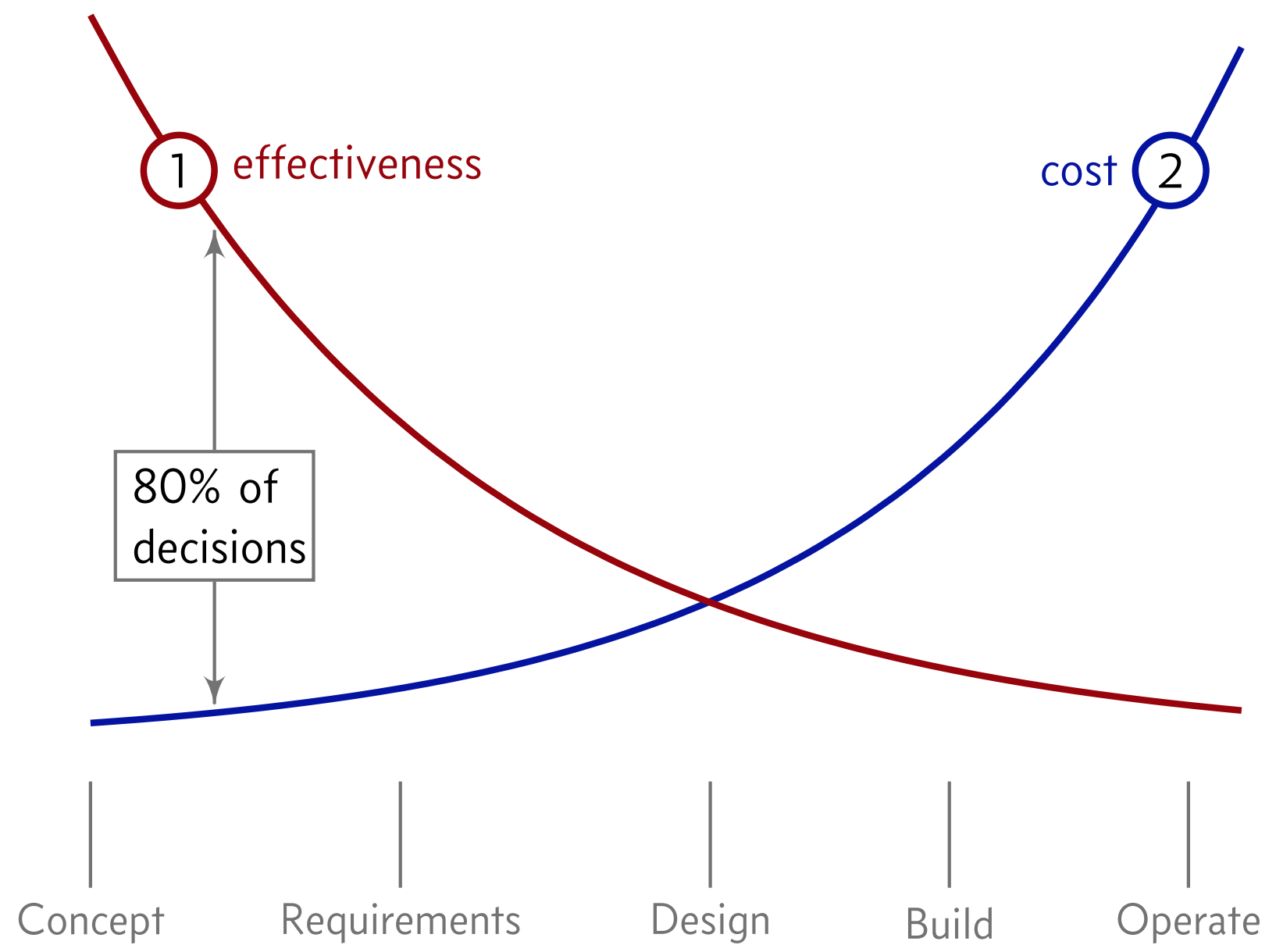

Figure 1.2: The ability to impact cost and performance is highest early in the system's lifecycle [56] (adapted from Strafaci [159]).

As a final observation, in practice there exists a gap between model solution, in-cluding their associated simulations and formal guarantees as compared to the eventual implementation of the system that will be manufactured, programmed, and deployed. Addressing composition between different modeling paradigms and tools is the first step for a fully interfaced toolkit that maintains simulation results and formal guarantees from behavioral models, to architectural designs, to finally synthesis on fabric. The definition of algebras and coalgebras within a categorical model has the potential of addressing this interoperability problem.

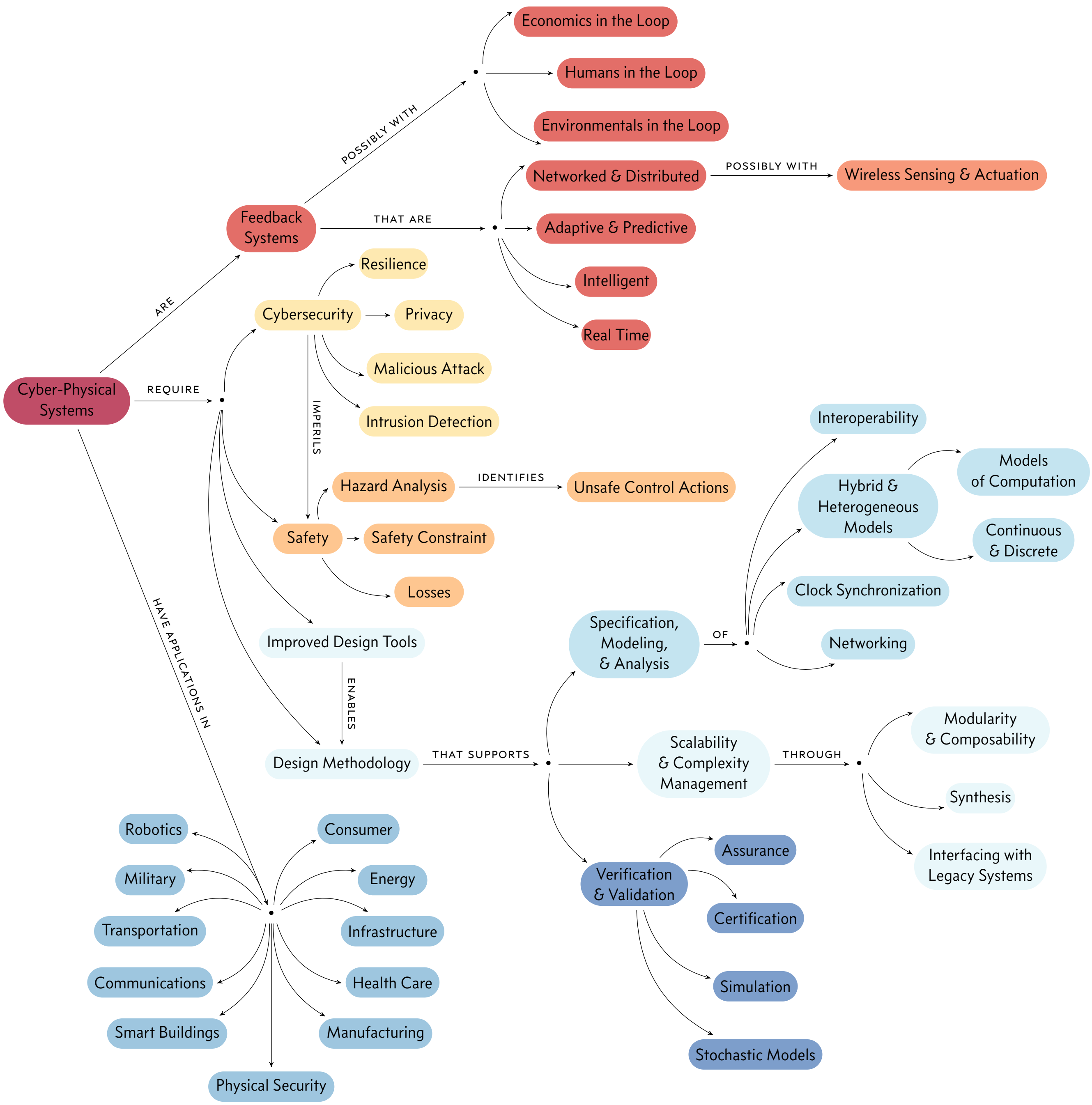

Figure 1.3: A cyber-physical systems concept map showing what these systems are, what they require, and where they are being used (adapted from Asare et al. [10]). Individual leaf nodes in this context map have been studied in detail but formal relationships between leaf node concepts are still sparse.

Specifically, the point of developing a compositional cyber-physical systems theory is to bridge together different models for the purpose of addressing the overlap between safety and security[3] by eventually producing new modeling languages. These modeling languages *must* have a strong notion of composition and increased semantic relatedness. Category theory, the mathematics of mathematics, has a number of appealing properties for this role. Namely that it is fundamentally based on composition of structures and that the behavior of components is defined by its relationship to other components instead of just examining components in isolation.

## 1.2 Assurance is about models, not systems

A model is any representation of a system but the thing itself. This includes what we might traditionally think of models but also intermediary representations that seem like they are the thing itself but in reality they are not, for example, software.

Design is the well-formed solution to a set of constraints. These constraints are defined through requirements. To construct a proposed design solution that adheres to the concrete requirements we first need to consult with the persons intending to use the system [130]. These persons are the stakeholders of the system which at a minimum include the system designers themselves, the operators, and owners. It is those differing views that lead to solid requirements and, therefore, to well-formed design solutions.

[3] As we deploy and further rely on cyber-physical systems for everyday tasks, this overlap becomes less narrow and more apparent.

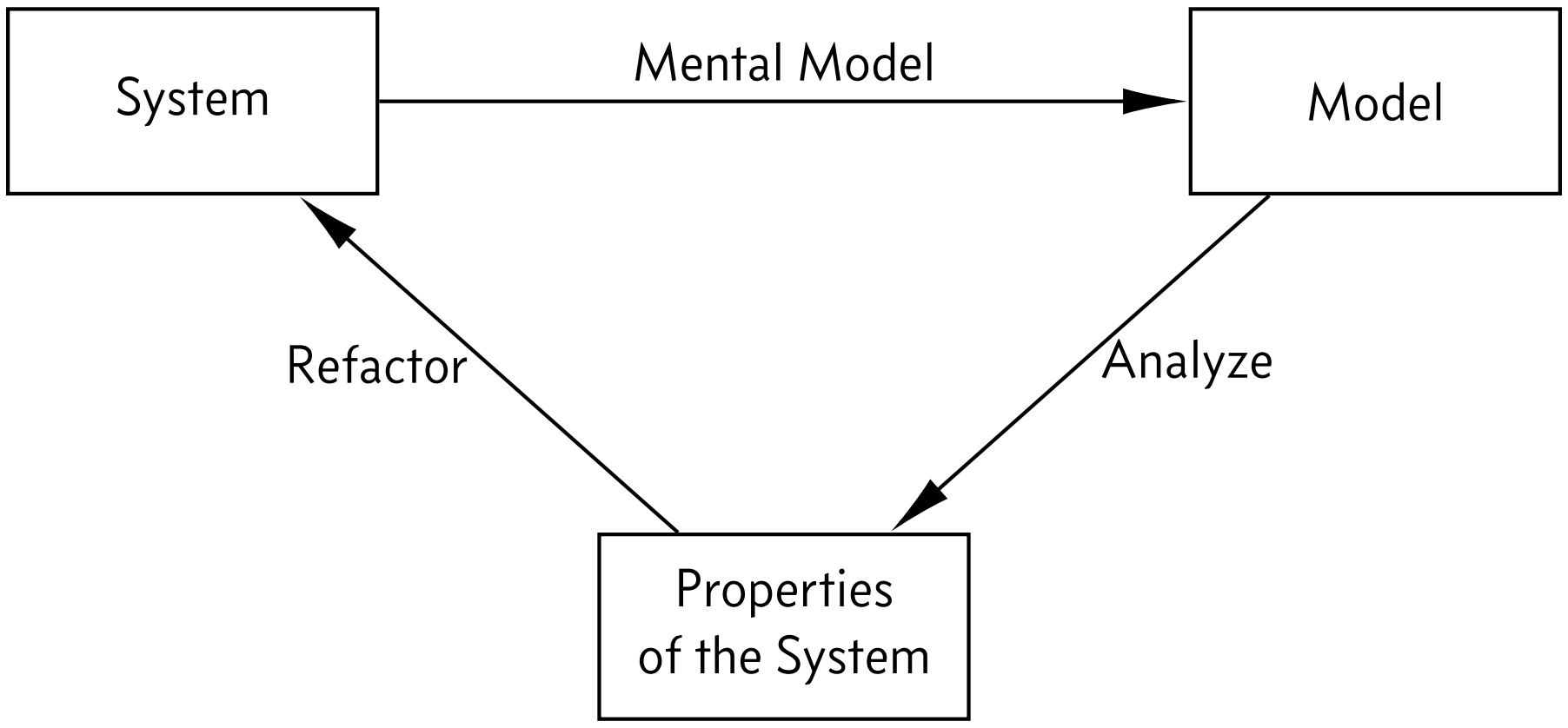

Figure 1.4: The standard modeling approach is based on analyzing a realized system with the goal of better understanding its behavior.

However, to reason about those differing views it is necessary to elicit information through structured methods [39]. Otherwise, the differing views confuse rather than elucidate. This leads to the fundamental challenge of design, which is to create and evaluate elicitation strategies that are inclusive of all stakeholders's views, such that the behavior and misbehavior of a system is correctly classified.

Further, any such strategy has to be able to build strong and nuanced mental models of the needs and behaviors any potential solution provides. This is a task that has been studied extensively when analyzing physical systems (figure 1.4). On the other hand, the construction of clearly defined requirements and unacceptable losses to those requirements has been less studied. The new reality and ubiquitous use of cyber-physical systems depend upon the refinement of requirements in the absence of a realized system (figure 1.5).

By using modeling techniques, it is possible to examine quickly, efficiently and cost effectively the different behaviors or responses a systems might exhibit. In terms of assurance that means that we can examine a reduced set of *edge-cases* based on the expected behavior of the system. The purpose of models for assurance is to test the assumptions and mental models that designers might have and assist with creating requirements regarding mitigation, raising security barriers, and/or adding resilience to address these cases [53].

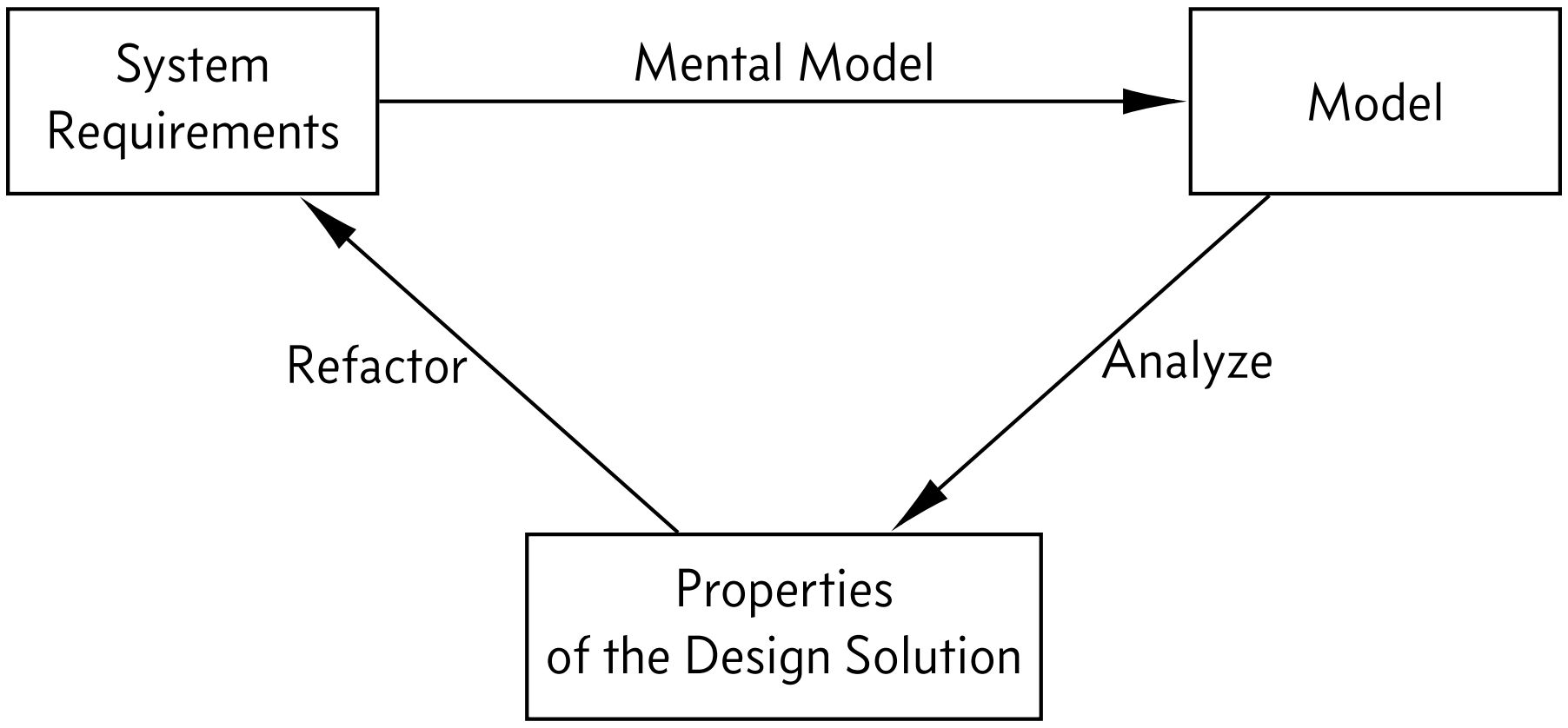

Figure 1.5: The majority of flaws are introduced in the early-concept to requirements phases of the system's lifecycle [102]. If we move the goalpost earlier we can produce concrete and helpful system specifications to assist in the implementation of safe and secure systems.

In the same sense, design too is a form and type of model that captures interfaces, connections, and dynamics. In that case, the model becomes the design or otherwise the blueprint of the eventual system. This is a natural way to think, a design is a model insofar as it represents a possible solution to the problem the system ought to fix when it is built.

As an additional benefit, models can be treated as living documents, maintained to reflect design choices and system revisions. Hence, they can be a valuable resource for the safety and cybersecurity specialists. Competent information technology (IT) professionals have long stressed the need for documenting the system designers's *intentions* versus simply stating the architecture of, for example, a network of computers [42]. Extending the current design practice to clearly define the set of high-level requirements that are traced down to a formal specification of a system documents all the *what's* that can be reassessed and revised throughout

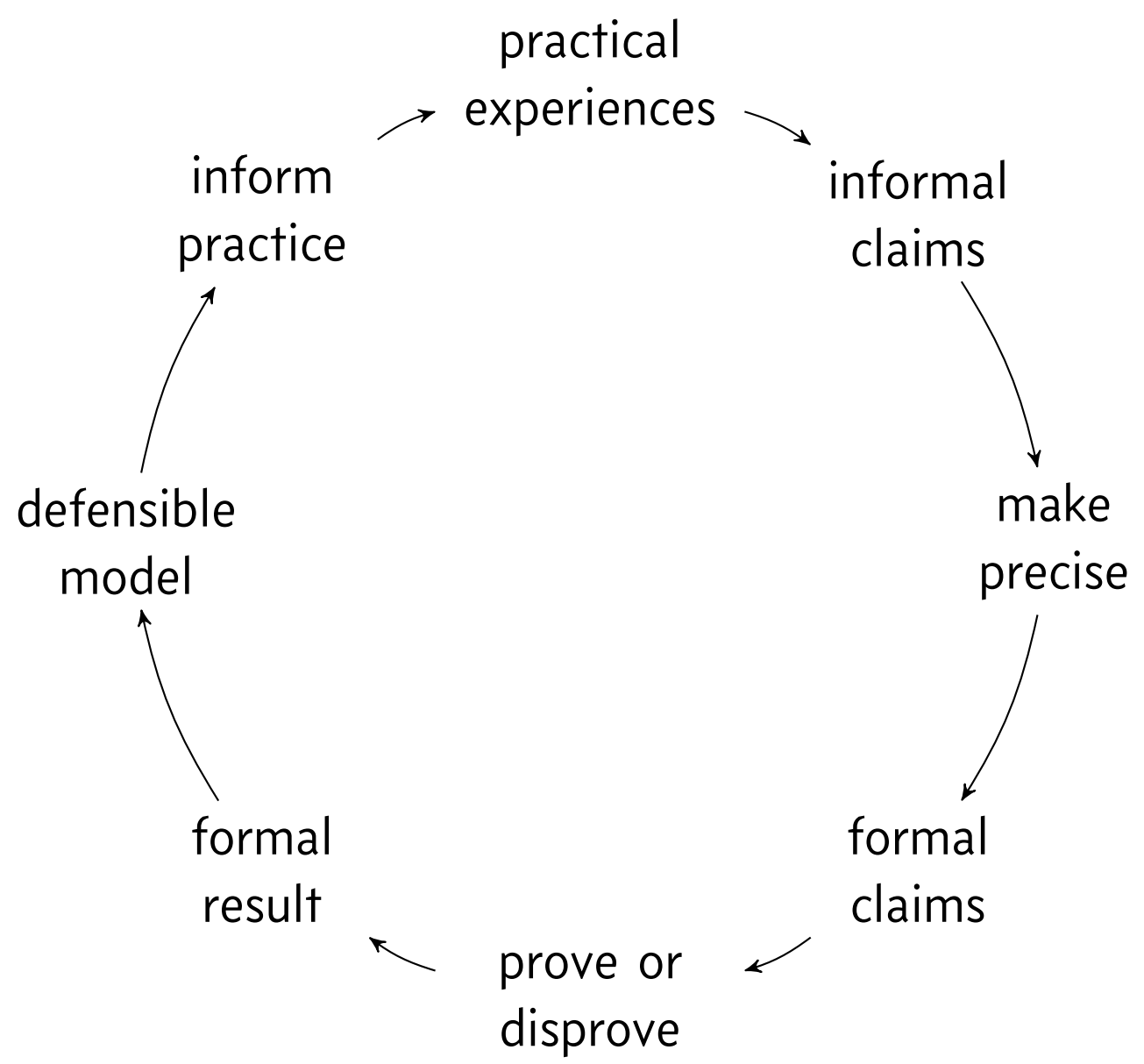

Figure 1.6: The formal informs the informal (adapted from Kuper [86]).

the system's lifecycle.

In this dissertation, we consider a *formal method* to the problem of composition in the modeling of cyber-physical systems. This formal method or model did not get developed in the absence of informal experience (figure 1.6). This historic detour to how the results presented in the subsequent chapters came to be is important to understand the context in which composition in modeling and design solutions became a central theme for our work.

Inconsistencies arise in much of the informal work necessary to design and assure safety-critical cyber-physical systems. Attempting to formalize some of these inconsistencies took us through a journey from linear logic to modal logic and then to graph theory. Logical inferences then made sense to think as connections of some form, which we get for free by using compositional methods, to use and develop for the design and assessment of systems. From another perspective, many of the concepts in system modeling were elusive from a philosophical perspective, for example, plagued by the question of what actually is a model? From this side too, capturing a formal notion of composition made sense because abstractions and refinements show up all over the place when modeling. But to define their types in any kind of meaningful way that was either theoretically sound or practically useful requires more than just putting lines between boxes. We can understand these things in a more systematic and rigorous way using compositional methods.

## 1.3 Cybersecurity is fundamental to safe behavior

In cyber-physical systems the security of the machine is tightly coupled with safety. This is due to the interaction of digital and often distributed systems with the physical environment. Hence, as with safety, security is an *emergent property* of a system [28, 40, 41, 169, 170], meaning that the safe or secure behavior of the components, cannot be used individually or in isolation to investigate the overall safety or security of the system [57, 135]. This is increasingly important for two reasons: these machines are ubiquitous and they are expected to make complex and timely decisions autonomously [123].

These decisions could be influenced by an attacker to cause physical harm. A successful exploitation of a cyber-physical system can lead to highly undesirable and unsafe behavior [108]. This is seen in several safety-critical domains that contain serious and exploitable vulnerabilities, including medical systems [3, 4, 65, 84], aviation [31, 74], automotive [43, 46, 77, 78, 81, 83], and electric power [85, 161], to name a few.

A new notion and potential *calculus* in the security of cyber-physical systems is then driven by the need to safely (and therefore securely) provide clearly defined system behaviors. While security can be a means to protecting private information, in the cojoint problem of security *and* safety, security is merely a tool towards the safe operation of the systems expected service and not a means in itself. Security for security's sake is not only unresourceful but also ineffective. This is because bolt-on security—security that is applied after a system is designed and

deployed—provides narrow options in the form of mitigation strategies. Instead, security by design that is cognizant of the systems's expected service and its potential unsafe behaviors, can better be implemented early, often, and throughout a system's lifecycle.

However, it is not always the case that designers work with newly constructed systems. More often than not, system designers are asked to incorporate new features and construct different behaviors using existing infrastructure or components. Therefore, security approaches require understanding the coupling of legacy hardware, firmware, and software (that might be more mature but potentially have less security consideration[4]) with new, better understood and designed subsystems as well as the interface between the two.

Indeed, security is the efficient use of resources to defend against reasonable threats [27, 170]. What constitutes a reasonable threat in a cyber-physical system can only be understood through its unsafe and undesirable behaviors that can be induced by a successful exploit. Unfortunately, today, security is often practiced in an unorganized piece-meal fashion, such that system's are secured almost exclusively through bolt-on solutions. These solutions by definition cannot avoid an increasing number of physical hazards since they are applied after the fact.

But how do physical hazards manifest? The operation of a cyber-physical system can be thought of as providing digital control; that is, computer states, through physical inputs and outputs; that is, the physical states. Specifically, the system's computer states are a complex set of interrelated hardware, firmware, and software states. Faults that occur in this layer include errors in sensor measurement, control algorithms, and state variables. The system's physical states capture the underlying concept, which includes the environmental conditions; that is, the activated fault from the system's computer states manifests in an unsafe control action. It is those conditions that might lead to a hazardous state that can lead to unacceptable losses and accidents (figure 1.7).

[4] It is often the case that stakeholders consider legacy software and hardware *more* secure than newly designed and implemented software and hardware. The misunderstanding seems to stem from the idea that the longer a system is in place, the less likely it is that vulnerabilities are going to survive. This notion is unequivocally wrong. For example, WinRAR contained a nineteen year old vulnerability that allows an attacker to take full control of the victims computer [67]. For firmware—especially in safety-critical applications—the situation is worse because even if patches exist; applying them might go against operational needs or patch distribution is handled incorrectly by the vendor [25]. ShadowHammer, for instance, abuses ASUS's own update channel and private signing key to distribute malicious patches [66, 87].

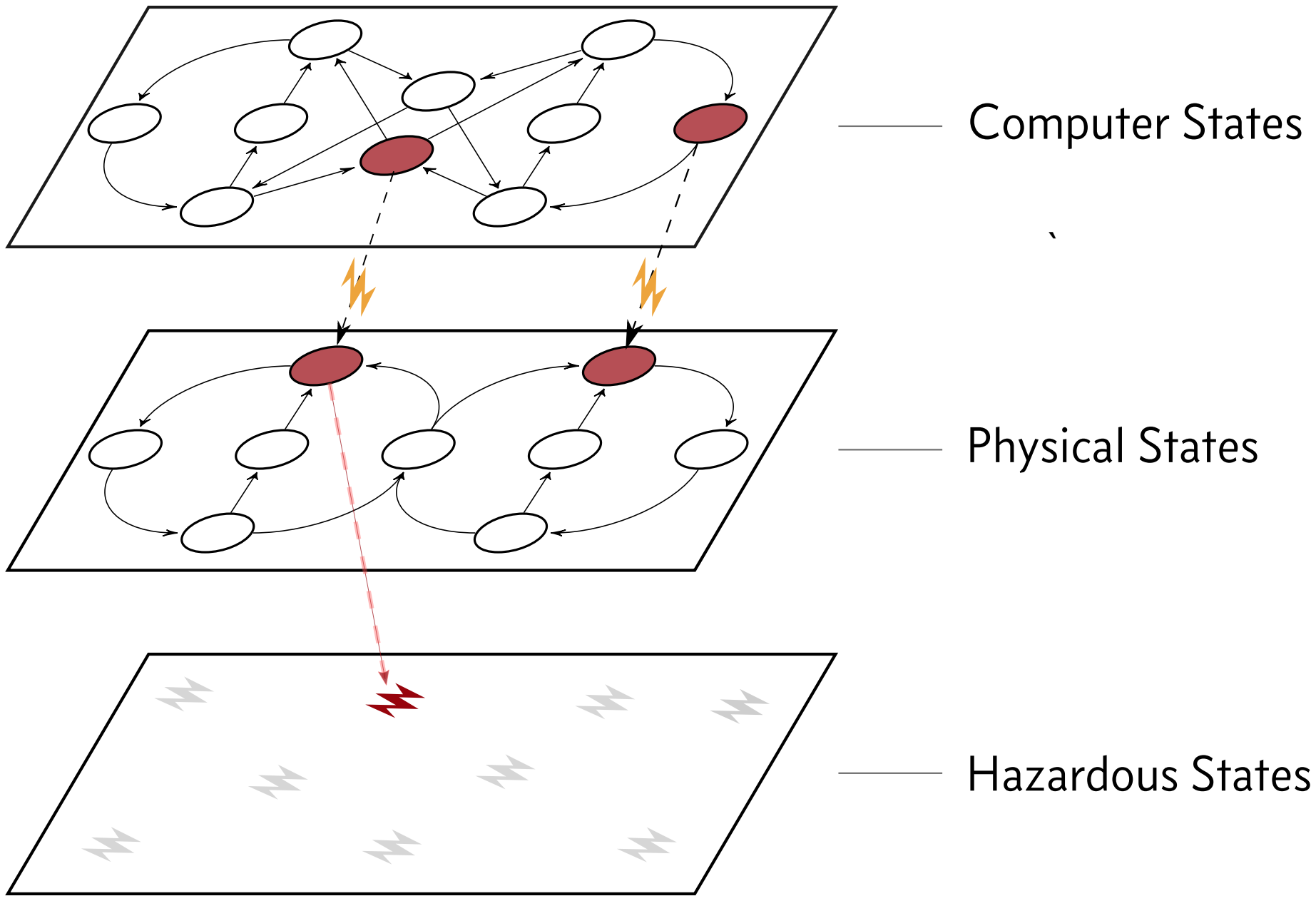

Figure 1.7: An activated erroneous state (⚡) in the computer layer can propagate to violate a physical state (●). In the right context the violation of a physical state can lead to a hazardous condition (⚡), which may lead to an accident. If the context is not right the hazardous state stays dormant (⚡).

To understand how these state layers interact with each other consider the operation of an adaptive cruise control (ACC). While driving the ACC software might issue an unintended acceleration command due to, for instance, calculating a wrong value or timing issue. If this happens in the right context, for example, this fault is activated while the vehicle is too close to the lead vehicle, the vehicle will not maintain a safe distance. This is an activated hazardous state that might, in turn, lead to a collision. The desirable behavior here is to always keep a safe distance from the lead vehicle, while the undesirable behavior is to unintentionally issue an acceleration command while close to another vehicle.

Once there exists a set of desirable and undesirable behaviors as well as an initial system design (or even a set of designs), it is important to specify them precisely. That is where a more expressive language rather than code needs to be used. This language is mathematics (or metamathematics) [90]. Through the language of mathematics we can start reasoning in terms of behaviors or dynamics instead of only implementations as is the case with only considering how to develop the code for a particular system. Reasoning about behavior before implementation is particularly important for cyber-physical systems, where code controls physical processes, such as actuators.

Different implementations might be mapped to the same set of system requirements. The language of mathematics, being agnostic to implementation, is much better equipped to evaluate and assure certain behaviors—even when we assume presence of attackers. Of course, the solution will eventually need to be coded to behave as we expect so mathematics and code must work in tandem. Compositional methods help with that too and code can be developed to implement the mathematical definition in terms of dependently typed programming languages.

## 1.4 Compositionality is subtle

In system design and engineering at large, compositionality refers to putting a system together from well-understood components to perform some behavior. In computer science, it refers to something more formal which is that the behavior of the whole program must follow from the inferences of each individual statement in sequence. Even though the term compositionality is elusive, it generally emerged to produce structure out of historically unstructured processes and that holds true both for engineering and computer science.

In a program we can see this development of compositional structures with the introduction of better abstraction and refinement features in programming languages. These take the form of, for example, additional features in the form of functions, classes, or even packages. By using these features programmers are more likely to create modular software but crucially be able to create interfaces. Today, programmers can work on huge programs without necessarily knowing the inner workings of most of the functions they are incorporate into their own code blocks. This has been such second nature that even short code might include libraries with thousands if not millions of lines of code being used without the programmer having ever examined what those are. Compositionality in this sense does not mean that any programmer will develop easy to understand and modular code—the opposite is often true—but that they have the tools to do so if they so desire.

In engineering, we have similar tools although the term compositionality is used less frequently to describe them. For example, in control we might view the different physical processes and systems in the form of a hierarchy (figure 1.8). This hierarchy means different things to different people but holds one thing in common, namely, the implicit existence of *interfaces*. A mechanical engineer might have complete familiarity with the physical actuation and at the same time only know the commands that come from a microcontroller. A computer engineer might instead know the opposite; know exactly how to produce the commands to move actuators but not how the physical dynamics of actuation actually work. Interfaces are then the glue between different components and expertise. A proper interface will help the mechanical engineer design an actuator while at the same time helping a computer engineer with fabricating hardware and developing software for the same actuator without necessarily knowing exactly how it works.

This is to say that compositionality is about *remembering* and *forgetting* [70], which map directly to *refinement* and *abstraction*. Remembering means adding or recovering more information about a particular component. Say something is wrong with the actuator, then the expertise of the mechanical engineer ought to be used to figure out why. But forgetting is equally important in designing increasingly complex systems which requires us to see the forest instead of the trees; that is, putting components together to form larger systems without necessarily being experts in every component that compose the systems we build.

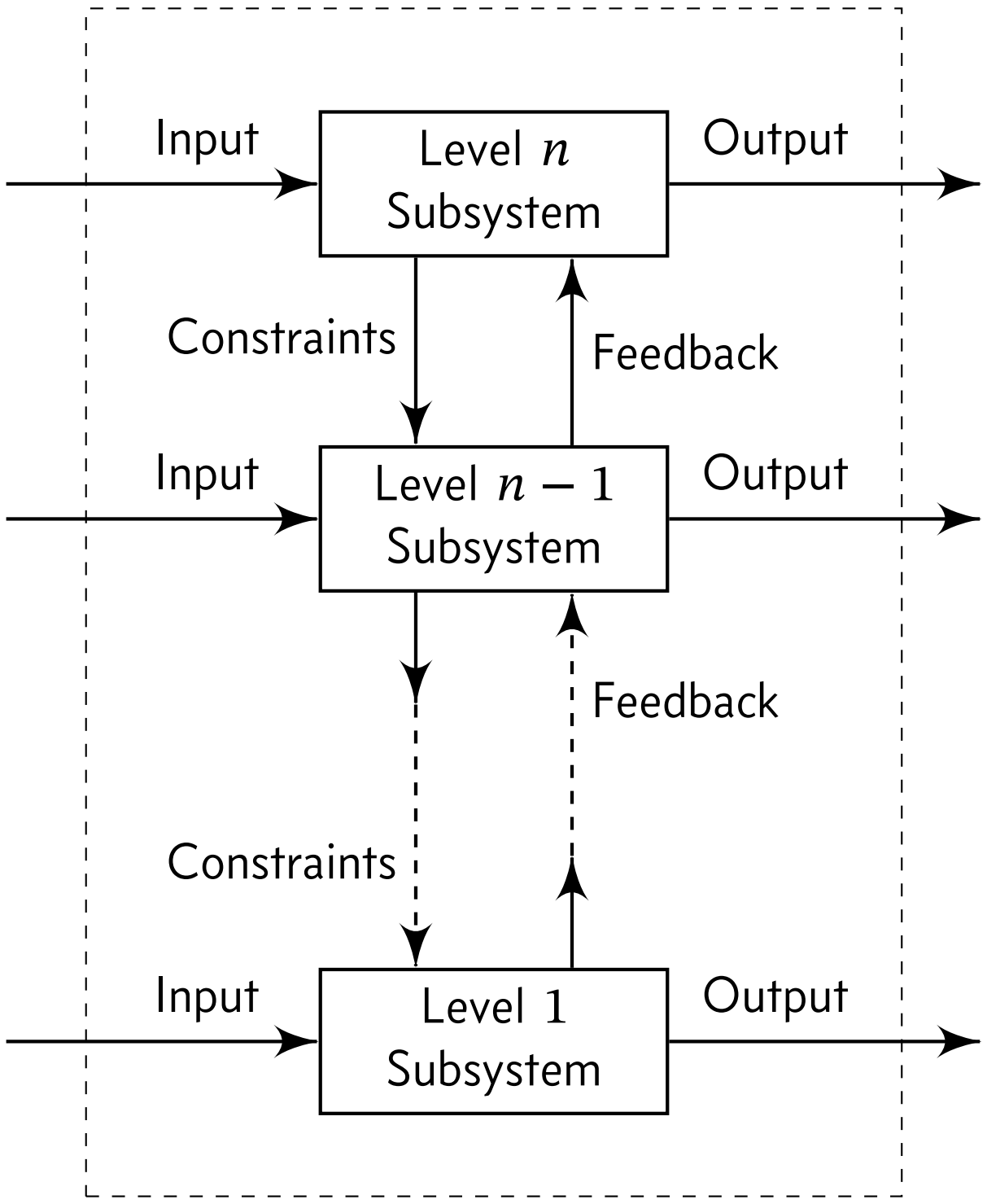

Figure 1.8: Hierarchical control implies the feedback relationship from bottom to top and constraint relationship from top to bottom (adapted from Mesarovic and Takahara [117]).

The opposing force to compositionality is that of emergent phenomena. A compositional model of a particular system must be sufficient to tell us how the behavior of the whole is produced by the behavior of their parts and connections between parts. It would be impossible to remove emergent phenomena and it is the case that often they are desired to achieve a particular behavior. However, the point of compositionality is not to eradicate emergence but rather to control it. Controlling emergence through means of compositionality requires us to add more information to models that are well understood. We will see for example, that in a controls model we will require the use of trivial functions for sensor and controller even though we already know how to calculate the overall behavior without them (by using the standard equations for dynamical behavior in those systems). That is what we mean by controlling emergent effects, not removing or reducing but rather being further mathematically explicit when needed to retain compositional properties in our models.

Most engineering work in compositionality centers around one formalism, for example, in hybrid systems [126], timed automata [34], linear temporal logic specifications [8], and contract theory [62]. Call this type horizontal composition; within one mathematical model we compose the same types of models to produce larger ones. This is generally an accepted line of work within one field but a perhaps more interesting rule would be vertical composition, which would relate or otherwise be able to enforce a hierarchy among multiple such formalisms. One way of achieving vertical composition is by using tools from category theory.

Category theory is instead the study of composition between mathematical structures. Logic and set theory created the scaffolding for the foundation of mathematics. Instead, category theory focused on showing how different mathematical concepts are related to each other. This mathematical study of relationship has positioned category theory beyond pure mathematics, the most visible application of which is in functional programming. This is to say that the earlier example of compositionality in computer science has been made formally explicit through the use of category theory. It goes further than that, there is a deep relationship between the fundamental concepts of category theory (objects and morphisms) and physics, logic, and computation—all of which are necessary to assure the behavior of cyber-physical systems (table 1.1). In this dissertation, we attempt to bring those ideas to the engineering of cyber-physical systems.

The general problem with compositionality is that is it subtle and finicky and does not always work in orthodox ways. This limitation can be partially overcome with the mechanization of the theory presented in this dissertation. Work in human-computer interaction and the increasing user friendliness of theorem provers give us indication that it is possible to enforce composition in new modeling languages regardless of how subtle doing so is when done with pen and paper. This does not necessarily solve the fact that composition, as is with all engineering design and code, is also an art. Therefore, the same problem persists as we have seen in programming. The theoretical tools are there for us to make modular programs but it is up to us to use them correctly.

Table 1.1: Category theory is an appropriate tool for studying the flow of information (adapted from Baez and Stay [12]).

| Category Theory | Physics | Logic | Computation |
|---|---|---|---|
| object | system | proposition | data type |
| morphism | process | proof | program |

## 1.5 Hypothesis

We thus far have described the problems we will consider in the abstract. It is infeasible for one piece of work to address all the problems that were presented above completely. The promise is that this dissertation touches on several different problems in the design of cyber-physical systems, including models of computation, composability, assurance for safety and security (figure 1.3), but in order to do so it borrows already developed theory. We do not see this as a limitation of the approach but rather as a benefit, in the sense that we know that the distinct theoretical models are already vetted and have been used in a practical setting. On top of these methods we provide a framework of formal composition that uses the algebras of each individual theory to provide vertical composition.

The *hypothesis* of this dissertation is that a compositional cyber-physical systems theory is desireable both because it transforms existing theory into one common language of composition within each mathematical model and because this common language can then be used to unify the associated model views across mathematical models. Concretely we can describe the potential benefits of such a theory through a number of desireable properties that stem from the previous sections.

| | |
|---|---|
| property 1 | Provide a common language for different models used in the design of cyber-physical systems, for example, models used in the leaf nodes of figure 1.3. |
| property 2 | Use that common language specifically to both prove and illustrate how different formalisms that describe a subset of requirements, system behaviors, and system architectures can be unified; that is can vertically compose in addition to horizontally. |
| property 3 | Elucidate how these horizontal and vertical composition rules can be practically used in the context of safety and security. |

We further hypothesize that these properties naturally emerge through a categorical theory of cyber-physical systems. To provide evidence that this hypothesis is correct we marry contract theory, linear time-invariant systems, and graph architectural models using category theory. We use these theories to construct safety contracts and apply component-based security analysis using a running example of an unmanned aerial vehicle (UAV).

The main result of the dissertation is twofold. First, we develop different algebras for requirements, system behaviors, and system architectures that preserve composition within each of the individual formalisms. Second, we develop the theoretical underpinning using fundamental concepts from category theory to provide verified composition or otherwise unify those distinct algebras in a way that changes in one algebra reflect some change on the other algebra. This theoretical machinery is both algebraic (and therefore mechanizable algorithmically) and can be represented using graphical tools (which are a familiar syntax to systems engineers) with the future goal of developing new modeling languages that preserve the property of composition across all modeling paradigms required to effectively engineer cyber-physical systems.

Using category theory to model systems has the potential of unifying even further paradigms that implement composition and often are as expressive as the associated categorical structure.

**Logic.** Logic-based approaches to cyber-physical system design, such as contract composition in the KeYmaera X tool [121] or hybrid Event-B [24]. Expressing design problems in logical inference rules further assist with the separation of computation and physics present in cyber-physical systems [111]. Logic has been consistently used to combat the problem of specification mismatching throughout the lifecycle of the system and its associated models [128]. Cyber-physical systems have also not gone through their "proofs as programs" paradigm shift [26] and, therefore, significant development can happen in this area by using other flavors of logic than linear temporal logic, such as intuitionistic logic or linear logic. Type theory [72] is also unexamined in the context of cyber-physical system design and it could arguably be used instead of category theory in this work.

**Hybrid systems.** Hybrid systems is a well-established version of computational dynamical systems theory [7] or more recently model interfaces [141]. Ames (as well as Tabuada et al. to some extent [162, 163]) did develop a categorical theory of hybrid systems [9] and therefore the results of this program are immediately transferable, in the sense that instead of linear time-invariant systems we could very well use the algebra of hybrid systems instead.

**Contracts.** The theory of contracts has had significant development, especially as applied to cyber-physical systems [30, 149], including notions of contract composition [136]. Recently there are also concrete applications in the form, for example, of a toolkit on top of SysML [50], which will make contracts increasingly accessible to system designers. Contracts have been implemented as an end-to-end requirements engineering framework, but more importantly have also been merged with linear temporal logic specifications that can compile down to contracts [134]; this idea could also be implemented into our compositional cyber-physical systems theory. Examples of synthesis from a contract-based design spec-ification [62], show that it is possible to use our generalized version of contracts to adapt control synthesis tools [76, 145] with our notion of modeling and simulation. Therefore, we would be able to not only have composition among requirements, system behaviors, and system architectures but we would also be able to produce a possible implementation that is compositionally constrained at any given level; this would represent an improvement over approaches that only consider the compositional verification of architecture models [45].

**Coalgebras.** Algebras are one view of the behavioral approach to systems. Another is that of coalgebras [148], which can often better mathematically describe *processes*. Both algebras and coalgebras can, for example, represent automata or streams. The main benefit of coalgebras is that bisimulation in process formalisms, say automata or streams, is precise enough for an algorithm to emerge immediately from the proof itself.

People think of abstraction as stripping away meaning.
Abstraction does the opposite—it enriches meaning.

— Francis Su

# 2 Category theory for the mathematical engineer

In recent years category theory has gained attention in engineering, particularly how it can be used at a practical setting [35, 88, 167, 171, 173]. Often, the question that might plague the rogue engineer that decides to try category theory relate to its utility. Let's get it out of the way, the promise of *applied* category theory is twofold. First, it claims to assist with *unification*. Supporting this statement is straightforward, functorial semantics allow us to manipulate data in ways that set theory or logic cannot but also still in ways that set theory and logic can. Second, it claims to provide *scalability*. This statement is harder to support and it will require future work, from developing tools familiar to engineers, to prototype implementations of categorical features and finally to validation through case studies.

Instead of describing category theory in full, we want the reader to focus on the categorical toolkit used in the rest of the dissertation so we will only cover what is later used directly. However, we refer to the many formal treatments on the topic, including books such as Fong and Spivak [54], throughout.

## 2.1 Categories relate different concepts

The bread-and-butter of category theory are

1. categories (hypergraph structures),
2. functors (morphisms between categories), and

3. natural transformations (morphisms of functors).

**Definition 2.1.1** (Category). A *category* **C** is composed of the following data

**objects** a collection of *objects*, denoted obj **C**

**morphisms** for each pair of objects $A, B$, a collection of *morphisms* from $A$ to $B$, denoted $\mathrm{Hom}_{\mathbf{C}}[A, B]$ [1]

**identity** for each object $A$, a morphism $A \xrightarrow{\mathrm{id}_A} A$

**composition** for each $A, B, C$ objects, a *composition* [2] operation

$$\circ_{A,B,C} : \mathrm{Hom}_{\mathbf{C}}[B, C] \times \mathrm{Hom}_{\mathbf{C}}[A, B] \to \mathrm{Hom}_{\mathbf{C}}[A, C].$$

To represent a category this data has to also respect the following *relations* for each $A \xrightarrow{f} B$, $B \xrightarrow{g} C$, and $C \xrightarrow{h} D$:

$$\mathrm{id}_B \circ f = f \qquad f \circ \mathrm{id}_A = f \qquad (h \circ g) \circ f = h \circ (g \circ f),$$

meaning that composing with the identity function either from the left or the right recovers the function itself (unitality) and the order of operations when composing functions does not matter as long as the order of operands is unchanged (associativity).

The simplest example of category, and the most important for engineering applications, is **Set**, the category whose objects are sets and morphisms are functions between them. For each set $A$, $\mathrm{id}_A$ is the identity function from $A$ to itself and composition is function composition. The category of sets and functions is not merely a way to manipulate a set of elements. It is, instead, a much more fundamental way of seeings sets as a whole where morphisms represent all functions that can take place between any numbers of sets.

[1] A morphism $f$ in $\mathrm{Hom}_{\mathbf{C}}[A, B]$ is usually denoted as $A \xrightarrow{f} B$.

[2] We usually omit the subscripts and just write $g \circ f$ to denote composition. Composition is also denoted diagrammatically as $A \xrightarrow{f} B \xrightarrow{g} C$. In the diagrammatic case, composition of morphisms is left-to-right instead of right-to-left when we syntactically write $g \circ f$.

Plenty of familiar structures in mathematics can be seen as categories. For instance, a monoid can be seen as a category with only one object, call it $*$. Any element of the monoid is interpreted as a morphism $* \to *$. The identity on $*$ is the monoid unit, and composition is the monoid operation. Indeed, categories can be thought of as generalized monoids with many objects. This might seem like an obfuscating example but it is actually a simple algebraic structure and in practice monoids equipped with actions recover the formalisms we use to describe any transition system.

Other familiar examples of categories include: groups and their homomorphisms, vector spaces and linear maps between them, topological spaces and continuous functions, and the category of states and transitions between them [49].

In general, the idea is that morphisms are transformations that preserve some properties possessed by the objects; properties that we care about.

A standard diagrammatic way to express composites is $X \xrightarrow{f} Y \xrightarrow{g} Z$ and equations via commutative diagrams of the following form.

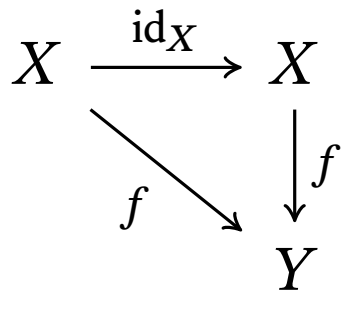

stands for $f \circ \mathrm{id}_X = f$

**Definition 2.1.2** (Isomorphism). Given a category **C**, a morphism $A \xrightarrow{f} B$ is called an *isomorphism* if there is a morphism $B \xrightarrow{f^{-1}} A$ such that the following square *commutes*, meaning that any two paths sharing the same start and end points define the same morphism.

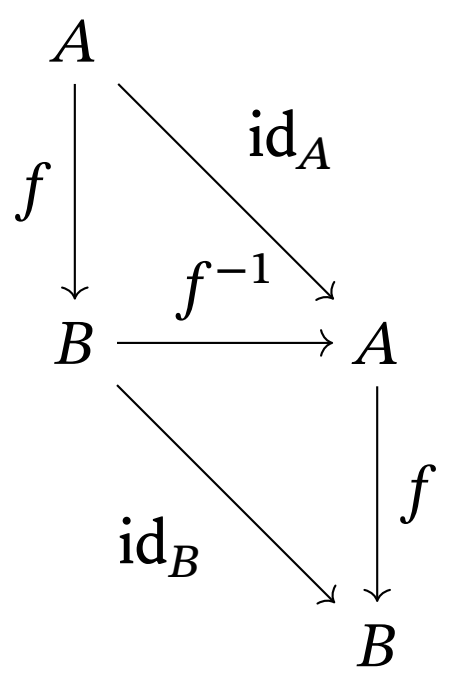

If there is an isomorphism $A \xrightarrow{f} B$, then we say that $A$ and $B$ are *isomorphic*, and write $A \simeq B$.

Two isomorphic objects in a category behave exactly in the same way *only* with respect to the structure captured by the category. For instance, $\mathbb{R}$ and $\mathbb{R}^2$ can be seen both as objects in **Set**, the category of sets and functions, and as objects in $\mathbf{Vect}_{\mathbb{R}}$, the category of real vector spaces and linear maps between them. They are isomorphic in **Set**, since bijective functions satisfy isomorphism (definition 2.1.2) and $\mathbb{R}$ and $\mathbb{R}^2$ are in bijection. Nevertheless, they are *not* isomorphic in $\mathbf{Vect}_{\mathbb{R}}$, since an isomorphism in this category has to be a *linear* bijection, which in particular has to preserve dimension. What is happening here is that since $\mathbf{Vect}_{\mathbb{R}}$ keeps track of more structure than what **Set** does, our ability to tell objects apart in $\mathbf{Vect}_{\mathbb{R}}$ is *finer* than in **Set**.

Functors are morphisms between categories. Functors too should preserve the properties we care about when we study categories. The properties that we care about in categories are similar to those of functors; merely identities and composition (definition 2.1.1). This gives us some insight to the definition of functors.

**Definition 2.1.3.** Given categories **C** and **D**, a functor $\mathbf{C} \xrightarrow{F} \mathbf{D}$ consists of the following information.

- A mapping obj $\mathbf{C} \xrightarrow{F}$ obj $\mathbf{D}$ (figure 2.1a).
- For each $A, B \in$ obj $\mathbf{C}$, a mapping (figure 2.1b)

$$\mathrm{Hom}_{\mathbf{C}}[A, B] \xrightarrow{F} \mathrm{Hom}_{\mathbf{D}}[FA, FB].$$

- We moreover require that the following two equations hold

$$F(\mathrm{id}_A) = \mathrm{id}_{FA} \qquad F(g \circ f) = F(g) \circ F(f).$$

We will often omit parentheses when not strictly necessary. As such, we will write $Ff$ instead of $F(f)$ to denote the application of a functor $F$ to a morphism $f$.

Functors are structure preserving maps that allow us to connect different model types by defining the particular semantics of transformations that are necessary to change the domain of discourse within a particular category, say **Set** → **Set**, or between different categories, say **C** → **Set**.

For example, there is a functor from $\mathbf{Vect}_{\mathbb{R}}$ to **Set** that *forgets* structure. Any real vector space is mapped to its underlying set, and any linear map between them is mapped to its underlying function between sets.

As we will see shortly, a useful functor in category theory is the hom-functor: fix an object $A$ in a category **C**. Then we can define a functor

$$\mathbf{C} \xrightarrow{\mathrm{Hom}_{\mathbf{C}}[A,-]} \mathbf{Set}$$

Which sends every object $B$ of **C** to the set of morphisms $\mathrm{Hom}_{\mathbf{C}}[A, B]$. A morphism $B \xrightarrow{g} C$ is sent to the function

$$\mathrm{Hom}_{\mathbf{C}}[A, B] \xrightarrow{\mathrm{Hom}_{\mathbf{C}}[A,g]} \mathrm{Hom}_{\mathbf{C}}[A, C].$$

Which acts by postcomposing; that is, a morphism $A \xrightarrow{f} B$ is sent to $A \xrightarrow{f} B \xrightarrow{g} C$. Functoriality follows from the composition and identity axioms of **C**.

The next fundamental concept of category theory is *naturality*.

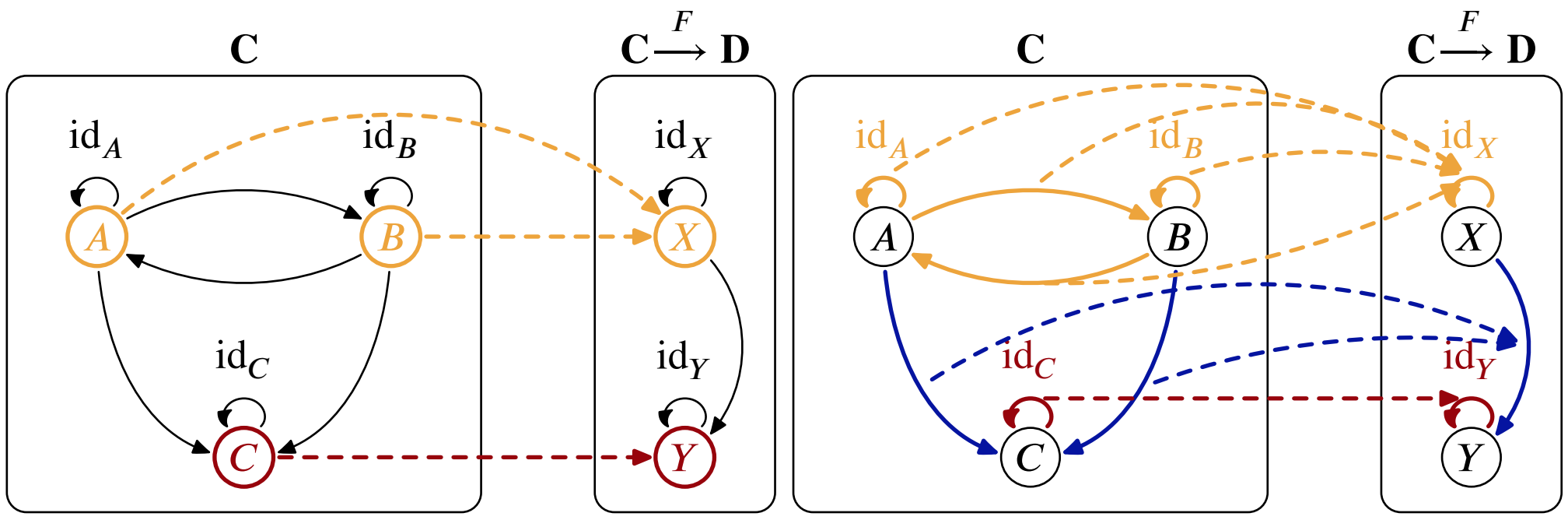

(a) A functor maps objects.

(b) A functor also maps morphisms.

Figure 2.1: A functor is a structure-preserving map.

**Definition 2.1.4** (Natural Transformation)**.** Given functors $\mathbf{C} \xrightarrow{F,G} \mathbf{D}$, a *natural transformation* $F \stackrel{\eta}{\Rightarrow} G$ consists, for each object $A$ of $\mathbf{C}$, of a morphism $FA \xrightarrow{\eta_A} GA$ in $\mathbf{D}$ such that, for each morphism $A \xrightarrow{f} B$ in $\mathbf{C}$,

$$\eta_B \circ Ff = Gf \circ \eta_A$$

This is often expressed diagrammatically by saying that the following square has to commute.

$$\begin{array}{ccc} FA & \xrightarrow{\eta_A} & GA \\ {\scriptstyle Ff}\downarrow & & \downarrow{\scriptstyle Gf} \\ FB & \xrightarrow[\eta_B]{} & GB \end{array}$$

The information contained in a natural transformation or algebra map $F \stackrel{\eta}{\Rightarrow} G$ is enough to guarantee that any commutative diagram made of images of things in $\mathbf{C}$ via $F$ can be turned into a diagram of images of things in $\mathbf{C}$ via $G$ without breaking commutativity. The commutativity condition of $\eta$ means that it doesn't matter in which order we will *walk through* these arrows, the result will be the same.

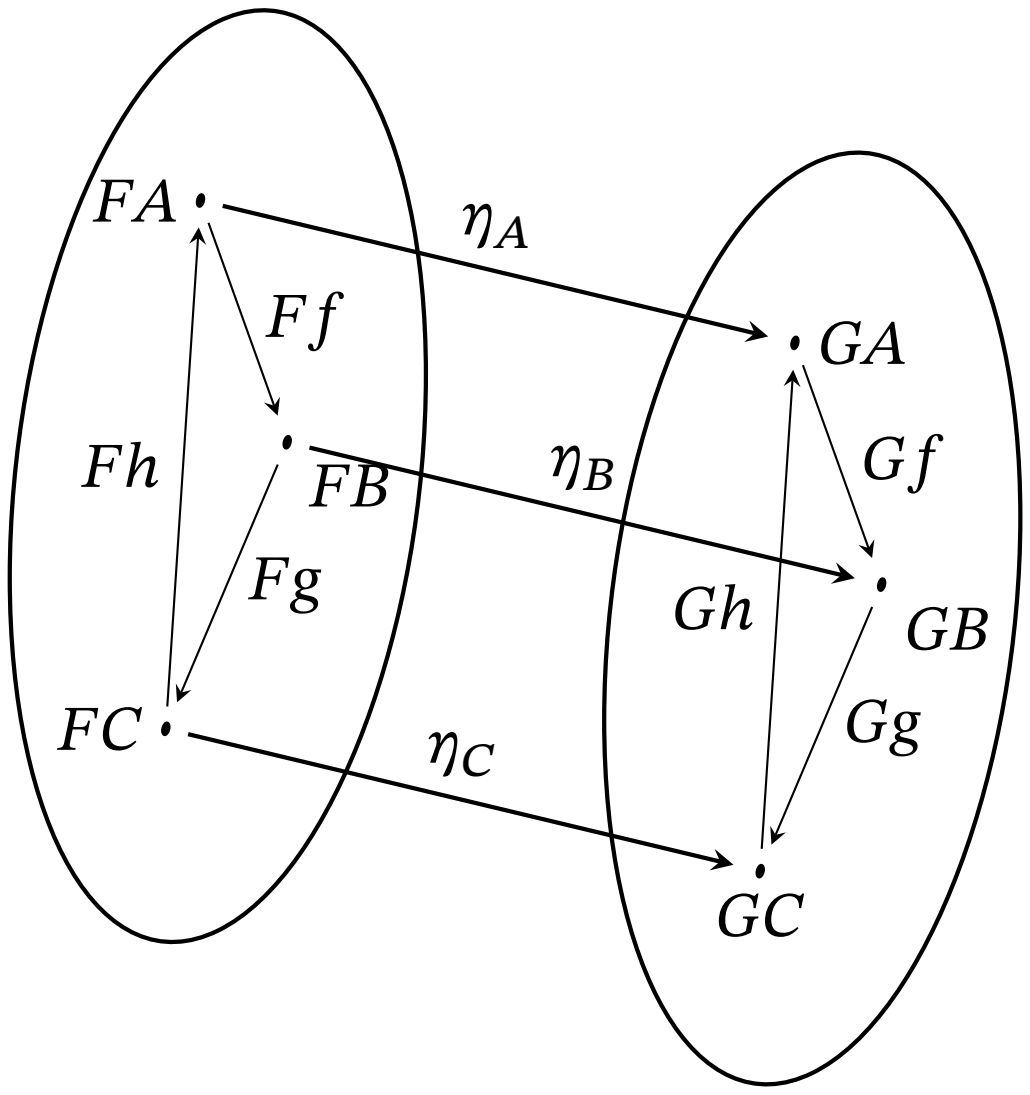

The Yoneda lemma [33] is arguably the most important result in category theory. Understanding the Yoneda lemma only requires the definitions above; categories and functors and natural transformations. It follows from a clever observation: consider a functor $\mathbf{C} \xrightarrow{F} \mathbf{Set}$ and an object $A$ of $\mathbf{C}$. By definition, any natural transformation $\mathrm{Hom}_{\mathbf{C}}[A, -] \overset{\eta}{\Rightarrow} F$ has as components functions between *sets* $\mathrm{Hom}_{\mathbf{C}}[A, B] \xrightarrow{\eta_B} FB$. Moreover, for each $A \xrightarrow{f} B$, the following diagram must commute because $\eta$ is a natural transformation.

$$\begin{array}{ccc} \mathrm{Hom}_{\mathbf{C}}[A, A] & \xrightarrow{\eta_A} & FA \\ {\scriptstyle \mathrm{Hom}_{\mathbf{C}}[A, f]}\downarrow & & \downarrow{\scriptstyle Ff} \\ \mathrm{Hom}_{\mathbf{C}}[A, B] & \xrightarrow[\eta_B]{} & FB \end{array}$$

**Lemma 2.1.1** (Yoneda Lemma)**.** *There is a bijection*

$$\mathrm{Nat}[\mathrm{Hom}_{\mathbf{C}}[A, -], F] \simeq FA$$

*where* $\mathrm{Nat}\,[F, G]$ *denotes the set of natural transformations between any two functors* $F, G$*. The homomorphism,* $\mathrm{Hom}_{\mathbf{C}}\,[-,-]$*, is a contravariant functor in the first component and a covariant functor in the second. It can be proven that the bijection is natural in F and A.*

Isomorphic objects in a category behave as if they were the same. Repetitively applying Yoneda lemma results in

$$FA \simeq \mathrm{Nat}\,[\mathrm{Hom}_{\mathbf{C}}\,[A, -]\,, F] \simeq \mathrm{Nat}\,[\mathrm{Hom}_{\mathbf{C}}\,[B, -]\,, F] \simeq FB$$

and, hence, $A \simeq B$ if and only if $FA \simeq FB$ for any $\mathbf{C} \xrightarrow{F} \mathbf{Set}$. So, with respect to whatever it is that we want to capture by defining a category $\mathbf{C}$, the Yoneda lemma affirms that we can *completely* characterize an object $A$ by studying its images $FA$ for any functor $F$ to $\mathbf{Set}$. We shall see how the Yoneda lemma can be used to test equivalence between system representations by using its essence, *if two objects agree under any possible test we can perform, then they behave the same.*

There are also different types of categories. The ones useful to engineers are called *monoidal* categories. Monoidal categories enable the study of parallel processes by equipping a tensor product to the base category (definition 2.1.1). Additionally, for the study of compositional structures in engineering there are several graphical representation of processes within a monoidal category [153].

**Definition 2.1.5** (Monoidal Category)**.** A *monoidal* category $\mathbf{V}$ is a category that comes equipped with a *monoidal product functor*

$$\mathbf{V} \times \mathbf{V} \xrightarrow{\otimes} \mathbf{V}$$

which can be thought of as multiplication of objects and morphisms, or more broadly as doing operations in parallel. We require that, for any objects $X, Y, Z$,

$$(X \otimes Y) \otimes Z \simeq X \otimes (Y \otimes Z),$$

meaning that multiplying objects in any order gives isomorphic results.

We also require the existence of a distinguished object $I$ of $\mathbf{V}$, called *monoidal unit*, such that

$$I \otimes X \simeq X \simeq X \otimes I$$

that is, $I$ acts like an identity for this multiplication.

All this data must satisfy certain axioms [75].

Widely used examples of monoidal categories include $(\mathbf{Set}, \times, \{*\})$ with the cartesian product of sets and the singleton, as well as $(\mathbf{Lin}, \otimes_k, k)$ with the tensor product of $k$-vector spaces. Moreover, $(\mathbf{Cat}, \times, \mathbf{1})$ with the cartesian product of categories (similarly to that of sets) and the unit category with a single object and single arrow forms a monoidal category. In fact, all these are examples of *symmetric* monoidal categories, which come further equipped with isomorphisms

$$X \otimes Y \cong Y \otimes X,$$

for example, for two sets $X \times Y \cong Y \times X$ via the mapping $(x, y) \mapsto (y, x)$.

Now that we defined monoidal categories, that are nothing but categories together with some additional structure for modeling parallel processes, we have to refine our notion of functor accordingly such that it works in the parallel case of operations too.

**Definition 2.1.6** (Lax Monoidal Functor)**.** A *lax monoidal* functor between two monoidal categories $F : (\mathbf{V}, \otimes_{\mathbf{V}}, I_{\mathbf{V}}) \to (\mathbf{W}, \otimes_{\mathbf{W}}, I_{\mathbf{W}})$ is a functor that preserves the monoidal structure in a lax sense (meaning not up to isomorphism). It comes equipped with two morphisms. A unitor

$$F(I_{\mathbf{V}}) \xrightarrow{\phi_0} I_{\mathbf{W}}$$

and a laxator

$$FX \otimes_{\mathbf{W}} FY \xrightarrow{\phi_{X,Y}} F(X \otimes_{\mathbf{V}} Y)$$

with $X, Y$ ranging over the objects of **V**, that express the relation between the image of the tensor and the tensor of the images inside the target category **W**. These also adhere to certain axioms [75].

Monoidal categories and lax monoidal functors also form a category of their own, denoted $\mathbf{MonCat}_{\text{lax}}$.

## 2.2 Wiring diagrams are an algebraic graphical language

We can work within a monoidal category diagrammatically by, for example, using the category **W** of *labeled boxes* and *wiring diagrams*. Wiring diagrams are a particularly interesting example of the congruence between category theory and model-based design. Wiring diagrams have been independently created by category theorists [146, 157, 165] but surprisingly look and *feel* similar to engineering block diagrams used as the basis diagrammatic framework for modeling, for example, the unified modeling language (UML), the systems modeling language (SysML), and a variety of tools from Mathworks including Simulink. These types of diagrams are increasingly part of various research directions in CPS, for example the Ptolemy project [36] or Möbius [115]. Engineering is a discipline where diagrammatic reasoning has long been considered an important element in managing complexity. But several challenges persist, for example using SysML for the analysis of systems designs means a scarcity of simulation capabilities, an increased modeling effort to capture different views of the system, and the need to maintain all these differing views concurrently even as they evolve asynchronously. While the approach using wiring diagrams has little tool support currently, as an intellectual framework they overcome these limitations by augmenting this diagrammatic reasoning with stronger mathematical semantics.

Additionally, the algebraic backend of wiring diagrams overcomes several challenges that arise in visual programming. Particularly, in Myers's seminal work [122] on taxonomies of visual programming, it is argued that visual programming languages have a number of issues in practice, some of which are enumerated in the following points.

| | |
|---|---|
| point 1 | the visual representation is always significantly larger than the text representation they replace |
| point 2 | visual languages lack formal specifications |
| point 3 | visualizations often are poor representations of the actual data |
| point 4 | implementations lack portability of programs |

Instead, within compositional cyber-physical systems theory, the wiring diagram formalism is grounded on preserving portability through interoperability of wiring diagrams and their algebras. We can equip several algebras to our wiring diagram formalism and, therefore, there is a precise mathematical *and* computational interpretation of those diagrams. This means that we can visualize the syntax and semantics of the particular model we are working with both with diagrammatic reasoning but also through a textual, mathematically precise definition. Through this congruence between diagrammatic reasoning and computational algebra it is possible to create hybrid languages, where at some level might look like SysML diagrams but at another might look like formal verification languages like the architecture analysis & design language (AADL), and therefore being able to partially combat the limitations of visual modeling languages.

Informally, the objects of this category are to be thought of as empty placeholders for processes, so far only specifying the types of the input and output data that they may receive. For example, an object $X$ is diagrammatically depicted as

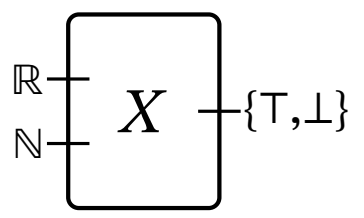

A process that can later be positioned inside this box is, for example, the function

$$f(r,n) = \begin{cases} \top & \text{if } r = n \\ \bot & \text{if } r \neq n \end{cases}.$$

To begin with, however, these boxes are uninhabited: they merely represent the architecture of a possible system. The two input wires above can be represented by a single wire typed $\mathbb{R} \times \mathbb{N}$.

These interfaces, with finitely many input and output wires along with their associated types, are essentially the building blocks for forming larger interfaces from smaller ones, and this is what is captured by the morphisms in the category **W**. For example, suppose $\mathbb{C}\!-\!\boxed{Y}\!-\!\mathbb{R}$ is another box. Intuitively, since the type of the output wire of $Y$ matches the type of one of the input wires of $X$, they could be linked along that wire

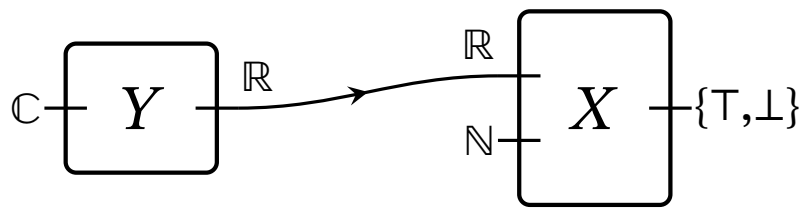

to provide a new interface that receives two inputs, one complex and one natural number, and outputs a true or false:

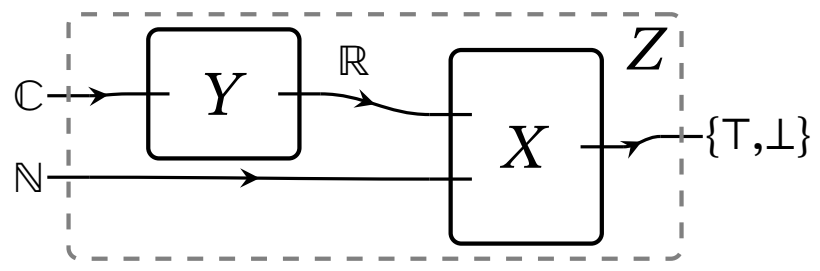

While combining interfaces together, we want to be able to express not only the new interface they form, which in the above example is $\begin{smallmatrix}\mathbb{C}\\ \mathbb{N}\end{smallmatrix}\!\boxed{Z}\!\{\top,\bot\}$, but also keep track of the intermediate wires. In our envisioned category, this will be expressed as a morphism from "$X$ and $Y$" into $Z$.

**Definition 2.2.1** (Wiring Diagram Category)**.** There is a category **W** with pairs of sets $X = (X_{\text{in}}, X_{\text{out}})$ as objects, thought of as the products of types of the input and output ports of an empty box as in

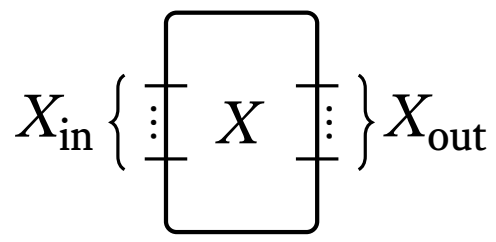

A morphism $f : X \to Y$ in this category is a pair of functions (in reality, these are not just arbitrary functions, rather generated by projections, diagonals and switchings [156, Def. 3.3])

$$\begin{cases} f_{\text{in}} : X_{\text{out}} \times Y_{\text{in}} \to X_{\text{in}} & \text{(2.1a)} \\ f_{\text{out}} : X_{\text{out}} \to Y_{\text{out}} & \text{(2.1b)} \end{cases}$$

thought of as providing the flow of information in a picture as follows which illustrates in diagrammatic view the system of equations (2.1a) and (2.1b), where the forks correspond to duplication and black bullets correspond to discarding. Information going through those wires can be anything insofar as the types match between ports. The wires of the external input ports $Y_{\text{in}}$ can only go to the internal input ports $X_{\text{in}}$ (equation (2.1a)), whereas the wires of the internal output ports $X_{\text{out}}$ can either be directed to the external output ports $Y_{\text{out}}$ (equation (2.1a)) or fed back to the internal input ports $X_{\text{in}}$ (equation (2.1a)).

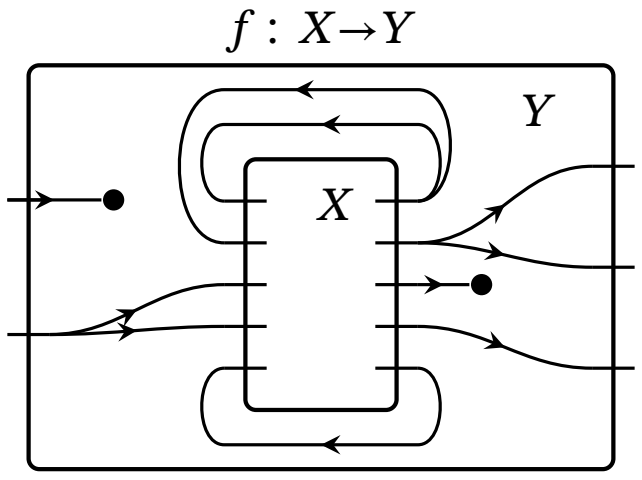

This is a monoidal category, where the tensor product of any two labelled boxes $X$ and $Y$ is $X \otimes Y = (X_{\text{in}} \times Y_{\text{in}}, X_{\text{out}} \times Y_{\text{out}})$ that represents the parallel placement of the two

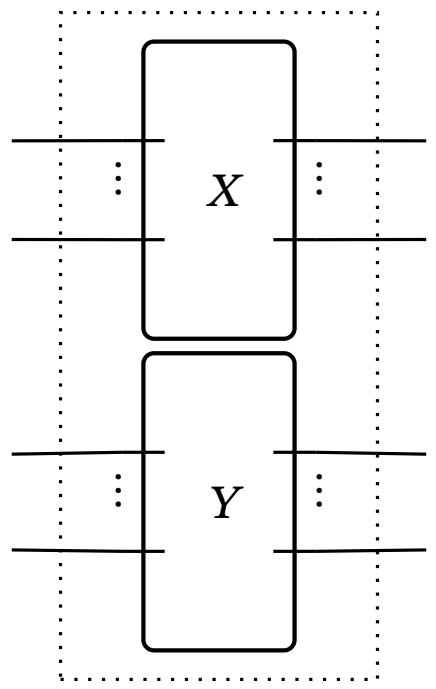

(2.2)

with input and output the (cartesian) product of the respective sets.

For simplicity, we often abstract the pictures for objects, morphisms and tensor in **W** to

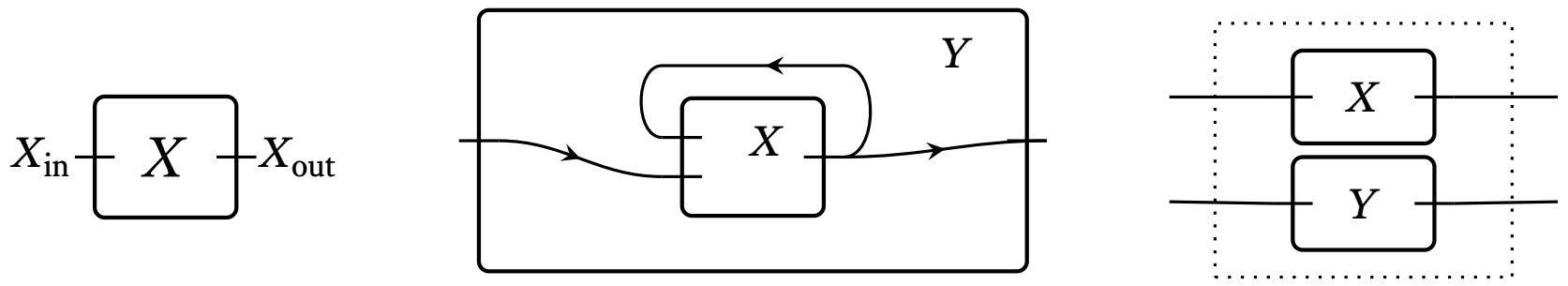

The composition in this category zooms two levels deep. For $f = (f_{\text{in}}, f_{\text{out}}) : X \to Y$ and $g = (g_{\text{in}}, g_{\text{out}}) : Y \to Z$ in the form of (2.1a), the new wiring diagram

$$g \circ f : X \to Z$$

consists of the functions

$$\big((g \circ f)_{\text{in}} : X_{\text{out}} \times Z_{\text{in}} \to X_{\text{in}}, (g \circ f)_{\text{out}} : X_{\text{out}} \to Y_{\text{out}}\big)$$

given by

$$(g \circ f)_{\text{in}}(x', z) = f_{\text{in}}(x', g_{\text{in}}(f_{\text{out}}(x'), z))$$
$$(g \circ f)_{\text{out}}(x') = g_{\text{out}}(f_{\text{out}}(x')).$$

The identity morphism on $X$ is

$$(\pi_2 : X_{\text{out}} \times X_{\text{in}} \to X_{\text{in}}, \text{id}_{X_{\text{out}}} : X_{\text{out}} \to X_{\text{out}}),$$

and the axioms of a category hold. Moreover, the monoidal unit is the box $\{*\} - \boxed{I} - \{*\}$ and the axioms of a monoidal category can also be verified to hold [165].

The category **W** as defined above is really **Set**-typed or *labeled*, namely the objects and morphisms are described using sets. However, the formalism allows to label the wires with any category equipped with finite products instead of $(\mathbf{Set}, \times, \{*\})$. For example, the types could be in linear spaces $\mathbb{R}^n$ or topological spaces $(X, \tau)$ or even more general time-related categories like lists of signals expressed as sheaves on real-time intervals [152]. Not only do these different types accommodate systems with such inputs and outputs, but also often provide a passage between different models on the same system by functorially changing the types.

The construction of this category allows us to formally give meaning to arbitrary wiring diagram pictures and as a result, coherently describe interconnections. As an example, consider three processes $X$, $Y$, and $Z$ (figure 2.2a).

The involved labelled boxes are $X = (\mathbb{R}, \mathbb{R})$, $Y = (\mathbb{R}, \mathbb{R})$ and $Z = (\mathbb{R}^3, \mathbb{R})$, which connected in the depicted way form the composite interface $A = (\mathbb{R}^3, \mathbb{R})$. Although $A$'s inputs and outputs are to the "outside world," they could also potentially interconnect to other boxes themselves.

To implement the above as a morphism in the category **W**, we first "align" the boxes such that the wires follow their input and output (figure 2.2b), which then forms a morphism from the tensor product of the three boxes $X \otimes Y \otimes Z$ (the dotted box) with input $\mathbb{R}^5$ and output $\mathbb{R}^3$, to the outside box $A = (\mathbb{R}^3, \mathbb{R})$ with explicit description

$$\begin{cases} f_{\text{in}}: & \overbrace{\mathbb{R}\times\mathbb{R}\times\mathbb{R}}^{(X\otimes Y\otimes Z)_{\text{out}}}\times\overbrace{\mathbb{R}\times\mathbb{R}\times\mathbb{R}}^{A_{\text{in}}} \to \overbrace{\mathbb{R}\times\mathbb{R}\times\mathbb{R}\times\mathbb{R}\times\mathbb{R}}^{(X\otimes Y\otimes Z)_{\text{in}}}, \\ & (x,y,z,w,u,v) \mapsto (w,u,x,y,v) \\ f_{\text{out}}: & \underbrace{\mathbb{R}\times\mathbb{R}\times\mathbb{R}}_{(X\otimes Y\otimes Z)_{\text{out}}} \to \underbrace{\mathbb{R}}_{A_{\text{out}}}, \\ & (x,y,z) \mapsto z \end{cases} . \tag{2.3}$$

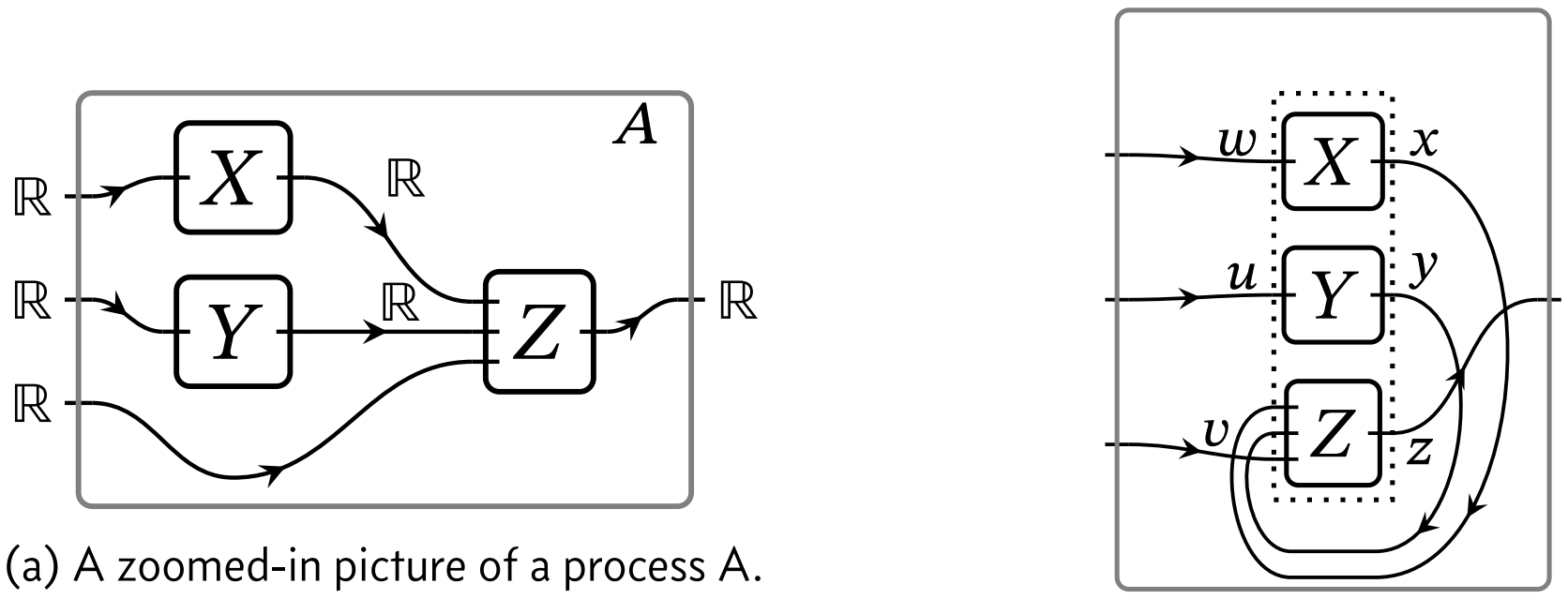

(a) A zoomed-in picture of a process A.

(b) View as a wiring diagram.

Figure 2.2: An example wiring diagram as a morphism in the category **W**.

The two functions, $f_{\text{in}}$ and $f_{\text{out}}$, specify which wires are connected to which; $f_{\text{in}}$ maps the three internal outputs $x, y, z$ together with the external inputs $w, u, v$ to the internal inputs, in the order determined by our, and $f_{\text{out}}$ projects out of the three internal outputs $x, y, z$ the third one $z$. We could choose a different alignment of the internal boxes, which would result to a different, but essentially equivalent, pair of functions. This would not affect our analysis.

To sum up, the category **W** provides a formal way of mathematically expressing any configuration at hand, with sole focus on the interconnection of vacant building blocks. The rest of this dissertation uses these fundamental constructions of categories, functors, and natural transformations in general and the wiring dia-gram formalism in particular to develop and equip algebras for verifying the composition and formal relationship between requirements, behavioral models, and system architectures.

Algebra is to the geometer what
you might call the Faustian offer.

— Michael Atiyah

# 3 Compositional systems modeling

We use the categorical formalism to unify between the disparate but necessary models used to assure correctness in cyber-physical systems, including requirements, system behaviors, and system architectures. Currently, these views are largely managed in an informal, piecemeal fashion, with no notion of formal traceability between different *types* of models, which could lead to designing and implementing systems that are ultimately unsafe due to inconsistencies between these three views. Model-based design attempts to address some of the aforementioned issues, but while it contains a notion of formal composition within each model view, it lacks a notion of formal composition *between* different model types.

The need for compositional theories to support the design of cyber-physical systems is a consistent theme in cyber-physical systems literature both in the general modeling sense [32] and particularly in contract-based design [30, 132]. First, the model-based design of cyber-physical systems can be greatly assisted by the composition of different *types* of models, which would provide traceability between the coupled physical and computational dynamics present in cyber-physical systems [5]. Second, the application of formal composition makes precise abstraction and refinement, which are necessary in model-based engineering [145]. Third, by investing in a compositional modeling paradigm we are better able to identify unsafe or uncontrolled interactions between subsystems [147]. Category theory and the wiring diagram formalism [156] provide an appealing framework to build and analyze compositional models of cyber-physical systems.

Wiring diagrams are a particularly interesting example of the congruence between category theory and model-based design. Wiring diagrams have been independently created by category theorists [146, 157, 165] but surprisingly look and *feel* similar to engineering block diagrams used as the basis diagrammatic framework for modeling, for example, the unified modeling language (UML), the systems modeling language (SysML), the generic modeling environment (GME), and a variety of tools from Mathworks including Simulink. These types of diagrams are increasingly part of various research directions in cyber-physical systems, for example the Ptolemy project [36] or Möbius [115]. Systems engineering is a discipline where diagrammatic reasoning has long been considered an important element in managing complexity. But several challenges persist, for example using SysML for the analysis of systems designs means a scarcity of simulation capabilities, an increased modeling effort to capture different views of the system, and the need to maintain all these differing views concurrently even as they evolve asynchronously. While the approach using wiring diagrams has little tool support currently, as an intellectual framework they overcome these limitations by augmenting this diagrammatic reasoning with stronger mathematical semantics.

The general phases of system design are often implicitly compositional.

**Requirements** describe the *what's* or inentions of system designers. They may include the purpose of the systems the compose the whole as well as the functions. Requirements are drafted early and modified throughout the system lifecycle and they are used to create the architecture and the components of the system.

**Specification** describes the *how's*; that is, it gives a precise definition on the *what's* and serves as the contract between customer and system designers.

**Architecture** desribes the *framework of the how's*, meaning how the system implements those specified functions. It includes a plan for the overall structure of the system that will be used later to design the components that make up the architecture.

**Realization** is the component design or selection. The component design effort builds the components in conformance to the architecture and specification, which often include both hardware and software modules.

**System integration** is a methodical process of realizing the architecture plan, including how to implement concretely the functional behavior that then can be tested for issues between requirements.

In this work we will go as far as architecture in the methodology and not consider realization or system integration. Although, we have indication that the categorical framework can assist with control and program synthesis that would tie all parts of design together, we leave that as future work.

## 3.1 Modeling cyber-physical systems compositionally

Assessing the correct behavior of cyber-physical systems requires several model views. Before discussing them, we must first clarify the meaning of the terminology that we will use. We choose to use the terminology of *requirements*, *system behavior*, and *system architecture* to describe the different diagrammatic abstractions of cyber-physical system models. We define requirements as constraints over system behavior and system architecture.[1] By system behavior we mean models of the form of automata or state space models that describe the systems states's evolution. By system architecture we mean models of candidate implementations that, in the case of cyber-physical systems, include hardware and software for the embedded system portion of the cyber-physical system and motors, control surfaces, actuators, and mechanical structure for the physical portion of the cyber-physical system. In the following formalism in general, we will view the individual diagram pictures as architecture, and the particular semantics that go into the boxes within this diagram as behavior, omitting the leading word 'system' when only discussing about the diagrammatic representation. Contracts that constraint both behavior and architecture in this sense will represent a subset of system safety requirements.

[1] Requirements don't only define constraints. There are permissible (what the system ought to do) and restrictive (what the system ought not to do) types of requirements. We can capture both in the contract framework but safety is associated with constraints so we will focus on that.

The categorical approach has the advantage of providing a *compositional* modeling and analysis, in which the composite system is completely and uniquely determined from its subsystems and their interconnections. This is achieved through the implementation of the formalism in two parts. The first is a behavior algebra that allows the hierarchical modeling between the abstraction of system behavior and system architecture, in a *zoom-in, zoom-out* approach [168], where each view may have distinct inputs and outputs. The second is a contract algebra that computes safety constraints to the system behavior.

An analogously high-level approach using monoidal categories and compositional techniques has already found success in categorical quantum mechanics [2, 44], where it has become the de facto language to describe and manipulate quantum processes diagrammatically. We posit that a similar innovation should take place in the design and assessment of safety-critical cyber-physical systems, due to the concerns raised by the intertwined nature of digital control with physical processes and the environment. We will view distinct but related system models, pertinent to assuring the correct behavior of cyber-physical systems, as *algebras* of the monoidal category of wiring diagrams.

The wiring diagram approach diverges from input-output models. While the diagrammatic syntax looks similar to such models, what is contained within the boxes need not be a mathematical function. It can instead be any sort of process, from very concrete descriptions like automata, to more abstract processes which could be deterministic or non-deterministic, to mere requirements of a mathematically unknown formula. Similarly, the arrows do not need to contain one piece of information, for example the input and output of a function; rather, arrows can carry arbitrary objects of a chosen category of types. Previous compositional modeling methods for cyber-physical systems are often limited to sets and functions or in the most general sense, relations. However, the state space of a controls system need not be the set $\mathbb{R}^n$, but could instead be a topological space like the line or circle $\mathbb{S}$. The rich interplay between topology and category theory positions category theory as a particularly good candidate for modeling dynamics, for example see Hansen and Ghrist [69] or earlier, in the more related area of hybrid systems, Ames [9] and Tabuada et al. [162, 163].

## 3.2 System behavior via algebras on the category W

The category of wiring diagrams does not populate the boxes with actual systems, for example, dynamical systems. This is instead done by developing extra structure on top of the boxes. By knowing the configuration of the component systems, the behavior of the composite system can then be uniquely determined.

Categorically, this is described as an *algebra* on $\mathbf{W}$ (definition 2.2.1), namely a lax monoidal functor $F\colon (\mathbf{W}, \otimes, I) \to (\mathbf{Cat}, \times, \mathbf{1})$. The idea is that each algebra assigns to a box $X = (X_{\text{in}}, X_{\text{out}})$ a category $FX$ of systems that can be placed in the box, and also assigns to a wiring diagram $f = (f_{\text{in}}, f_{\text{out}})$ a functor $Ff\colon FX \to FY$ that, given a system $s$ inhabiting the internal box of a wiring diagram, produces the *composite system* $F(f)(s)$ inhabiting the external box.

$$
\begin{array}{rcll}
F\colon \mathbf{W} & \longrightarrow & \mathbf{Cat} & \\
X{=}(X_{\text{in}}, X_{\text{out}}) & \longmapsto & FX & \text{subsystems category} \\
f\downarrow & & \downarrow F(f) & \text{composite system functor} \\
Y{=}(Y_{\text{in}}, Y_{\text{out}}) & \longmapsto & FY &
\end{array}
$$

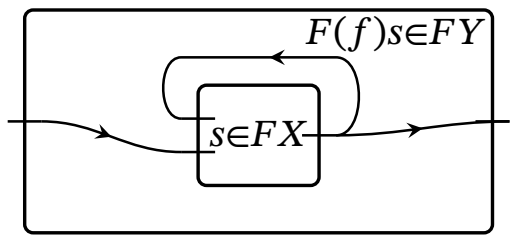

Intuitively, the object assignment $FX$ and $FY$ gives semantics to arbitrary boxes through the subsystems category while the composite system functor $Ff$ assem-bles the composite operations of the overall system behavior. The monoidal structure of the functor via the laxator $\phi_{X,Y} : FX \times FY \to F(X \otimes Y)$ ensures that for given systems inside parallely placed boxes, we can always determine a system inhabiting their tensor product

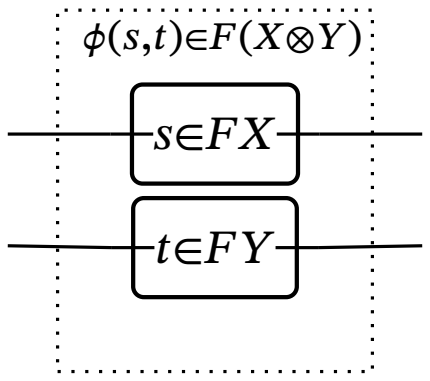

The categorical formulation allows us to use a number of algebras according to our purposes. Below we describe two such algebras of discrete dynamical systems, and later we will examine the algebra of contracts (chapter 4). There exist also other algebras, describing systems behaviors that are not like difference equations, such as, algebras for abstract total or deterministic machines [156].

The diagrammatic representation via wiring diagrams for system modeling and analysis is rather straightforward, particularly because wiring diagrams are similar to engineering block diagrams and, hence, the visual syntax is equivalent to existing cyber-physical systems design tools. However, the current diagrammatic representation is mathematically richer and more concrete—it also accounts for actual composition computations as we will see below. Another important factor specifically for cyber-physical systems is the richness of other possible algebras or semantics that one can develop and assign in these boxes backed up by the notion of the monoidal category.

### 3.2.1 MOORE MACHINES

As an illustrative example on how to develop and use the behavior algebra on an architecture in **W**, we will position the familiar Moore machines inside the boxes $X$, $Y$ and $Z$ of figure 2.2a. This is a simple yet useful demonstration of the algebra machinery because Moore machines model discrete dynamical systems. To concretely describe the systems composite, we first need to verify that Moore machines form a **W**-algebra. Indeed, there is a monoidal functor

$$\mathcal{M} : \mathbf{W} \to \mathbf{Cat}$$

which maps each $(X_{\text{in}}, X_{\text{out}})$ to the category $\mathcal{M}(X_{\text{in}}, X_{\text{out}})$ where

- objects are triples $(S, u, r)$ where $S$ is the *state space* set, $u : S \times X_{\text{in}} \to S$ is the *update function* and $r : S \to X_{\text{out}}$ is the *readout function*;
- morphisms $(S, u, r) \to (S', u', r')$ are functions $f : S \to S'$ between the state spaces that commute with the update and readout functions, namely $f(u(s, x)) = u'(fs, x)$ and $f(r(s)) = r'(fs)$.

Hence, $\mathcal{M}(X_{\text{in}}, X_{\text{out}})$ this is the category of Moore machines with fixed input and output alphabet $X_{\text{in}}$ and $X_{\text{out}}$ respectively. For example, an object of the category $M(\{0, 1\}, \{0, 1\})$ with inputs and outputs the booleans, is the "`not`" finite state machine with state space $S = \{s_1, s_2\}$ and update and readout functions depicted in the above state diagram, for example, $u(s_1, 0) = s_2$ (middle top edge) and $r(s_2) = 1$ (bottom part of $s_2$-node).

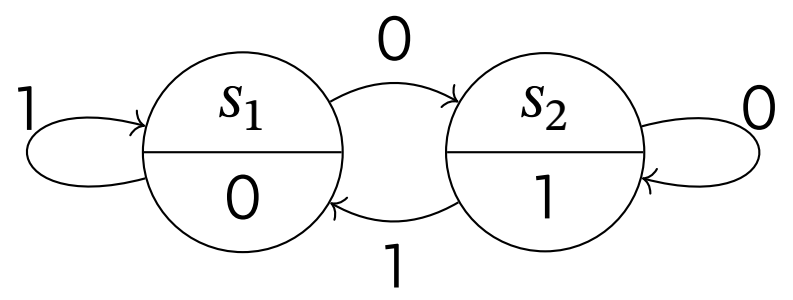

Having defined the categories of systems that can inhabit boxes in wiring diagram pictures for this specific Moore machine model, we proceed to define the composite system functor $\mathcal{M}(f) : \mathcal{M}X \to \mathcal{M}Y$, given a wiring diagram $f = (f_{\text{in}}, f_{\text{out}}) : X \to Y$. Explicitly, this functor maps a Moore machine $(S, u, r)$ with input and output $X_{\text{in}}, X_{\text{out}}$ to a Moore machine $(S, u', r')$ with input and output $Y_{\text{in}}, Y_{\text{out}}$ having the *same* state space $S$, but with new update and readout functions formed as follows

$$\begin{aligned} u' : Y_{\text{in}} \times S \to S, \qquad & u'(y,s) = u(f_{\text{in}}(y, r(s)), s) \\ r' : S \to Y_{\text{out}}, \qquad & r'(s) = f_{\text{out}}(r(s)) \end{aligned} \tag{3.1}$$

Finally, we need to specify the monoidal structure of $\mathcal{M}$ by providing functors $\mathcal{M}(X) \times \mathcal{M}(Y) \to \mathcal{M}(X \otimes Y)$. Given two Moore machines $(S_X, u_X : X_{\text{in}} \times S_X \to S_X, r_X : S_X \to X_{\text{out}})$ and $(S_Y, u_Y : Y_{\text{in}} \times S_Y \to S_Y, r_Y : S_Y \to Y_{\text{out}})$, we construct a new Moore machine with space set $S_X \times S_Y$ and update and readout functions

$$\begin{aligned} u : X_{\text{in}} \times Y_{\text{in}} \times S_X \times S_Y \to S_X \times S_Y, \qquad & u(x,y,s,t) = (u_X(x,s), u_Y(y,t)) \\ r : S_X \times S_Y \to X_{\text{out}} \times Y_{\text{out}}, \qquad & r(s,t) = (r_X(s), r_Y(t)) \end{aligned} \tag{3.2}$$

It can been be verified that with the above assignments, Moore machines satisfy the axioms of a wiring diagram algebra [156, §2.3]. We can therefore arbitrarily interconnect such systems, in particular as in figure 2.2a, and produce a new such system with a description only in terms of its components and their wiring. Suppose we have Moore machines in the boxes $\boxed{X}$, $\boxed{Y}$, $\boxed{Z}$, all with $\mathbb{R}$-valued wires, with state spaces $S_X$, $S_Y$ and $S_Z$ and update and readout functions respectively as in

$$\begin{cases} S_X \times \mathbb{R} \xrightarrow{u_X} S_X \\ S_X \xrightarrow{r_X} \mathbb{R} \end{cases} \quad \begin{cases} S_Y \times \mathbb{R} \xrightarrow{u_Y} S_Y \\ S_Y \xrightarrow{r_Y} \mathbb{R} \end{cases} \quad \begin{cases} S_Z \times \mathbb{R}^3 \xrightarrow{u_Z} S_Z \\ S_Z \xrightarrow{r_Z} \mathbb{R}. \end{cases}$$

The algebra machinery described in formulas (3.1) and (3.2) for the specific wiring diagram 2.3 produces the composite Moore machine which inhabits the outer box $\boxed{A}$ with state space $S_X \times S_Y \times S_Z$, readout function $r : S_X \times S_Y \times S_Z \to \mathbb{R}$ given by $(s,t,p) \mapsto r_Z(p)$ and update function $S_X \times S_Y \times S_Z \times \mathbb{R}^3 \to S_X \times S_Y \times S_Z$ given by

$$(s,t,p,w,u,v) \mapsto (u_X(s,w), u_Y(t,u), u_Z(p, r_X(s), r_Y(t), v)).$$

In general, the composite system is produced using the algebra machinery, no matter how complicated the systems or the wiring diagram is: given any type-respecting interconnection involving arbitrary feedback loops or parallel/serial arrangements, the monoidal functor will determine a result. Therefore, this functoriality alleviates some of the scalability issues present in other formalisms. As we will see later, often some pre-existing knowledge of the desired behavior of a composite system can possibly inform not only the components' behavior but also the choice of wiring.

### 3.2.2 LINEAR TIME-INVARIANT SYSTEMS

There is a sub-algebra of the algebra of Moore machines, for *linear time-invariant systems* or linear discrete dynamical systems per Spivak [156]. In fact, the Moore machines model is an algebra of $\mathbf{W}_{\mathbf{Set}}$, where the types of wires are sets and the wiring diagrams are given by functions, whereas the linear time-invariant system model is an algebra of $\mathbf{W}_{\mathbf{Lin}}$, where the types are given by **Lin**, the category of linear spaces and linear maps.

Explicitly, there is a monoidal functor $\mathcal{L} : \mathbf{W}_{\mathbf{Lin}} \to \mathbf{Cat}$ that assigns to any box $X_{\text{in}} \!-\!\square\!-\! X_{\text{out}}$ a category $\mathcal{L}(X_{\text{in}}, X_{\text{out}})$ of systems $(S, u : S \times X_{\text{in}} \to S, r : S \to X_{\text{out}})$ like before, but where all $S, X_{\text{in}}, X_{\text{out}}$ are linear spaces and both update and readout functions $u$ and $r$ are linear functions expressed as

$$\begin{aligned} u(s,x) &= \mathscr{A} \cdot s + \mathscr{B} \cdot x = \begin{pmatrix} \mathscr{A} & \mathscr{B} \end{pmatrix} \begin{pmatrix} s \\ x \end{pmatrix} \\ r(s) &= \mathscr{C} \cdot s \end{aligned}$$

where $\mathscr{A}$, $\mathscr{B}$ and $\mathscr{C}$ are matrices of appropriate dimension. For example, if the input, output and state spaces are $X_{\text{in}} = \mathbb{R}^k, X_{\text{out}} = \mathbb{R}^\ell$ and $S = \mathbb{R}^n$, then

$$\begin{cases} \mathscr{A} \in {}_nM_n & \text{represents a linear transformation } \mathbb{R}^n \to \mathbb{R}^n \\ \mathscr{B} \in {}_nM_k & \text{represents a linear transformation } \mathbb{R}^k \to \mathbb{R}^n \\ \mathscr{C} \in {}_\ell M_n & \text{represents a linear transformation } \mathbb{R}^n \to \mathbb{R}^\ell. \end{cases} \tag{3.3}$$

Now given an arbitrary wiring diagram $f = (f_{\text{in}}, f_{\text{out}})\colon (X_{\text{in}}, X_{\text{out}}) \to (Y_{\text{in}}, Y_{\text{out}})$ as formalized in the system of equations (2.1a), where for $Y_{\text{in}} = \mathbb{R}^{k'}$ and $Y_{\text{out}} = \mathbb{R}^{\ell'}$ both linear functions of the wiring diagram are also expressed as corresponding matrices $f_{\text{in}} = \left({}_k(\mathcal{A}^f)_\ell \quad {}_k(\mathcal{B}^f)_{k'}\right)$ and $f_{\text{out}} = {}_{\ell'}\mathcal{C}^f_\ell$, the functor $\mathcal{L}(f)$ maps some system $(S, \mathcal{A}, \mathcal{B}, \mathcal{C})$ in $\mathbb{R}^k \boxed{X} \mathbb{R}^\ell$ to the system

$$(S, \mathcal{A} + \mathcal{B} \cdot \mathcal{A}^f \cdot \mathcal{C},\ \mathcal{B} \cdot \mathcal{B}^f,\ \mathcal{C}^f \cdot \mathcal{C}) \tag{3.4}$$

in $\mathbb{R}^{k'} \boxed{Y} \mathbb{R}^{\ell'}$. The earlier-used term *sub-algebra* precisely means that this formula is a special case of equation (3.1) when the functions involved are of this specific form.

Finally, the monoidal structure of this assignment $\mathcal{L}\colon \mathbf{W_{Lin}} \to \mathbf{Cat}$ is given by functors $\mathcal{L}(X) \times \mathcal{L}(Y) \to \mathcal{L}(X \otimes Y)$ that map any two systems $(S_X, \mathcal{A}_X, \mathcal{B}_X, \mathcal{C}_X)$ and $(S_Y, \mathcal{A}_Y, \mathcal{B}_Y, \mathcal{C}_Y)$ inhabiting parallel boxes as in wiring diagram (2.2) give rise to a parallel composite system

$$\left(S_X \times S_Y, \begin{pmatrix} \mathcal{A}_X & 0 \\ 0 & \mathcal{A}_Y \end{pmatrix}, \begin{pmatrix} \mathcal{B}_X & 0 \\ 0 & \mathcal{B}_Y \end{pmatrix}, \begin{pmatrix} \mathcal{C}_X & 0 \\ 0 & \mathcal{C}_Y \end{pmatrix}\right).$$

### 3.2.3 FUNCTIONS (AS A NON-EXAMPLE)

If we would like to populate the boxes of a wiring interconnection with mathematical functions, namely assign to some $X_{\text{in}} \boxed{X} X_{\text{out}}$ a function $h\colon X_{\text{in}} \to X_{\text{out}}$, there is no natural way to make this assignment into an algebra $\mathbf{W} \to \mathbf{Cat}$. The main reason this fails is the existence of the feedback loop.

However, we can incorporate functions into other existing models, for example Moore machines. It is possible to express a function $h\colon X_{\text{in}} \to X_{\text{out}}$ as an object of $\mathcal{M}(X_{\text{in}}, X_{\text{out}})$, with state space the domain $X_{\text{in}}$ and update and readout functions $\pi_2\colon X_{\text{in}} \times X_{\text{in}} \to X_{\text{in}}$ projecting the second variable and $h\colon X_{\text{in}} \to X_{\text{out}}$ applying the said function. The resulting finite state machine at each round replaces the old input with the new input, and outputs the function application on it. Analogously, a linear function can be viewed as a linear time-invariant system if we set $\mathcal{A} = 0$ the zero matrix, $\mathcal{B} = I$ the unit matrix and $\mathcal{C} = h$ the matrix represents the given linear transformation.

As a result, functions can be indeed used to populate boxes, and wired with other functions or Moore machines they produce a composite Moore machine using the algebra $\mathcal{M} : \mathbf{W} \to \mathbf{Cat}$. It is also the case that sometimes, wiring two functions using the Moore machine algebra machinery, we end up with another function and not a more general Moore machine—this usually happens in serial-like wirings without loops.

The take away message of these examples is the following: the starting point is the category of wiring diagrams **W** with no processes inside the boxes. We can then assign the behavior of Moore machines inside the boxes using the corresponding **W**-algebra $\mathcal{M}$, or the behavior of linear discrete dynamical systems using the subalgebra of linear time-invariant systems $\mathcal{L}$, which recovers the standard model of state-space representation in control albeit with a slightly different syntax. By composing behaviors this way, we recover a block-diagonal state space model, a useful representation for modeling the control portion of cyber-physical systems.

### 3.2.4 COMPOSITIONAL STATE-SPACE MODELS

At the moment, we have illustrated a couple of examples of behavioral models that can inhabit the boxes, but the algebraic machinery is not limited by those. We could, for example, inhabit the boxes with hybrid systems, while at the same time ensuring composition. In the absence of the time element, behaviors are to be observed in an instantaneous way, for example, feedback loops do not produce delay effects. Such issues shall be tackled when the wires carry time-sensitive data.

We algebraically recover a standard controls model *compositionally* in the behavior algebra of the form

$$s_{k+1} = \mathcal{A} s_k + \mathcal{B} c_k,$$

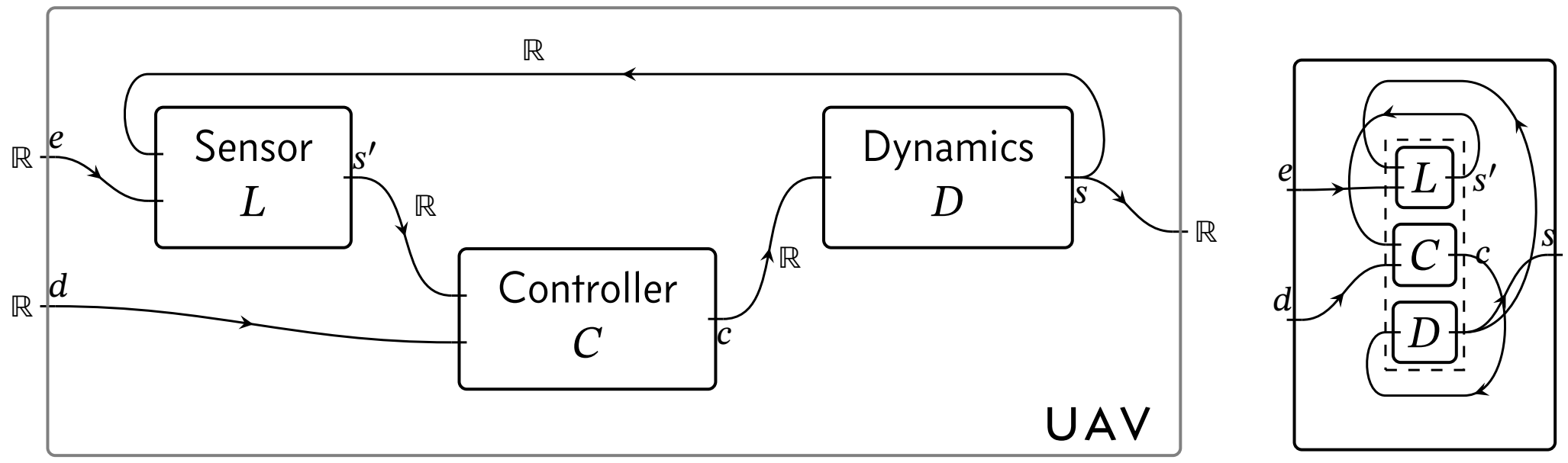

$$f_{\text{in}} : \mathbb{R}^3 \times \mathbb{R}^2 \to \mathbb{R}^5, \ (s', c, s, e, d) \mapsto (s, e, s', d, c)$$
$$f_{\text{out}} : \mathbb{R}^3 \to \mathbb{R}, \ (s', c, s) \mapsto s$$

Figure 3.1: The physical decomposition of the UAV, where $d$ denotes the desired state, $s'$ the predicted state, $c$ the control action, $s$ the current state, and $e$ the environmental inputs.

where $s_k \in \mathbb{R}^n$ is the discrete time state, $s_{k+1}$ (also denoted $\dot{s}$ or $u(s, c)$ using the earlier update function notation) is the subsequent time-step state and $c_k \in \mathbb{R}^n$ is the control signal/output, and

$$y_k = \mathscr{C} s_k + \mathscr{D} c_k$$

is the measurement, which is also in $\mathbb{R}^n$. We assume $\mathscr{D} = 0$.

We are going to illustrate the algebra machinery using longitudinal equations of motion for a fixed-winged aircraft represented in the following state-space model [119]

$$\begin{pmatrix} \dot{a} \\ \dot{q} \\ \dot{\theta} \end{pmatrix} = \begin{pmatrix} -0.313 & 56.7 & 0 \\ -0.0139 & -0.426 & 0 \\ 0 & 56.7 & 0 \end{pmatrix} \begin{pmatrix} a \\ q \\ \theta \end{pmatrix} + \begin{pmatrix} 0.232 \\ 0.0203 \\ 0 \end{pmatrix} \begin{pmatrix} \delta \end{pmatrix} \tag{3.5}$$

$$y = \begin{pmatrix} 0 & 0 & 1 \end{pmatrix} \begin{pmatrix} a \\ q \\ \theta \end{pmatrix},$$

where $a$ is the angle of attack, $q$ is the pitch rate, $\theta$ is the pitch angle and $\delta$ is the elevator deflection angle. This behavior is the *composite* one, built up from the subcomponents behavior and their wiring (figure 3.1).

Working with the linear time-invariant system algebra $\mathcal{L} : \mathbf{W_{Lin}} \to \mathbf{Cat}$ (section 3.2.2), suppose $(S_L, \mathscr{A}_L, \mathscr{B}_L, \mathscr{C}_L)$, $(S_C, \mathscr{A}_C, \mathscr{B}_C, \mathscr{C}_C)$ and $(S_D, \mathscr{A}_D, \mathscr{B}_D, \mathscr{C}_D)$ are three linear systems inhabiting the respective boxes of figure 3.1, with

$$\begin{aligned} u_L(s_L, s, e) &= \mathscr{A}_L \cdot s_L + \mathscr{B}_L \cdot (s, e) & r_L(s_L) &= \mathscr{C}_L \cdot s_L \\ u_C(s_C, d, s') &= \mathscr{A}_C \cdot s_C + \mathscr{B}_C \cdot (d, s') & r_C(s_C) &= \mathscr{C}_C \cdot s_C \\ u_D(s_D, c) &= \mathscr{A}_D \cdot s_D + \mathscr{B}_D \cdot (c) & r_D(s_D) &= \mathscr{C}_D \cdot s_D. \end{aligned}$$

Using the algebra machinery for the specific wiring diagram (figure 3.1) given by matrix transformations

$$\begin{cases} f_{\text{in}} = \begin{pmatrix} 0 & 0 & 1 & 0 & 0 \\ 0 & 0 & 0 & 1 & 0 \\ 1 & 0 & 0 & 0 & 0 \\ 0 & 0 & 0 & 0 & 1 \\ 0 & 1 & 0 & 0 & 0 \end{pmatrix} = \begin{pmatrix} {}_5(\mathscr{A}^f)_3 & {}_5(\mathscr{B}^f)_2 \end{pmatrix} \\ f_{\text{out}} = \begin{pmatrix} 0 & 0 & 1 \end{pmatrix} = \mathscr{C}^f \end{cases}$$

we can compute the composite linear dynamical system that inhabits the box UAV from formula (3.4). Its state space is $S_L \times S_C \times S_D$, and its update and readout linear functions are

$$\begin{aligned} u_{\text{UAV}} : \ & S_L \times S_C \times S_D \times \mathbb{R}^2 \to S_L \times S_C \times S_D, \\ & (s_L, s_C, s_D, d, e) \mapsto (\mathscr{A}_L s_L + \mathscr{B}_L \begin{pmatrix} \mathscr{C}_D s_D \\ e \end{pmatrix}, \\ & \qquad \mathscr{A}_C s_C + \mathscr{B}_C \begin{pmatrix} \mathscr{C}_L s_L \\ d \end{pmatrix}, \\ & \qquad \mathscr{A}_D s_D + \mathscr{B}_D \mathscr{C}_C s_C) \\ r_{\text{UAV}} : \ & S_L \times S_C \times S_D \to \mathbb{R} \\ & (s_L, s_C, s_D) \mapsto \mathscr{C}_D s_D. \end{aligned}$$

We assume, for simplicity, that the state spaces of the sensor and controller are in $\mathbb{R}^2$. Knowing that only the dynamics $D$ actually relate to the triplet $(a, q, \theta)$, we deduce that $S_D$ is in $\mathbb{R}^3$ which results in a composite state space $S_{\text{UAV}}$ in $\mathbb{R}^2 \times \mathbb{R}^2 \times \mathbb{R}^3 \cong \mathbb{R}^7$. Moreover, from the shape of the boxes according to formula (3.4) we deduce that the matrices $\mathscr{A}_L$, $\mathscr{A}_C$, $\mathscr{B}_L$ and $\mathscr{B}_C$ are two-by-two, $\mathscr{C}_L$ and $\mathscr{C}_C$ are one-by-two, whereas $\mathscr{A}_D$ is three-by-three, $\mathscr{B}_D$ is three-by-one and $\mathscr{C}_D$ is one-by-three.

Unraveling the above update and readout functions of the composite linear time-invariant system denoted by UAV, the only output of the composite system behavior is that of the dynamics $D$, because by formula (3.4)

$$\mathscr{C}_{\text{UAV}} = \mathscr{C}^f \cdot \mathscr{C}_{L\otimes C\otimes D} = \begin{pmatrix} 0 & 0 & {}_1(\mathscr{C}_D)_3 \end{pmatrix}.$$

Hence for obtaining equation (3.5), in the specific example we deduce that $\mathscr{C}_D = \begin{pmatrix} 0 & 0 & 1 \end{pmatrix}$ meaning only $\theta$ is output to the outside world as desired.

For an element of the state space $\mathbb{R}^7$ of the form $(\vec{s}_L, \vec{s}_C, \overbrace{a, q, \theta}^{\vec{s}_D})$, isolating the first two variables we obtain

$$\dot{\vec{s}}_L = \mathscr{A}_L \vec{s}_L + {}_2(\mathscr{B}_L)_2 \begin{pmatrix} \overbrace{\mathscr{C}_D \vec{s}_D}^{\theta} \\ e \end{pmatrix} \quad \text{and} \quad \dot{\vec{s}}_C = \mathscr{A}_C \vec{s}_C + {}_2(\mathscr{B}_C)_2 \begin{pmatrix} \mathscr{C}_L \vec{s}_L \\ d \end{pmatrix},$$

which could be viewed as some extra information of the composite system relating to the behaviors of the sensor and controller, not appearing in equation (3.5) but still part of the total system's behavior.

Now isolating the last three variables we obtain a description

$$\begin{pmatrix} \dot{\alpha} \\ \dot{q} \\ \dot{\theta} \end{pmatrix} = {}_3(\mathscr{A}_D)_3 \begin{pmatrix} \alpha \\ q \\ \theta \end{pmatrix} + {}_3(\mathscr{B}_D)_1 \mathscr{C}_C \vec{s}_C.$$

Comparing with the desired equation (3.5), the elevator deflection angle $\delta$ is the output of the controller $\mathscr{C}_C s_C$ which matches the physical reality, and the $\mathscr{A}_D$, $\mathscr{B}_D$ are completely determined by the composite description, namely

$$\mathscr{A}_D = \begin{pmatrix} -0.313 & 56.7 & 0 \\ -0.0139 & -0.426 & 0 \\ 0 & 56.7 & 0 \end{pmatrix} \qquad \mathscr{B}_D = \begin{pmatrix} 0.232 \\ 0.0203 \\ 0 \end{pmatrix}.$$

The remaining data for $\mathscr{A}_{L,C}, \mathscr{B}_{L,C}, \mathscr{C}_{L,C}$ depend on engineering and physical parameters based on design choices.

We were thus able to partly reverse-engineer a given composite system behavior described by equation (3.5), where for the given system architecture (figure 3.1) we completely identified the behavior of the linear time-invariant system $D$ by determining $S_D, \mathscr{A}_D, \mathscr{B}_D, \mathscr{C}_D$. We also obtained certain information about the other two subcomponents $C$ and $L$, for example, two possible behaviors could be the linear functions (such as signal concatenations) $s' = s + e$ for the sensor $L$ and the linear function $c = s' + d$ for the controller $C$. Expressing those as linear time-invariant systems, we obtain the following description

$$(S_L, \mathscr{A}_L, \mathscr{B}_L, \mathscr{C}_L) = \left( \mathbb{R}^2, \begin{pmatrix} 0 & 0 \\ 0 & 0 \end{pmatrix}, \begin{pmatrix} 1 & 0 \\ 0 & 1 \end{pmatrix}, \begin{pmatrix} 1 & 1 \end{pmatrix} \right),$$
$$u_L(\vec{s}_L, s, e) = (s \;\; e), \;\; r_L(\vec{s}_L) = s_L^1 + s_L^2$$

$$(S_C, \mathscr{A}_C, \mathscr{B}_C, \mathscr{C}_C) = \left( \mathbb{R}^2, \begin{pmatrix} 0 & 0 \\ 0 & 0 \end{pmatrix}, \begin{pmatrix} 1 & 0 \\ 0 & 1 \end{pmatrix}, \begin{pmatrix} 1 & 1 \end{pmatrix} \right),$$
$$u_C(\vec{s}_C, s', d) = (s' \;\; d), \;\; r_C(\vec{s}_C) = s_C^1 + s_C^2.$$

Then the composite system's update function is explicitly computed, using for-

mula (3.4), as

$$\begin{cases} \dot{\vec{s}}_L = \begin{pmatrix} \theta \\ e \end{pmatrix} \\ \dot{\vec{s}}_C = \begin{pmatrix} s_L^1 + s_L^2 \\ d \end{pmatrix} \\ \dot{a} = -0.313a + 56.7q + 0.232s_C^1 + 0.232s_C^2 \\ \dot{q} = -0.0139a - 0.426q + 0.0203s_C^1 + 0.0203s_C^2 \\ \dot{\theta} = 56.7q \end{cases}$$

where $s_C^1$ and $s_C^2$ are essentially the previous desired state $s'$ and input $d$, producing the deflection angle $\delta$ that appears in formula (3.5). The first two equations give the functions of $L$ and $C$ (whose states are placeholders for their inputs at each instance), whereas the last three give the dynamics $D$ as before. This shows the interplay between what the system is sensing, what its desired operating state is, and how it must react. If there were more information about the elevator deflection angle $\delta$, that would restrict the possible behaviors for $C$ appropriately.

From a more categorical perspective, the above process is summarized as follows: given an algebra $\mathcal{L}$ and a wiring diagram $f : L \otimes C \otimes D \to \textsc{uav}$ in $\mathbf{W_{Lin}}$ (figure 3.1), as well as an object of the target category $\mathcal{L}(\textsc{uav})$, namely a specific linear system as in equation (3.5) inhabiting the outside box $\textsc{uav}$, the goal is to find an object in the pre-image of the given system under the composite functor

$$\mathcal{L}(L) \times \mathcal{L}(C) \times \mathcal{L}(D) \xrightarrow{\phi_{L,C,D}} \mathcal{L}(L \otimes C \otimes D) \xrightarrow{\mathcal{L}(f)} \mathcal{L}(\textsc{uav}).$$

Such a problem certainly does not have a unique solution, namely a unique description of the three systems that form the composite, but for example in this specific case due to the form the wiring diagram, the component system

$$(S_D, \mathscr{A}_D, \mathscr{B}_D, \mathscr{C}_D)$$

was completely determined by the composite behavior. Further work would aim to shed light on possible shapes of wiring diagrams that have better identifiable solutions under algebras of interest.

## 3.3 System architecture via hierarchical decomposition

We have shown how to model a behavioral response of the system as it pertains to its physics. But how might we connect this behavioral understanding to a concrete implementation, in particular how can we zoom-in on each of the boxes to create an *architectural implementation* that will yield the wanted behaviors? Of course, there is no one solution for how to implement a cyber-physical system, so how might we categorically capture different architectural solution with an explicit relationship to the wanted set of behaviors?

Starting with a cyber-physical system from a designer point of view, we now might want to model a candidate system architecture. In general, decomposing a cyber-physical system in certain sub-components and using a specific wiring between them follows some choices based on the physical reality, experience, purpose and access to particular components at the time. Having formalized an agnostic process interface where various descriptions could live on as an object in the category of wiring diagrams $\mathbf{W}$, as well as arbitrary zoomed-in pictures of a system as a morphism in $\mathbf{W}$, we have now access to all necessary tools to realize the above system architecture design process using the general notion of a *slice category*.

For any category $\mathbf{C}$ and a fixed object $C \in \mathbf{C}$, the slice category $\mathbf{C}/C$ has as objects $\mathbf{C}$-morphisms with fixed target $C$, for example $f : A \to C, g : B \to C, \cdots$. The arrows in that category from some $f$ to some $g$ are $\mathbf{C}$-morphisms $k : A \to B$ between the domains, making the formed triangle commute, namely $g \circ k = f$. This data forms a category, which also illustrates the abstract nature of the initial category definition (definition 2.1.1): objects and arrows can be of any sort (in this case objects are morphisms of a certain shape in some fixed category, and arrows are also morphisms that satisfy a property) as long as they satisfy the axioms of a category.

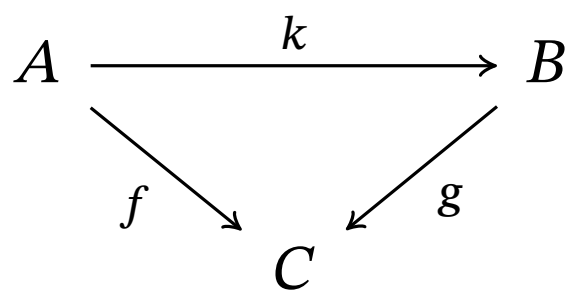

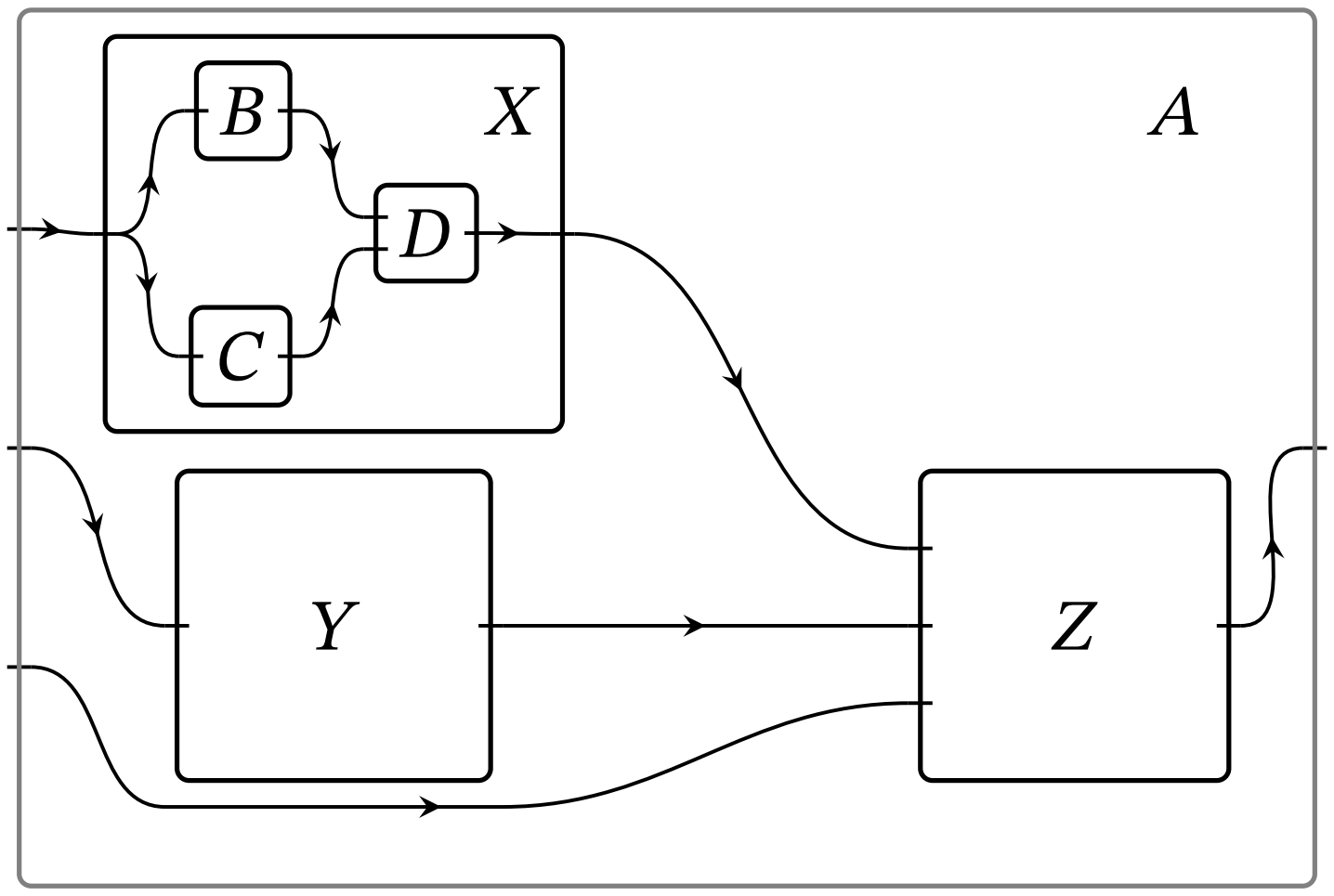

Figure 3.2: A two-level zoomed-in picture of a process A.

For our wiring diagram category $\mathbf{W}$, where a morphism $f : X \to Y$ can be thought of as an implementation of an interface $Y$ into sub-interface(s) $X$ wired in a specific manner, the slice category $\mathbf{W}/Y$ of all arrows mapping into the chosen object $Y$ contains all possible design choices available to a system engineer. This formally captures the possibility of implementing a system in multitudes of ways.

Concretely, suppose we have a system with $\mathbb{R}^3$-inputs and $\mathbb{R}$-outputs, namely inhabiting a box $\mathbb{R}\,\mathbb{R}\,\mathbb{R}\ \boxed{A}\ \mathbb{R}$. How can we decompose it into sub-processes, and how can they be interconnected to form the given system? All the possible decompositions can thus be thought of as the objects of the slice category $\mathbf{W}/A$. For example, (figure 2.2a) depicts one of these choices, namely the specific wiring diagram $f : X \otimes Y \otimes Z \to A$.

Now suppose we make another implementation choice to further decompose the

box $X$ as in

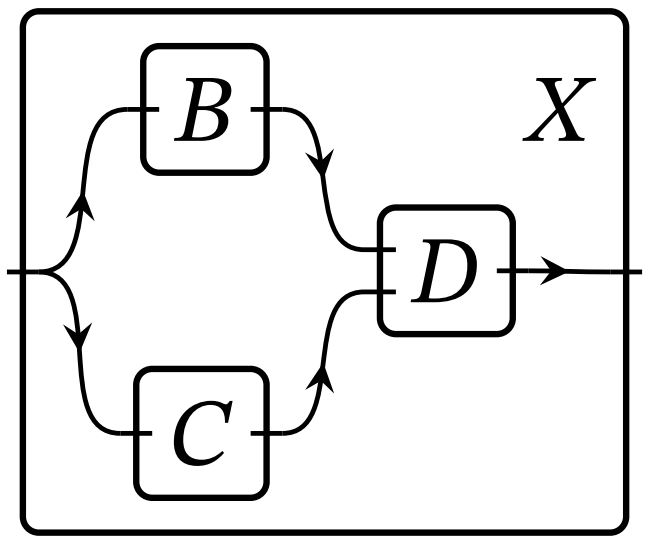

meaning we choose a specific wiring diagram $g : B \otimes C \otimes D \to X$. This constitutes another level of zoom-in for the process in $A$, at least for the subcomponent $X$ (figure 3.2). This allows us to implement the behavioral approach to modeling systems by breaking the morphisms within the category of wiring diagrams into any possible implementations. The implication of this statement is that eventually we might be able to synthesize actual hardware and software solutions using composition along different model views. In this instance we show how a flavor of formal requirements can be related to a behavioral model and finally to an architecture, thereby putting all the pieces together for an algorithm that can synthesize all this information into fabric. In terms of category theory this represents the composite morphism $f \circ (g \otimes \mathrm{id} \otimes \mathrm{id})$, the dashed arrow in the following commutative diagram

$$\begin{array}{ccc} (B \otimes C \otimes D) \otimes Y \otimes Z & \xrightarrow{g \otimes \mathrm{id} \otimes \mathrm{id}} & X \otimes Y \otimes Z \\ & \searrow \; A \; \swarrow_{f} & \end{array}$$

where the top arrow uses the morphism $g$ as the implementation of $X$ and identity morphisms on $Y$ and $Z$ (as trivial implementations), and $f$ is the earlier $A$-implementation (figure 2.2a). In the end, we can disregard the borders of the interface $X$ and map directly from the subcomponents $B \otimes C \otimes D \otimes Y \otimes Z$ to $A$ without passing through $X$ at all if desired. As a result, we are free to use hierarchical decomposition of processes for any sub-component (or for many simultaneously) and each time, these architectural choices add one more composite morphism

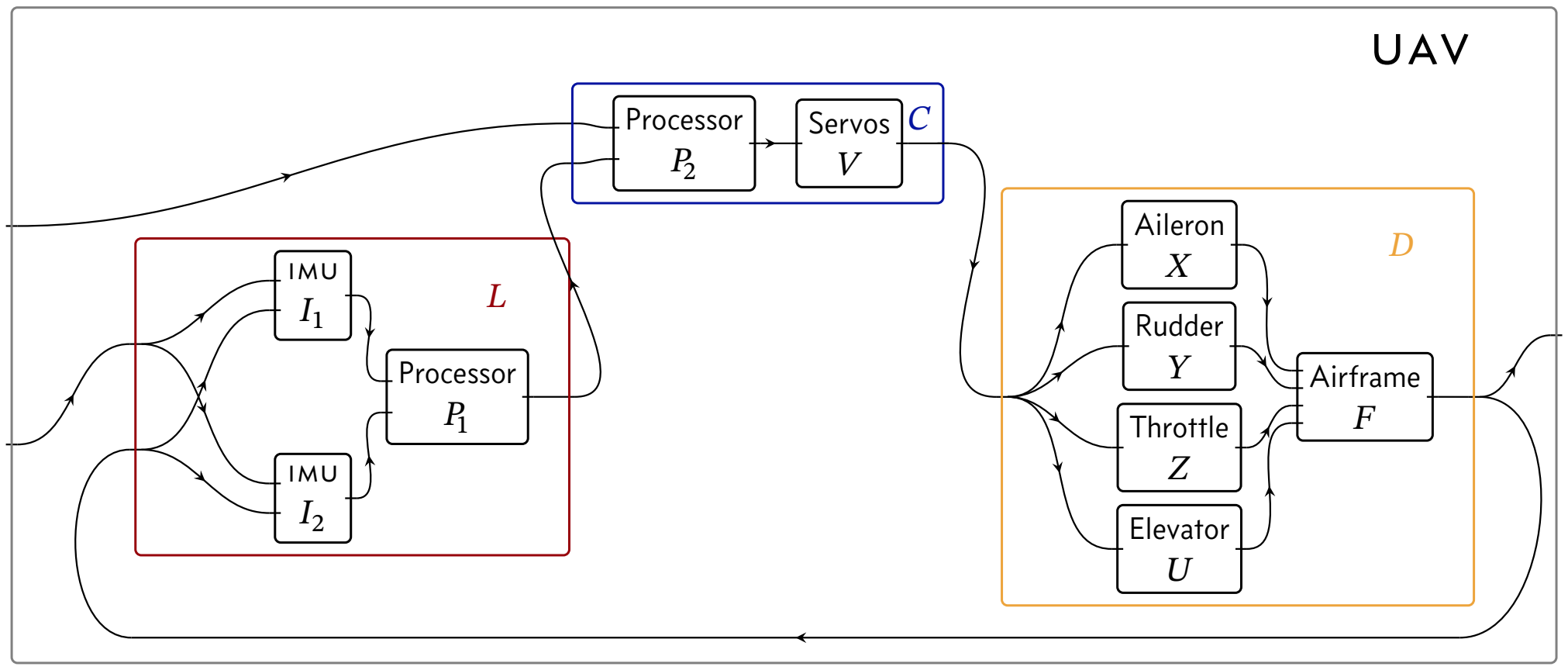

Figure 3.3: Any decomposition, including the previous one (figure 3.1) resides within the slice category $\mathbf{W}/$UAV. In this case, the slice category contains all possible design decisions that adhere to the behavioral model; we pick one such design choice.

to the resulting wiring diagram that expresses an implementation of the outmost system process.

One of the important advantages of expressing system decompositions as a morphism in the category $\mathbf{W}$ is that we can perform further zoomed-in decompositions as desired in a *hierarchical* way, and these are all realized as composite morphisms in the wiring diagram category. We should examine here what we mean by hierarchy in the context of system design. If we see the slice category as all possible architectural choices, we can also examine the opposite, namely that many different architectures can relate to the same behavior. In that sense the behavioral interpretation of the same concept is an authority level above and can be perceived as a useful simplification of what all possible architectures could be. This aligns with our normal understanding of hierarchy in system design [82] and it is one way to make this process of simplification formally explicit.

For example, consider a possible UAV architecture (figure 3.1). We may further choose to implement the sensor box $L$ using two IMU units $I_1, I_2$ and a processor $P_1$ in a certain interconnection. Expressing this as a morphism with target $L$ (an object in the slice category $\mathbf{W}/L$) namely $g : I_1 \otimes I_2 \otimes P \to L$ means that we can compose this with the original one-level implementation $f$ to obtain a two-level zoomed-in decomposition

$$(I_1 \otimes I_2 \otimes P_1) \otimes C \otimes D \xrightarrow{\mathrm{g} \otimes \mathrm{id} \otimes \mathrm{id}} L \otimes C \otimes D \xrightarrow{f} \text{UAV}$$

that only "opens-up" the box $L$. We could moreover implement the control as well as the dynamics box, and decompose them in a choice of subcomponents and wires between them. An example where the control box is decomposed into $P_2$ followed by $V$ in a serial composition, and the dynamics box is decomposed into four parallel boxes, $X$, $Y$, $Z$ and $W$ followed by $F$ amounts to choosing a specific $h\colon P_2 \otimes V \to C$ in $\mathbf{W}/C$ and a specific $k\colon X \otimes Y \otimes Z \otimes W \otimes F \to D$ in $\mathbf{W}/D$. Combining all these morphisms we have the composition (figure 3.3)

$$(I_1 \otimes I_2 \otimes P_1) \otimes (P_2 \otimes V) \otimes (X \otimes Y \otimes Z \otimes W \otimes F) \xrightarrow{\mathrm{g} \otimes h \otimes k} L \otimes C \otimes D \xrightarrow{f} \text{UAV}$$

that can be considered as a single morphism from the tensor of all second-level sub-components to the box UAV. Pictorially, this would be realized by erasing the intermediate colored dashed boxes.

Finally, Stephen Albert said: "In a guessing game to which
the answer is chess, which word is the only one prohibited?"
I thought for a moment and then replied: "The word is chess."

— Jorge Luis Borges

# 4 The algebra of safety contracts

The concept of a *contract* is another example of an algebra for the monoidal category of labeled boxes and wiring diagrams **W** (definition 2.2.1). For any labeled box $X = (X_{\text{in}}, X_{\text{out}})$, a contract is defined to be a relation

$$R \subseteq X_{\text{in}} \times X_{\text{out}}$$

expressing the *allowable* tuples of input and output behaviors of a process. Such a description is one among the most widespread abstract systems modeling notions, see for example Mesarovic and Takahara [118, §2]. We make a distinction between the explicit defining process of a system; that is, the behavior assigned to a wiring diagram, and the system behavior. However, abstractly a system *is* its behavior and therefore modeling a system in the wiring diagram paradigm makes those two notions equivalent. The distinction is however useful for separating the *behavior* algebra from the *contracts* algebra, which are formally related but can be used independently of each other.

## 4.1 Static contracts

The algebra of *static contracts* is a variation of the algebra originally developed by Schultz et al. [156, §4.5]. Explicitly, the functor $\mathcal{C} : \mathbf{W} \to \mathbf{Cat}$ bound to express conditions on inputs and outputs in a time-less manner, assigns to a box $X_{\text{in}} \boxed{X} X_{\text{out}}$ the category $\mathcal{C}(X_{\text{in}}, X_{\text{out}})$ of binary relations; that is, subsets $i : R \hookrightarrow$

$X_{\text{in}} \times X_{\text{out}}$, with morphisms $f : R \to P$ being subset inclusions of the form

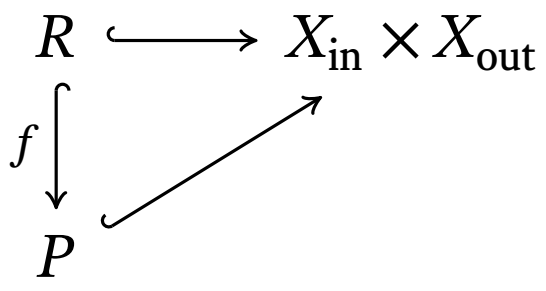

For a given contract $R_X \subseteq X_{\text{in}} \times X_{\text{out}}$ and a wiring diagram $(f_{\text{in}} : X_{\text{out}} \times Y_{\text{in}} \to X_{\text{in}}, f_{\text{out}} : X_{\text{out}} \to Y_{\text{out}})$, the application of the functor $\mathcal{C}(f)$ on $R_X$ is the contract $R_Y \subseteq Y_{\text{in}} \times Y_{\text{out}}$ described by

$$\begin{aligned} R_Y = &\{(y_1, y_2) \in Y_{\text{in}} \times Y_{\text{out}} \mid \exists x_2 \in X_{\text{out}} \\ &\text{such that } (f_{\text{in}}(x_2, y_1), x_2) \in R_X \text{ and } f_{\text{out}}(x_2) = y_2\}. \end{aligned} \tag{4.1}$$

This formula arises categorically via a pullback. The functor $\mathcal{C}f : \mathcal{C}(X) \to \mathcal{C}(Y)$ for a wiring diagram $f : X \to Y$, described by interface formulas (2.1a), assigns a contract $R_X \subseteq X_{\text{in}} \times X_{\text{out}}$ on the inside box to a contract $R_X \subseteq X_{\text{in}} \times X_{\text{out}}$ on the outside box, following a two-step procedure:

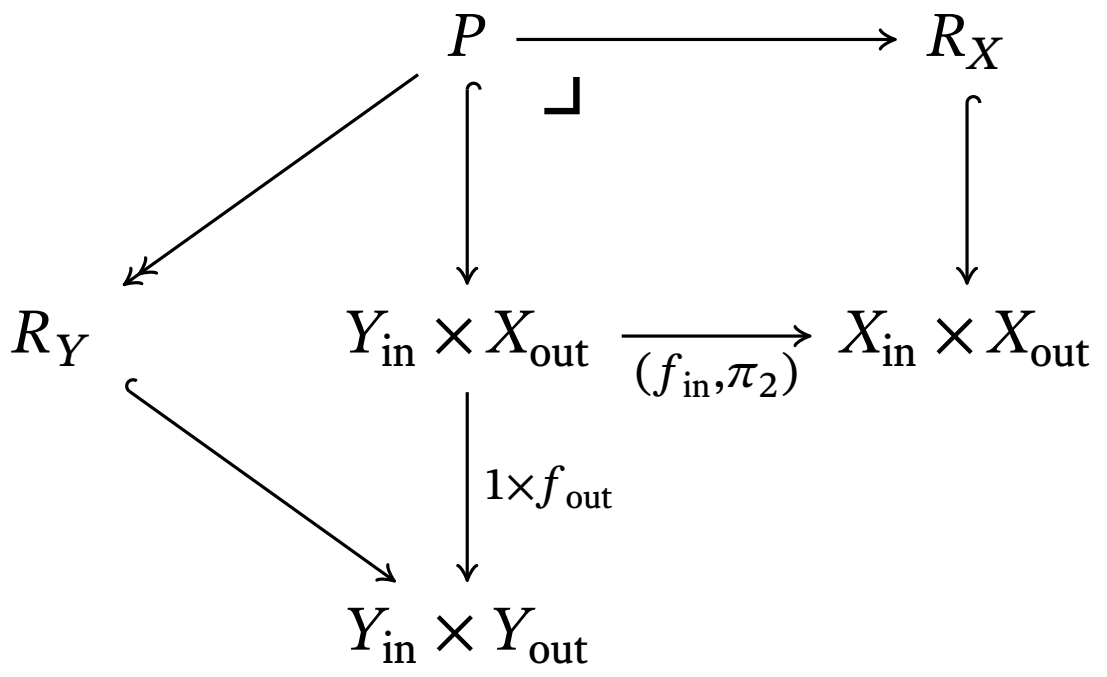

(4.2)

First, we compute the pullback—a limit of a diagram of two morphisms with common codomain [98, 5.1.16]—of the relation $R_X$ along the function $(f_{\text{in}}, \pi_2)$ which is defined by $Y_{\text{in}} \times X_{\text{out}} \ni (y, x') \mapsto (f_{\text{in}}(y, x'), x') \in X_{\text{in}} \times X_{\text{out}}$. The explicit description of that pullback in **Set** is

$$P = \{(y, x') \mid (f_{\text{in}}(y, x'), x') \in R_X\}$$

namely those pairs of $Y$-inputs and $X$-outputs which the bottom function actually maps to elements of the contract $R_X$. Second, we take the image factorization of the inclusion $P \subseteq Y_{\text{in}} \times X_{\text{out}}$ post-composed with the function $1 \times f_{\text{out}}$ that maps some $(y, x')$ to the pair $(y, f_{\text{out}}(x'))$. The image of a function is the subset of its codomain where all elements of the domain get mapped to, namely for an arbitrary $g : A \to B$, $\text{Im}(g) = \{b \in B \mid \exists a \in A \text{ such that } g(a) = b\}$. In the end, using the above constructions of the two-step process exhibited in (4.2), the explicit description of the resulting contract is precisely equation (4.1). In various examples, this composite contract may be expressed in more elementary terms depending on the form of the component contracts $R_X$ and the given wiring diagram.

For the monoidal structure of the functor, suppose we have two parallel boxes (2.2) with contracts $R_X \subseteq X_{\text{in}} \times X_{\text{out}}$ and $R_Y \subseteq Y_{\text{in}} \times Y_{\text{out}}$. The laxator $\phi_{X,Y} : \mathcal{C}(X) \times \mathcal{C}(Y) \to \mathcal{C}(X \otimes Y)$ induces a contract on the box $(X_{\text{in}} \times Y_{\text{in}}, X_{\text{out}} \times Y_{\text{out}})$ which is merely the cartesian product

$$R_X \times R_Y \hookrightarrow X_{\text{in}} \times X_{\text{out}} \times Y_{\text{in}} \times Y_{\text{out}} \xrightarrow{\cong} X_{\text{in}} \times Y_{\text{in}} \times X_{\text{out}} \times Y_{\text{out}}$$

that essentially switches the two middle variables,

$$\phi_{X,Y}(R_X, R_Y) = \{(x_1, y_1, x_2, y_2) \mid (x_1, x_2) \in R_X \text{ and } (y_1, y_2) \in R_Y\}.$$

As an example, suppose we ask that some process in $X$ (figure 2.2a) satisfies the contract $R_X \subseteq \mathbb{R} \times \mathbb{R}$, some process in $Y$ satisfies the contract $R_Y \subseteq \mathbb{R} \times \mathbb{R}$ and some process in $Z$ satisfies the contract $R_Z \subseteq \mathbb{R}^3 \times \mathbb{R}$. The fact that contracts form an algebra on **W** ensures that the composite process in $A$ will necessarily satisfy a contract formed only in terms of $R_X, R_Y$ and $R_Z$ and their interconnection $(f_{\text{in}}, f_{\text{out}})$, and specifically

$$\begin{aligned} R_A = &\{(w, u, v, z) \in \mathbb{R}^4 \mid \exists (x, y) \in \mathbb{R}^2 \\ &\text{such that } (w, x) \in R_X, (u, y) \in R_Y, (x, y, v, z) \in R_Z\}. \end{aligned} \tag{4.3}$$

The algebra machinery produces a contract that matches our intuition: whenever the interconnected composite in figure 2.2a receives three real numbers $(w, u, v)$ as inputs, it must produce an output $z$ which is $R_Z$-allowable by (that is, related to) $(x, y, v)$, for some real $x$ which is $R_X$-allowable by $w$ and some real $y$ which $R_Y$-allowable by $u$. Not all inputs of this composite $A$ will have an allowable output, and that completely depends on the contracts of components $X$, $Y$ and $Z$.

As another example, which highlights the strong connection between the contract algebra machinery and the usual relation operators, consider a simple wiring diagram with $\mathbb{R}$-typed wires on the left, expressing serial composition of two boxes

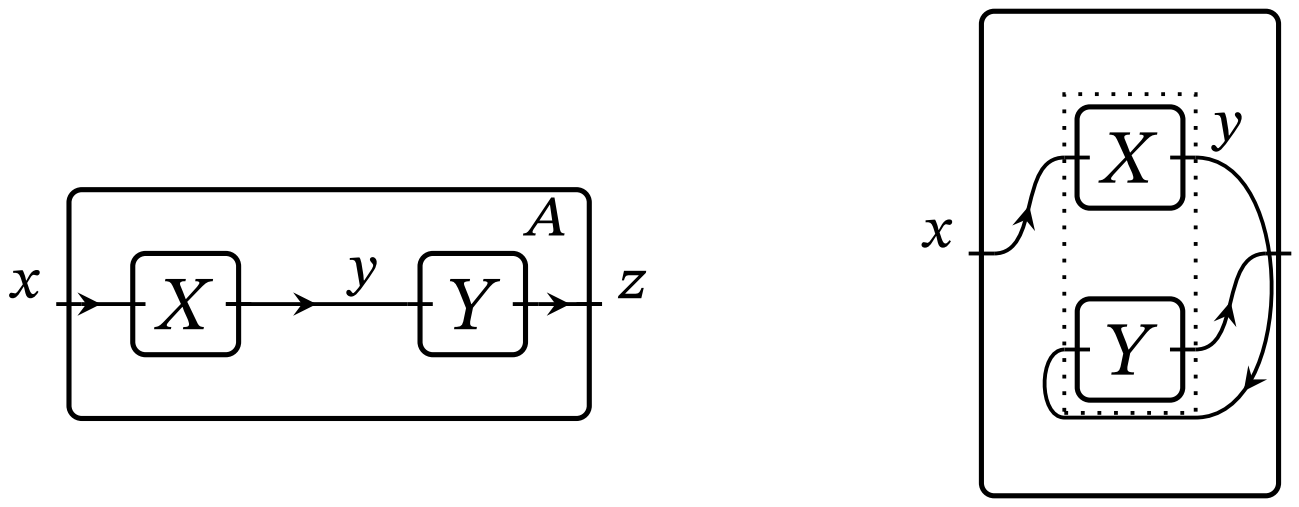

This morphism $f : X \otimes Y \to A$ is described by $(f_{\text{in}}(y, z, x) = (x, y), f_{\text{out}}(y, z) = z)$ according to its equivalent arrangement on the right, and given two contracts $R_X$ and $R_Y$ the formula (4.1) produces the composite contract

$$R_A = \{(x, z) \mid \exists y \text{ such that } (x, y) \in R_X \text{ and } (y, z) \in R_Y\}$$

which is the usual composition of binary relations.

What is interesting about this algebra of contracts is that it is "agnostic" to the exact specification of the systems. This means that although categorically it is expressed the same way as, for example, Moore machines, it is of a quite different flavor: we are not interested in giving explicit functions that describe the composite process, but in expressing all the possible (input,output) pairs that can be observed on it. This is very convenient especially when connecting systems of different models, for example, a Moore machine with an "abstract machine" [156, §4]. Even if we cannot compose them in the previous sense, since they form distinct algebras (that is, they are described by different functors $\mathbf{W} \to \mathbf{Cat}$), we can still compose and examine the requirements the composite satisfies, in this relational sense.

## 4.2 Independent contracts

We will also be interested in a subclass of static contracts, called *independent*, of the form

$$I = I^1 \times I^2 \subseteq X_{\text{in}} \times X_{\text{out}}.$$

These contracts capture cases like "inputs are always in range $I^1$ and outputs are always in range $I^2$, independently from one another".[1] Of course this is only a special case of arbitrary relations $R \subseteq X_{\text{in}} \times X_{\text{out}}$, since not all subsets of cartesian products are cartesian products of subsets, as a simple argument in the finite case shows: $|\mathcal{P}(X_{\text{in}} \times X_{\text{out}})| = 2^{n \cdot m}$ whereas $|\mathcal{P}(X_{\text{in}})| \cdot |\mathcal{P}(X_{\text{out}})| = 2^{n+m}$. For example, the contract $\{(x, y) \mid x < y\} \subseteq \mathbb{R} \times \mathbb{R}$ is not independent.

One could expect that these contracts form themselves an algebra, namely any wiring composite of independent contracts will also be an independent, rather than a general contract itself. However this is not the case in general: although the parallel placement of boxes with $I_X = I_X^1 \times I_X^2 \subseteq X_{\text{in}} \times X_{\text{out}}$ and $I_Y = I_Y^1 \times I_Y^2 \subseteq Y_{\text{in}} \times Y_{\text{out}}$ produces the independent contract $(I_X^1 \times I_Y^1) \times (I_X^2 \times I_Y^2)$ on $X \otimes Y$, closure under feedback fails. Explicitly, for an independent contract $I_X^1 \times I_X^2 \subseteq X_{\text{in}} \times X_{\text{out}}$ on $X$, and a wiring diagram $(f_{\text{in}}, f_{\text{out}}) \colon X \to Y$, the formula (4.1) produces the simpler composite contract

$$\begin{aligned} R_Y = \{(y_1, y_2) \in Y_{\text{in}} \times Y_{\text{out}} \mid \exists x_2 \in I_X^2 \\ \text{such that } f_{\text{in}}(y_1, x_2) \in I_X^1 \text{ and } f_{\text{out}}(x_2) = y_2\} \end{aligned} \tag{4.4}$$

which shows that $y_1$ and $y_2$ are not independent in general, hence $R_Y$ is not of the form $I_Y^1 \times I_Y^2$.

In certain examples, $R_Y$ can indeed be written as a product itself, for example, when $(f_{\text{in}}, \pi_2)$ is of the form $k \times s$ for two functions $k, s$. Even more interestingly, due to the special form of morphisms in the wiring diagram category (where they are only made up from projections, diagonals and duplications) in our examples below we will be able to write $R_Y$ as an independent contract itself.[2]

[1] These independent contracts in reality are even more special than that: not only are input restrictions separate from output restrictions, but also each individual *wire* has an associated subset of allowed values on it.

## 4.3 Relation to assume-guarantee contracts

Systems theory and design has long recognized the need for a formal requirement engineering through mathematical models and formal analysis techniques [29]. As part of contract-based design, there have been multiple efforts to formalize and analyze *assume-guarantee* contracts [142] and incorporate them in the design as a fundamental concept. We here discuss such examples and how they fit to the previously described static contract model.

Given a box 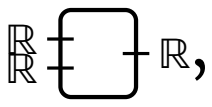, an example of an assume-guarantee contract (adapted from Benveniste et al. [29, § IV]) is

$$R_1 : \begin{cases} \text{variables:} & \text{inputs } x, y;\ \text{outputs } z \\ \text{types:} & x, y, z \in \mathbb{R} \\ \text{assumptions:} & y \neq 0 \\ \text{guarantees:} & z = \dfrac{x}{y} \end{cases} \tag{4.5}$$

This explicitly makes the assumption that the environment (namely the inputs coming either from the external world or from other component systems) will never provide the input $y = 0$, essentially leaving the behavior for that input undefined. In our formalism, we can express this contract as

$$R_1 = \{(x, y, z) \mid y \neq 0 \wedge z = \frac{x}{y}\} \subseteq \mathbb{R} \times \mathbb{R} \times \mathbb{R}$$

indicating the fact that the input $y = 0$ will never occur on the input wire of the box; and if it did, the contract is violated. A different choice we could make,

[2] It can be shown that independent contracts indeed form an algebra on **W** due to the special morphisms that generate it; the proof is beyond the scope of this section.

assuming the initial assume-guarantee contract is really expressing a conditional ("if-then") requirement, is

$$R_1' = \{(x, y, z) \mid y \neq 0 \Rightarrow z = \frac{x}{y}\} \subseteq \mathbb{R} \times \mathbb{R} \times \mathbb{R}$$

which is a different subset of allowable values on the wires. For example, $(3, 0, 25) \in R_1'$ whereas $(3, 0, 25) \notin R_1$.

We now consider a standard assume-guarantee contract operator called *contract composition and system integration*, and we realize it from the perspective of the wiring diagram algebra machinery—consequently a more general setting. Explicitly, the assume-guarantee contract composition operator as described for example by Benveniste et al. [29, § IV.B] or Le et al. [93], takes two assume-guarantee contracts $R_1 = (A_1, G_1)$ and $R_2 = (A_2, G_2)$ and produces a new assume-guarantee contract $R_1 \otimes R_2$ (this is a completely different use of our earlier monoidal product symbol $\otimes$) with

$$\begin{aligned} G_{R_1 \otimes R_2} &= G_1 \wedge G_2 \\ A_{R_1 \otimes R_2} &= \max\{A \mid A \wedge G_2 \Rightarrow A_1, A \wedge G_1 \Rightarrow A_2\} \end{aligned}$$

only when $R_1$ and $R_2$ are *compatible*, namely $A_{R_1 \otimes R_2} \neq \emptyset$. Since $A_{R_1 \otimes R_2}$ is the weakest assumption such that the two referred implications hold, if non-empty it ensures that there exists some environment in which the two contracts properly interact: when put in the context of a process that satisfies the first contract, the assumption of the second contract will be met and vice-versa. This definition looks "symmetric," since it considers a certain compatibility of output guarantee/input assumption in both directions, but in reality this is not quite the case.

One issue with the above assume-guarantee contract composition is that the *names* of the variables and not only the types of the wires need to match, in order to connect along them [29, 133]. For example, the contract $R_1$ as in formula (4.5) can be

composed with the contract on

$$R_2 : \begin{cases} \text{variables:} & \text{inputs } u; \text{ outputs } x \\ \text{types:} & u, x \in \mathbb{R} \\ \text{assumptions:} & \top \\ \text{guarantees:} & x > u \end{cases}$$

not along any wire, as could be deduced by noticing that all wire types are $\mathbb{R}$, but specifically along the wire with variable name $x$. Pictorially, we can realize them as inhabiting boxes wired as

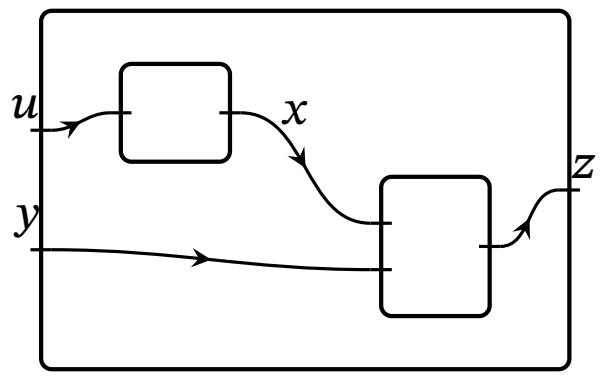

and using the formulas (4.3) we obtain

$$A_{R_1 \otimes R_2} = \max\{A \mid (A \wedge (x > u) \Rightarrow y \neq 0) \wedge (A \wedge (z = x/y) \Rightarrow \top)\} = (y \neq 0)$$
$$G_{R_1 \otimes R_2} = (x > u) \wedge (z = x/y).$$

On the other hand, composing $R_1$ and $R_2$ using the static contract algebra (section 4.1) for the above wiring diagram ($f_{\text{in}}(x, z, u, y) = (u, x, y), f_{\text{out}}(x, z) = z$), we obtain the composite contract

$$R = \{(u, y, z) \in \mathbb{R}^3 \mid \exists x \in \mathbb{R} \text{ such that } y \neq 0 \wedge x > u \wedge z = x/y\},$$

which could be written in assume-guarantee form as $A = \{(u, y) \mid y \neq 0\}$ and $G = \{z \mid \exists x > u \text{ such that } z = x/y\}$. The contract algebra machinery does not present this variable-match problem, since it does not prevent us from composing along the second input wire of $X_1$ or even do first $X_1$ and then $X_2$ in the opposite order, since all types of wires are real numbers. In all these cases, it would be possible to compute appropriate composite contracts in this uniform way.

The second issue, which can also be seen from the above calculation, is that the assume and guarantee of the composite contract include information that mix the variables of the resulting input and output wires. For example, using the assume-guarantee formalism, the variables of $R_1 \otimes R_2$ are set to be $\{u, y\}$ for inputs and $\{x, z\}$ for outputs, therefore this operation behaves as if the intermediate wires of a system composition can be extracted as extra output wires to the outside world

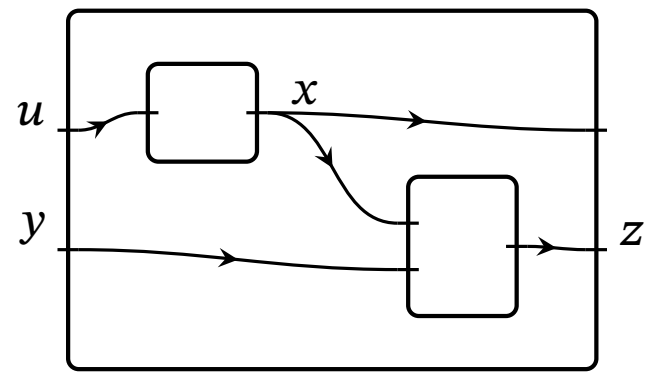

This "choice" does not agree with the wiring diagram formalism, and moreover is somewhat ad-hoc given that it could potentially add arbitrary many wires to the composite system, essentially according to the result of the contract composition. Adding extra wires is of course possible for the algebra formalism, but corresponds to a choice of architecture on how we decide to wire the subcomponents together, rather than a necessity that arises from dealing with contracts.

Finally, the assume-guarantee formalism asks that compositions $(R_1 \otimes R_2) \otimes R_3$ and $R_1 \otimes (R_2 \otimes R_3)$ give equivalent contracts, and that so do $R_1 \otimes R_2$ and $R_2 \otimes R_1$. In the contract algebra formalism, the first statement follows for any $\mathbf{W}$-algebra: consider a possible wiring of three boxes, each inhabited with a contract (and/or a behavior formalism)

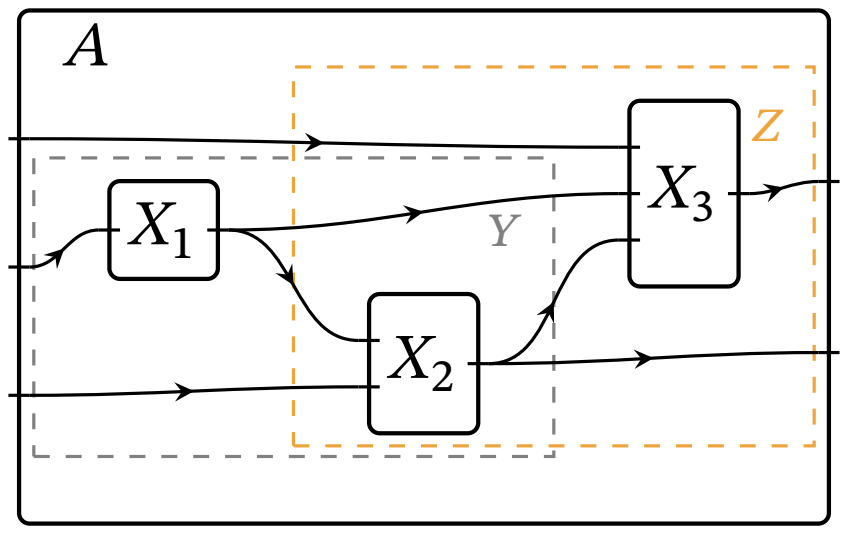

First composing the contracts $R_1$ and $R_2$ and then the result with $R_3$ comes from the application of the functor $\mathcal{C} : \mathbf{W} \to \mathbf{Cat}$ on a wiring diagram morphism

$$(X_1 \otimes X_2) \otimes X_3 \to Y \otimes X_3 \to A$$

whereas the other way around comes from the application of the functor $C$ on the morphism

$$X_1 \otimes (X_2 \otimes X_3) \to X_1 \otimes Z \to A$$

which both express the same morphism $X_1 \otimes X_2 \otimes X_3 \to A$ in $\mathbf{W}$ as an implementation of $A$ (section 3.3).

Regarding the second statement about $R_1 \otimes R_2$ and $R_2 \otimes R_1$, in the assume-guarantee formalism this can indeed be proved due to the symmetric formulation of composition described by formula (4.3) as observed earlier. However, this refers more to the earlier variable-sharing clause (which would not allow the composition along arbitrary wires therefore in arbitrary order) and less to composition intuition: changing the order of two boxes and expecting the same behavior or requirements is something highly non expected, from a categorical but also a design point of view due to the input-output directionality. As a result, commutativity in this assume-guarantee setting is slightly misleading, since it is just a technical term relevant to the constructed formula (it does not really have an effect on the operation) rather to an actually commuting composition which is not expected to hold—and does not, in the algebra formalism.

We will use the algebra of static contracts $\mathcal{C} : \mathbf{W} \to \mathbf{Cat}$, where all requirements are expressed as subsets of the cartesian product of input and output types. Consider the original system decomposition to sensor, controller, and dynamics boxes (figure 3.1) and suppose we have certain contracts on these components given by

$$R_L \subseteq \mathbb{R}^2 \times \mathbb{R}, \quad R_C \subseteq \mathbb{R}^2 \times \mathbb{R}, \quad R_D \subseteq \mathbb{R} \times \mathbb{R}.$$

These contracts could be any subsets, from the extreme case of equality which means that *all* combinations of inputs and outputs are allowed, to some specific requirement imposed to the example at hand, or in certain cases some *maximal* contract dictated by a discrete dynamical system (governed by a difference equation) that actually inhabits the box.

The contract algebra applies to the wiring diagram of figure 3.1 and based on the formula (4.1) produces a contract $R_{\mathrm{UAV}} \subseteq \mathbb{R}^2 \times \mathbb{R}$ on the composite system, with the following explicit description

$$\mathbb{R}^3 \supseteq R_{\mathrm{UAV}} = \{(a_1, a_2, a_3) \in \mathbb{R}^3 \mid \exists (x, y) \in \mathbb{R}^2 \text{ such that } (a_3, a_1, x) \in R_L,\ (x, a_2, y) \in R_C,\ (y, a_3) \in R_D\}.$$

Further, we could assume that all contracts are independent, namely they can be written as products of subsets of each wire type independently, like

$$R_L = R_L^1 \times R_L^2 \times R_L^3, \quad R_C = R_C^1 \times R_C^2 \times R_C^3, \quad R_D = R_D^1 \times R_D^2$$

where all components are subsets of $\mathbb{R}$ meaning the allowed values on each wire are completely unrelated to one another. Then the composite contract (4.4) takes the following, also independent contract form

$$R_{\mathrm{UAV}} = \begin{cases} R_L^2 \times R_C^2 \times (R_L^1 \cap R_D^2) & \text{if } R_L^3 \cap R_C^1 \neq \varnothing \text{ and } R_C^3 \cap R_D^1 \neq \varnothing \\ \varnothing & \text{if } R_L^3 \cap R_C^1 = \varnothing \text{ or } R_C^3 \cap R_D^1 = \varnothing \end{cases}$$

The above formula expresses that the allowable tuples that can be observed on the composite system are the $L$- and $C$-external input contracts for the two input wires, along with an intersection of contracts for the output wire, subject to whether there exists a scenario where the contracts of the intermediate wires match: if their intersection is non-empty, there exist appropriate values that work for both contracts and the total system "runs." Otherwise the composed contracts are incompatible and the composite system fails to adhere to a contract, namely there is no guarantee about its observable input and output values (expressed by the empty set contract) and possibly the whole process fails.

We now proceed to a similar process to what we have seen before (section 3.2), which in a sense reverse-engineered the behavior of the subcomponents, given a composite behavior of the total system using the system behavior algebra machinery. In this setting, given a specific desired requirement $R_{\text{UAV}}$ on the composite system, we will identify possible contracts on the components that produce that specific composite; once again we do not expect unique solution to this problem.

Suppose the envisioned composite contract on the behavioral representation of our example UAV (figure 3.1) is

$$R_{\text{UAV}} = [0, 100] \times [-20, +20] \times [-35, +35].$$

This contract represents a possible requirement that the desired UAV pitch is no more or less than 20 degrees and the plane really must not pitch more or less than 35 degrees for a hypothetical safe flight. As hypothetical environmental conditions, we assume air speed does not exceed 100 km/h.

Comparing the above composite contract against equation (4.3), we can first of all deduce that

$$R_L^2 = [0, 100] \qquad R_C^2 = [-20, +20]$$

namely the external inputs for $L$ and $C$ are necessarily constrained by the ranges of the given composite contract on those wires. Moreover we have that $R_L^1 \cap R_D^2 = [-35, 35]$ and also that necessarily the intersections $R_L^3 \cap R_C^1$ and $R_C^3 \cap R_D^1$ are non-empty—since the composite contract is indeed non-empty. All these intersections correspond to specific wiring connections or splittings we performed between subcomponents for the initial UAV's implementation.

Given these restrictions, we are free to choose contracts that satisfy them, for example

$$R_D^2 = [-35, +35], \; R_L^1 = R_L^3 = R_C^1 = R_C^3 = R_D^1 = \mathbb{R}$$

The above choices are made to also dispose of bad scenarios for the given interconnection of the boxes. For example, choosing the opposite contracts for $R_D^2$ and $R_L^1$ would be mathematically correct because their intersection is still $[-35, 35]$, but could lead to a real value of, say, 40 degrees entering the sensor $L$, which would then violate its contract (that said "all my inputs on the first wire will be less than 35"). Although in general, processes can be wired together as long as types match, in the contract algebra setting it is implied (by the algebra machinery) that the only values passing through an interconnected wire are those in the intersections of the individual (independent) contracts—so long as the composite system does not "break." It is important to realize that the contract algebra *describes* the observable inputs and outputs on a running composite machine, rather than *ensures* that the process runs: this has to be safeguarded by the designer also. This discussion relates to developing "total" or "deterministic" contracts.

## 4.4 Time in compositional systems modeling

Thus far, we are able to compositionally express the process of building larger systems from smaller, depending on their interconnections. In the earlier examples, this process is somewhat static and more related to the classical model of computation, for example, given functions, we express their composite function that given some inputs values produces a specific output value. How would we express that composite as a system that over time, it receives and produces *sequences* of inputs and outputs via its interactions of subsystems and the environment? This in particular entails moving to a *reactive* model of computation, a key characteristic of cyber-physical systems [6, §1.2].

For this reason, any sufficiently capable modeling language needs to address the notion of time. A major challenge in modeling is the cojoint verification of computational models, which are deterministic, and physical models, which are not necessarily deterministic [96]. To incorporate time into our model in such a way, we need to change the type of information that the wires hold from a mere set like $\mathbb{R}$ to something else that captures the necessary time-related structure.

Previously we restricted the definition of the category of labeled boxes and wiring diagrams **W**, to the category **Set**, where $X_{\text{in}}, X_{\text{out}}$ are sets; that is, objects in the category **Set**, and $f_{\text{in}}, f_{\text{out}}$ are functions; that is, morphisms in **Set**. Therefore there is a clear sense in which the category **W** depends on the category **Set**, pointed out also at that time. Let us change our perspective and consider a new category $\mathbf{W}_N$ which is described using a different category than sets and functions **Graph** of (directed, multi-) graphs [158, Chapter 4]. Explicitly, the objects are given by a set of vertices $V$, a set of edges $E$ and a source $s$ and target $t$ function

$$E \underset{t}{\overset{s}{\rightrightarrows}} V$$

and morphisms are pairs of functions $(f, g)$ that map vertices to vertices and edges to edges by respecting their sources and targets; that is, making both squares commute, meaning in this case that $g(s(e)) = s'(f(e))$ and $g(t(e)) = t'(f(e))$ for any edge $e \in E$.

$$\begin{array}{ccc} E & \underset{t}{\overset{s}{\rightrightarrows}} & V \\ {\scriptstyle f}\downarrow & & \downarrow{\scriptstyle g} \\ E' & \underset{t'}{\overset{s'}{\rightrightarrows}} & V' \end{array}$$

The motivation for labeling wires with graphs rather than with sets like before is to capture all possible "streams" of data that can flow through a wire during any discrete interval of time $[0, n]$ as *paths* of length $n$.

Therefore, at time 0 when the observer looks at a **Graph**-labeled box $G\!-\!\boxed{X}\!-\!H$, both wires have a 0-path length on them namely a vertex of each graph: it is as if $X = (V_G, V_H)$ in the previous **Set**-labeled notation. At time 1, each wire has a path of length one, namely an edge consisting of two vertices: the one corresponding to clock-tick 0 and the other corresponding to clock-tick 1. As every time-unit passes by, one edge (starting at the latest vertex) and one vertex (the new edge's target) are added, and the *total* information flowing on the wire from time 0 is packaged in the form of a path of the graph labeling that wire.

Hence, the data flowing on the wires at any given interval $[0, n]$ can be thought of as $n+1$ elements of the vertex set connected by $n$ continuous edges. An important characteristic of this formalism is that, given a path of length $n$ for the time $[0, n]$ and a path of length $m$ for the time $[n, n+m]$ that share their end and start vertex respectively, their concatenation is a path of length $n + m$ arising on the time interval $[0, n + m]$. Another is that given a path over some period of time say $[0, 7]$ and a chosen sub-interval say $[2, 5]$ of it, we can always *restrict* in a coherent way to a path of length 3. These traits, although straightforward in this graph-time framework, come from the initial *sheaf* formulation of the time-model [152] which necessary to extend to the continuous time-model.

To understand the equivalent sheaf formulation of time on the wires, consider a category **DInterv** of *discrete intervals*. Its objects are natural numbers $n$, thought of as lengths of discrete time intervals $n = [a, b]$ if $a, b \in \mathbb{N}$ with $a \leq b$ and $b - a = n$. In this category, $[3, 5] = [36, 38] = [0, 2]$ because all those intervals are represented by the object 2. Morphisms $p \colon m \to n$ are all $p \leq n - m \in \mathbb{N}$, thought of as "translations-by-$p$"

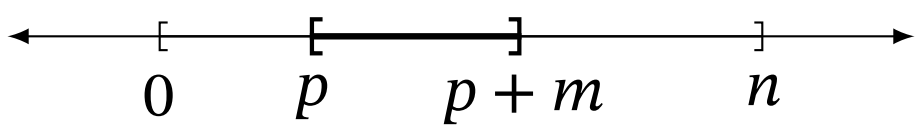

These morphisms are the ways of positioning a smaller interval within a larger interval, for example, the morphism $2 \colon 5 \to 8$ expresses the interval inclusion $[2, 7] \subseteq [0, 8]$ of a 5-length interval within an 8-length interval starting at its second clock-tick.

Any graph $G$ can equivalently be viewed as a functor

$$\hat{G} \colon \mathbf{DInterv}^{\mathrm{op}} \longrightarrow \mathbf{Set}$$

that maps any discrete length $n$ to the set $\hat{G}(n)$ of length-$n$-paths of $G$, thought of the set of all possible signals that may flow on the wire at *any* interval of length $n$. Any interval inclusion $p \colon m \to n$ is mapped to a function $\hat{G}(p)$ from $n$-signals to $m$-signals that cuts off the longer path at the $p$-th vertex as described earlier. Below, the path $f.g.h$ belongs to the set $\hat{G}(3)$, and there is a well-defined function $\hat{G}(1) \colon \hat{G}(3) \to \hat{G}(2)$ which slices a given 3-path to a 2-path starting at 1, namely $g.h \in \hat{G}(2)$:

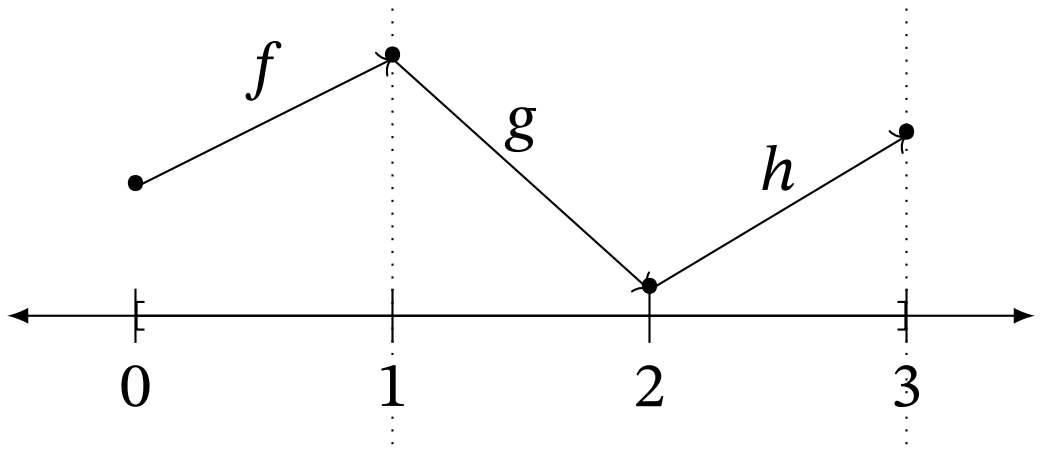

This operation $\hat{G}$ is in fact not only a functor, but also can be represented as a *sheaf* [113]. Informally, this ensures that any $n$-signal also understood as the path $x = a_1 \to \cdots \to a_n$ over an interval of length $n$, and any $m$-signal $y = b_1 \to \cdots \to b_m$ that happen to agree on their left and right restriction, namely $a_n = b_1$, can be *glued* into a $(m+n)$-signal $x \star y$, their concatenation.

Wiring diagrams can in fact be viewed in this new time-sensitive category $\mathbf{W}_N$ with objects pairs of graphs $(X_{\text{in}} = G, X_{\text{out}} = H)$ and morphisms pairs of graph transformations. This is due to the existence of the 'complete graph' functor

$$\begin{aligned} K : \mathbf{Set} &\longrightarrow \mathbf{Graph} \\ A &\longmapsto (A \times A \rightrightarrows A) \end{aligned} \tag{4.6}$$

on the labels of the wires, that maps every set $A$ to the graph $K(A)$ with vertices the elements of $A$ and edges pairs of elements in $A$, namely there exists a unique arrow from each vertex to another. The source and target functions $A^2 \to A$ are the two projections on the first and second variable, and the paths of lengths $n$ of the graph $K(A)$ are precisely tuples in $A^{n+1}$. A possible "section" of data flowing on a wire labeled with $\mathbb{R}$ as earlier would now look like $(a_0, a_1, \dots, a_n) \in \mathbb{R}^{n+1}$ for a time interval of length $n$. At any specific time unit $i$, only one $a_i$ would be transmitted at that time, but all its history is available (as well as its future possibilities determined by the whole graph).

If we think about the boxes having wires typed with those graphs or equivalently discrete interval sheaves, in particular we recover "window frames" of other stan-dard time models [80]. Therein, time is encapsulated using a product space

$$X[\cdot] = \prod_{k \in I} X$$

over a time interval $I = [a, b]$ of natural numbers with $a \leq b$. To connect that to our setting, if $n = b - a$ is the length of that interval, indeed $X[\cdot] = X^{n+1}$ is the product of $X$ over all possible natural numbers in the interval, counting both endpoints $a$ and $b$. Therefore, this approach provides the same $n$-paths of a graph $G = K(X)$ coming from a set $X$, or equivalently sets $\hat{G}(n)$ of a discrete interval sheaf $\hat{G}$. However, there is no operation of restricting signals over larger intervals into signals over smaller subintervals, or the gluing condition, both of which are indispensable for our current formalism.

This change of wire types gives us the flexibility to include time, but can also make the algebras on $\mathbf{W}_N$ more involved. For example, Schultz et al. [152, 5.1.1] describe a mapping from the $\mathbf{W}$-algebra of Moore machines to a $\mathbf{W}_N$-algebra of *discrete machines*, which transforms any Moore machine

$$(S, u\colon\ S \times \mathbb{R} \to S, r\colon\ S \to \mathbb{R})$$

to a process whose inputs over time-lengths $n$ are lists $(a_0, \ldots, a_n)$ in $\mathbb{R}^n$ as described above, and whose outputs are lists in $\mathbb{R}^n$ of the form

$$(b, r(u(s_0, a_0)), r(u(s_1, a_1)), \ldots)$$

namely all the outputs of the updated state by the given inputs, step-by-step.

As another example, the contracts introduced earlier, in the absence of time provide the boxes with a constant behavior $R \subseteq X_{\text{in}} \times X_{\text{out}}$. Without Moore machine objects, we can still make a static contract into a time contract by applying the functor $K$ from equation (4.6) to both input/output wires but also to the relation

$$KR \subseteq KX_{\text{in}} \times KX_{\text{out}}$$

which says that in any time interval of length $n$, $R^{n+1} \subseteq X_{\text{in}}^{n+1} \times X_{\text{out}}^{n+1}$ and so all instantaneous input and output pair will be related via $R$. An example of such a time contract for a box with inputs and outputs in $\mathbb{R}$ could be "my inputs are within range $[2, 3]$ and I will transmit outputs within range $[10, 11]$." An allowable signal or section over, say, the interval [0,3] would be

$$(2, 2.5, 2.7, 3, 10, 11, 11, 11) \subseteq [2, 3]^4 \times [10, 11]^4 \tag{4.7}$$

However, we can also consider new algebras on $\mathbf{W}_N$ that don't come from static ones like the above examples of Moore machines and contracts. For example, general time contracts do not have to be of the above restricted form. Indeed, motivated by the above generalization, we can define an algebra

$$\mathbf{TimeContr} \colon \mathbf{W}_N \to \mathbf{Cat}$$

that to each **Graph**-labeled box $G \boxed{X} H$ where $G$ and $H$ can equivalently be viewed as sheaves $G, H \colon \mathbf{DInterv}^{\text{op}} \to \mathbf{Set}$, assigns a *subfunctor* of $G \times H$

$$C \subseteq_f G \times H \colon \mathbf{DInterv}^{\text{op}} \to \mathbf{Set}$$

This only means that for every discrete interval of time $n \in \mathbf{DInterv}$, the contract of allowable behaviors over that interval is any subset

$$C(n) \subseteq G(n) \times H(n) \tag{4.8}$$

Moreover, for any time inclusion $m \to n$, the restriction of an allowable $n$-behavior is still an allowable behavior over the smaller $m$. Notice how we could have asked for $C$ to be a *subsheaf* of $G \times H$; that is, to also have the gluing property, but that could create unreasonable characteristics on the contract machinery. For example, the concatenation of two allowable behaviors could potentially violate a time-sensitive contract [152, §4.5].

An example of such a flexible time contract for a process with input graph and output graphs $K\mathsf{Bool}$; that is, streams of booleans, could be "if I receive two trues in a row, I will output a false within 5 seconds.” The allowable behaviors by such a contract over an interval of length $n$ are formally expressed as

$$\begin{aligned} C(n) = \{ & (a_0, \dots, a_n, b_0, \dots, b_n) \in \mathsf{Bool}^{n+1} \times \mathsf{Bool}^{n+1} \mid \\ & \text{for all } i \geq 0, \text{ if } i + 6 \leq n \text{ and } a_i = a_{i+1} = T \text{ then} \\ & \text{there exists a } j \text{ such that } i + 2 \leq j \leq i + 6 \text{ and } b_i = \bot \} \end{aligned}$$

Notice the strong similarity between the sets of signals of a time contract described by equation (4.8) and a standard expression of a dynamical system [80], as well as an input-output specification, as a relation between input and output signals

$$\Sigma, \phi \subseteq G^{n+1} \times H^{n+1}$$

when the inputs and outputs come from fixed sets. A system *satisfies* $\phi$ just when $\Sigma \subseteq \phi$. The contracted behaviors in our approach are of this form for any fixed $n$, but we also have the freedom to restrict to smaller intervals and the extra, reasonable requirement that any restricted behavior is still part of the contract.

Furthermore, relatively to the *assume-guarantee* contracts expressed as a pair of contracts $(\phi_a, \phi_g) \subseteq G^{n+1} \times H^{n+1}$ such that $\Sigma \cap \phi_a \subseteq \phi_g$, in set-theoretic notation this corresponds to $\Sigma \subseteq (\phi_a \Rightarrow \phi_g)$ therefore it can form a new contract of the previous form. For example, a contract that says “if I receive inputs within [2,3] I will transmit outputs within [10,11]” where $\phi_a = [2, 3] \times \mathbb{R}$ can be thought of as the assumption and $\phi_g = \mathbb{R} \times [10, 11]$ as the guarantee in the standard notation [80], can here be expressed as

$$\begin{aligned} C(n) = \{ & (a_0, \dots, a_n, b_0, \dots, b_n) \in \mathbb{R}^{n+1} \times \mathbb{R}^{n+1} \mid \\ & \text{for all } i, \text{ if } a_i \in [2, 3] \text{ then } b_{i+1} \in [10, 11] \} \end{aligned}$$

Compared to its static version from equation (4.7), in this case “unsatisfied assumptions do not trigger an obligation to satisfy a guarantee.” Also, the $(i + 1)$-index of the output in the contract above is in a sense optional (meaning that the output is not necessarily instantaneous) and relates to the *real-time computation* feature of cyber-physical systems [6].

There is nothing as practical
as a good theory.

— Kurt Lewin

# 5 The algebra of security tests

It is possible to use the algebra of contracts to model security requirements and potentially synthesize them in a traceable manner [144]. However, it is desirable to discuss security as its own entity with its own analysis methods. Of course, any of the theory we will develop in this section can work in tandem with security contracts and, more generally, it ought to be seen as a complimentary view to other methods in the field of model-based security.

In the past decade there has been significant effort in adding formal underpinnings to security [91, 139]. This trajectory is evidenced by the NSF/IARPA/NSA workshop on the science of security, which underlines that there are three areas in need of innovation: metrics, formal methods, and experimentation [52]. In addition, there is still increasing need for defining (in)security as a modeling problem [11, 19, 129]. We develop a *formal method* for *modeling* attacker actions at the abstraction level of component-level system models. Specifically, the categorical result of the Yoneda lemma is used to formally show that if two different architectural implementations of a system agree on every test, then they are behaviorally equivalent. The implication of this result is that an attacker can still effectively exploit a system, even with an inaccurate knowledge base of the architecture, using rewiring and/or rewriting attacks.

Security engineering has moved from the paradigm of securing a list of assets to modeling in graphs of networked components [16], which is more congruent with attacker behavior [89]. These graphs are useful for analyzing the system's secu-rity posture [17]; however, we can improve them further using added structure that comes with categorical models of component-level system models, which are by definition compositional; a useful property for security modeling [48]. The functorial semantics that come with this category-theoretic modeling framework also explicitly relate several views that are essential for the modeling and analysis of cyber-physical systems, where continuous and discrete behaviors interrelate to produce a total behavior. The following formal method is applicable to information technology systems too but the added complexity of cyber-physical systems is useful as a demonstration and to examine formally, in the future, how attacks can lead to hazards and misbehaviors.

Particularly, this structure comes in the form of formal decomposition rules between *system behavior* and *system architecture*, which improves upon current practice where system behavior is disjoint from system architecture. Additionally, this approach provides for *early* security modeling, where engineering decisions are most effective [18, 159], by operating on models instead of implementations. This is possible because categorical semantics reside in a higher level of abstraction than, for example, attack graphs [154], which work best after source code is available. The implementation of this category-theoretic modeling method is what allows us to use the Yoneda lemma to show the impact of exploitable vulnerabilities from an attacker perspective over a system model, which can be incomplete or even partially erroneous compared to the system under attack.

This chapter develops the foundations of security within compositional cyber-physical systems theory [20], a flavor of what Lee calls dynamical computational systems theory [95]. Formal composition rules could overcome some of the new challenges cyber-physical systems introduce to security [37, 64]. Specifically our contributions are in the domain of formal methods for security to assist model-based design of cyber-physical systems using category theory. These contributions are still bound by well-known problems of the foundations and general *science of security*, such as the lack of a well-defined common language [71].

In this chapter we first describe how fundamental concepts in category theory, such as functors to the category of sets, can be interpreted as testing procedures, which we previously used in the context of modeling safety requirements (chapter 4), system behaviors (section 3.2), and architectures (section 3.3). Additionally, results like the Yoneda lemma can be used to infer similarities between systems given the similarities between their test outcomes. Second, we use these insights to model the most common phases of an attack, that consist of learning first, and on the attack itself afterwards, thereby modeling from the attacker perspective formally. Third, we extend the approach to cyber-physical systems by wiring diagrams and their algebras (chapter 3) to provide compelling examples of how our mathematical formalization works in practice, which gives rise to a categorical formal methods for cyber-physical systems security.

## 5.1 Attacks change system behavior

Having fixed some system, by an *attack* we mean any procedure intended to change the system behavior. This definition is very broad, and ranges from privilege escalation in a computer system to sabotaging a car. We consider an *attacker* any actor who performs any such process to degrade a systems behavior.

Traditionally, attacking a system is considered to be an art, and as such it requires a certain degree of heuristics. Our goal is to formalize these heuristics mathematically, using the attacker point of view.

From an intuitive standpoint, it makes sense to divide an attack into two distinct phases: *learning*, where the attacker gathers information about the target system in the hope to find a weakness, and *hijacking*, where the attacker leverages on a given weakness changing the behavior of some component, which reverberates on the system as a whole. The particulars of how these phases are exercised and how they feedback with each other depend on the capabilities and goal of the attacker (in defender terms this would be detailed in the threat model).

To model the first phase, we need a mathematical description of what probing a system for information means. To model the second phase, we need to mathematically express the fact that the attacker can act on a system to change it.

Crucially, there is an intermediate step between these two phases, which we do not describe. We do not describe explicitly how the weakness is turned into an exploiting procedure to change the behavior of some subsystem. Instead, we only describe how the weakness is found, and how the exploiting procedure reverberates in the system as a whole. The reason why this assumption is acceptable is twofold. First, turning a weakness into a viable exploit is heavily dependent on implementation details, for example, the Spectre and Meltdown exploits [109], and our mathematical framework is too general to capture these details successfully. Second, the largest class of attacker actions involves already developed exploits, either internally or through a marketplace, deployed in some sequence to degrade the behavior of the system, without necessarily the attacker precisely knowing how they work [1].

For instance, an attacker may find out by testing (phase 1) that a given laptop uses some particular model of WiFi card. The attacker can then purchase – if it exists – an exploit for the given card, and deploy it. At this point, the behavior of the system as a whole will change (phase 2), for example by giving to the attacker the possibility to run any code on the machine. This example is taken straight out of real experience [116].

Importantly, we make the claim that an attacker is able to hijack a system even having a incomplete view of it, as long as the system and the attacker's mental model are behaviorally equivalent. Mathematically, this relies on the assumption that the exploit being used is invariant under isomorphism of behavior.

ASSUMPTIONS In practice, this means the following: we represent behaviors of systems using the concept of *categorical semantics* (chapter 2). This categorical semantics can be more or less granular, depending on how low-level we want our descriptions to be. For instance, consider specifying behavior in terms of au-tomata. We can represent automata as theoretical objects, but we can also take into account their implementation details, meaning that we not only describe the system behavior under some formalism but are also able to relate it to concrete elements; as a reductionist example, an automaton can be implemented in either a system-on-chip or a field-programmable gate array (FPGA). These would amount to different choices of categorical semantics.

Two automata may be isomorphic in the former setting but not in the latter: this could result in having two different implementations of the same theoretical concept (for example, this is the case in considering the same automaton implemented in two different programming languages). An attack exploiting the automaton design will be isomorphism-invariant in both settings: such an attack exploits the idea that it is possible to start from a state of a given automaton and end up in another state via a legitimate sequence of moves. An attack exploiting the automaton implementation (for example, some weakness of the programming language the automaton is implemented in) will be isomorphism-invariant only in the latter setting: at the theoretical level, our categorical semantics simply is not able to "see" implementation differences.

This is important, because it amounts to saying that "an attack can be carried out even if the attacker has an erroneous view of the system, as long as it is behaviorally equivalent to the system itself" holds only if the categorical semantics developed for application is granular enough to faithfully model the level of generality on which a given exploit acts.

## 5.2 The Yoneda lemma formalizes learning and hijacking

In our effort to formalize attacker actions, we embrace the perspective of the attacker. For us, an attacker is simply an actor wanting to influence or change the behavior of some given system. We do not distinguish between attacks aiming at taking control of the system (as it is common in computer hacking) and attacks aiming at just influencing its behavior to obtain a particular effect (as in sabotage).

To describe the full cycle of an attack, we split it into two main phases:

phase 1 — *Learning (exploration)*, or information gathering which is concerned with probing/eavesdropping the system to discern its behavior. We further divide this activity into *general learning*, where the attacker focuses on gathering a general understanding of the system at hand, and *specific learning*, where the attacker probes the system to understand some specific component design choices.

phase 2 — *Hijacking (exploitation)*, where the attacker, having learned enough information to understand the weak spots of a system, *deploys an exploit* which takes advantage of a found architectural flaw to influence the behavior of the system in some way the attacker desires.

**Example 5.2.1.** Consider an attacker wanting to take control (or to sabotage) a UAV. The attacker starts by learning; that is, gathering information, about the target UAV (phase 1). General learning here can be the attacker trying to understand if the UAV has a GPS module on board. In case the answer is positive, specific information gathering consists in understanding how the GPS module communicates with surrounding units.

This general-specific pattern of learning can be arbitrarily repeated: the attacker could now focus on the GPS module itself to understand if some particular kind of integrated circuit is used in its schematic.

Once learning is completed, the exploit can be deployed (phase 2). In our example, this may be rewriting the firmware of the GPS module *over the air* or, supposing that the attacker has physical access to the UAV, manually rewiring the module or replacing some integrated circuit in it.

One key assumption is that we do not specifically model how a given exploit is developed but only how it is administered, and give a compositional recipe to describe how this change of behavior propagates to the whole system. In practice, this means postulating that the attacker already has access to a knowledge database of tools made to take advantage of a given structural flaw in a given (sub)system. This postulate is not preposterous: indeed, an attacker can hijack a system without personally creating the exploit. This is common in hacking, where *zero day exploits* – that is, exploits of a given flaw that are still unknown to the public, including the target manufacturer – can be commonly bought over the web [1].

### 5.2.1 ATTACKER LEARNING (EXPLORATION)

First we model general learning. Assuming the perspective of an attacker, we want to model what an attacker does to understand the kind of system being dealt with. In practice this includes things such as scanning a computer for open ports, firewall and operating system running on it, probing a piece of hardware to obtain information about the integrated circuits being used, etc.

Algebras on $\mathbf{W}$ take the form of functors $\mathbf{W} \xrightarrow{\mathcal{B}} \mathbf{Cat}$, represent wiring diagrams together with a semantics linking any diagram to the category describing its possible behaviors. It follows that while a wiring diagram is just an object $X$ in $\mathbf{W}$, $\mathcal{B}X$ is instead a category: objects are taken to be general behavior assignments for $X$, while morphisms are taken to be mappings between them that preserve properties we care about. When focusing on a system we are not just fixing a wiring diagram $X$, but also one of the many possible behaviors in the category $\mathcal{B}X$. Hence,

*a system is a pair* $(X, S)$*, with* $X$ *an object of* $\mathbf{W}$ *and* $S$ *an object of* $\mathcal{B}X$.

We suppose we have access to $X$ (this amounts to saying that we are able to distinguish the type of inputs and outputs that our system has) and to $\mathcal{B}X$ (we know what type of system we are dealing with), but not to $S$ (we do not know the specific behavior of the system at hand). The first goal of the attacker is to infer $S$ to the degree that an attack is viable.

First things first, we select a subset of the objects of $\mathcal{B}X$, denoted $K_{\mathcal{B}X}$ (from "known"), representing the systems that the attacker knows or is familiar with. The knowledge base $K_{\mathcal{B}X}$ should not, in general, lift to a functor $\mathbf{W} \to \mathbf{Cat}$, since we do not assume the attacker knowledge to be compositional. For the same reason, given that there will be an injection $K_{\mathcal{B}X} \hookrightarrow \mathcal{B}X$, which represents how the smaller universe of systems known by the attacker *embed* in some bigger universe of the systems being considered. We do not necessarily assume the attacker to have knowledge of this embedding.

**Definition 5.2.1** (Knowledge Database)**.** Given a $\mathbf{W}$-algebra $\mathbf{W} \xrightarrow{\mathcal{B}} \mathbf{Cat}$ and an object $X$ in $\mathbf{W}$, a *knowledge database for* $\mathcal{B}, X$ is a subset $K_{\mathcal{B}X}$ of the objects of $\mathcal{B}X$.

Next, we consider functors $\mathcal{B}X \xrightarrow{\Theta} \mathbf{Set}$. These are interpreted as *tests*, or *probes*:

- Given $S$ in $\mathcal{B}X$, $\Theta S$ represents the information we get in probing $S$ with a test $\Theta$. For instance, $S$ may represent a machine on a network, while $\Theta S$ could represent the output one gets by running `nmap` on $S$.
- If $S \xrightarrow{f} S'$ is a morphism of $\mathcal{B}X$, then $\Theta f$ is a way to transform the information in $\Theta A$ to information in $\Theta B$. This expresses the fact that our tests are well suited to detect the properties we care about, that are preserved by morphisms of $\mathcal{B}X$.
- In this interpretation functoriality holds on the nose: transforming a system by "doing nothing" (identity morphism) should give the same test outcome (functor identity law). Moreover, composing transformations should amount to composing outcomes of the testing.

We package all this information as follows:

**Definition 5.2.2** (Security Test)**.** Given a **W**-algebra $\mathbf{W} \xrightarrow{\mathcal{B}} \mathbf{Cat}$ and an object $X$ in **W**, a *test for* $\mathcal{B}, X$ is a functor $\mathcal{B}X \to \mathbf{Set}$.

Again, the attacker does not have access to $S$ but has access to $\Theta S$ for some tests $\Theta$. This represents the ability of the attacker to perform tests on the system. These tests can be thought of as a persistent reconnaissance mission, which often is the step that takes the longest time and resources of the attacker. The goal of the attacker is to prove in some sense that $S \simeq S'$, for some $S'$ in $K_{\mathcal{B}X}$. This amounts to say that the system $S$ is an instance of a system $S'$ the attacker is familiar with. Given our assumption, we can then postulate that the attacker knows an exploit for $S'$ to move to phase 2.

We make the assumption that, for any $S' \in K_{\mathcal{B}X}$, the attacker has access to $S'$. This assumption is natural, since $S'$ is by definition in the knowledge base of the attacker. In particular, we assume that the attacker is able to perform *any* test to *any* known system, hence

$$\textit{for any } S' \in K_{\mathcal{B}X} \textit{ and } \mathcal{B}X \xrightarrow{\Theta} \mathbf{Set}\textit{, the attacker has access to } \Theta S'.$$

The Yoneda lemma says that if $\Theta S \simeq \Theta S'$ for all $\Theta$, then $S \simeq S'$. Let us try to interpret what this means in our framework, considering some corner cases.

- Suppose that for some object $S$ in $\mathcal{B}X$ there is an object $S'$ in $K_{\mathcal{B}X}$ such that $S \simeq S'$. If the attacker has access to $\Theta S$ for any $\Theta$, then the attacker will be able to conclude $S \simeq S'$ from $\Theta S \simeq \Theta S'$. That is, if the attacker is free to perform any form of testing and possesses a vast knowledge database, then $S$ can be determined with absolute precision.
- If the attacker has access to any $\Theta$, but there is no $S'$ in $K_{\mathcal{B}X}$ such that $S \simeq S'$, then the attacker won't be able to conclude $S \simeq S'$: Tests can be arbitrarily precise, but the attacker has no ability to interpret them.

- If there is an object $S'$ in $K_{\mathcal{B}X}$ such that $S \simeq S'$, but the attacker has no access to all $\Theta$, then it won't be able to conclude *with certainty* that $S \simeq S'$, because the Yoneda lemma does not hold in this setting. Still, after performing enough tests, the attacker may be prone to *infer* that $S \simeq S'$ if $\Theta S \simeq \Theta S'$ for *enough* $\Theta$ ran. This inference comes with a degree of uncertainty, which is exactly what makes information gathering an art more than a science.

That is, the Yoneda lemma provides a formal justification of the fact that testing a system extensively is enough to sufficiently characterize its behavior. We call this heuristic *Yoneda reasoning.*

Some tests are more informative than others. For instance, the "terminal test" $\Theta$ sending any $X$ to the singleton set $\{*\}$ is maximally uninformative: the result of this test is the same for any system. On the contrary, a functor that is *injective on objects* lifts to a test that yields the conclusion $X = Y$ from $\mathcal{B}X = \mathcal{B}Y$. Further formalizing the possible spectrum of tests, hopefully weighting them with probability distributions to model their reliability, is an ongoing direction of future work. Now we suppose that the attacker pinned down the target system $S$ with some precision. The next step of an attack is harvesting information about the architectural design choices implementing the system. That is, after we know how $S$ works, we need to find out what $S$ is made of.

Previously, we saw that a system is made of an object $X$ in $\mathbf{W}$ together with an object $S$ of $\mathcal{B}X$, for some $\mathbf{W}$-algebra $\mathcal{B} : \mathbf{W} \to \mathbf{Cat}$.

Now we consider the *category of architectural choices for* $X$, that is, the slice category $\mathbf{W}/X$. Objects of this category are morphisms $\bigotimes_i X_i \xrightarrow{\phi} X$, while morphisms are morphisms of wiring diagrams making the following triangle commute.

$$
\begin{array}{ccc}
\bigotimes_i X_i & \xrightarrow{\quad f \quad} & \bigotimes_j X_j \\
& \phi \searrow \quad \swarrow \psi & \\
& X &
\end{array}
$$

Architectural choices form a category, so we can repeat the reasoning done in the last section using $\mathbf{W}/X$ as the category on which we probe. Again using Yoneda, the attacker is able to ascertain that a given system $(X, S)$ is actually made of subsystems $(X_i, S_i)$, tensored and wired together by $\phi$. At this stage the attacker still does not know anything about the $S_i$, so the process must be repeated cyclically.

Summarizing, phase 1 is modeled as follows:

1. The attacker uses tests on $\mathcal{B}X$ and Yoneda-reasoning to find the system $S$ representing the semantics of $X$.
2. The attacker uses tests on $\mathbf{W}/X$ and Yoneda-reasoning to find the wiring $\bigotimes_i X_i \xrightarrow{\phi} X$ representing the implementation of $X$.
3. The attacker repeats step 1 on any $\mathcal{B}X_i$ of interest to find the precise behavior of the subsystem marked with $X_i$.
4. The attacker repeats step 2 on $\mathbf{W}/X_i$ to obtain more information about the subsystems making up $X_i$.
5. These steps are iterated cyclically until the attacker has gathered enough information to move to exploiting.

In practice, tests will not have to be materially re-run on every $\mathcal{B}X_i$: it is very likely that the attacker has only access to $\Theta S$—every $(X_i, S_i)$ being a subsystem of $(X, S)$ that may not necessarily be exposed to external testing. Nevertheless, it will always be the case that $\Theta(\mathcal{B} \bigotimes_i X_i) \xrightarrow{\Theta\mathcal{B}\phi} \Theta\mathcal{B}X$, meaning that the outputs of tests over every $S_i$ will have to be reconstructed from tests over $S$. This adds another layer of uncertainty for the attacker, who has to devise tests for which the mapping $\Theta\mathcal{B}\phi$ acts as transparently as possible. Again, this backs up intuition, going back to example 5.2.1, if some system $(X, S)$ comprises a subsystem $(X_i, S_i)$ (say, a GPS module), then we could devise a test on $\mathcal{B}X$ such that in $\Theta S$ the behavior of the subsystem $S_i$ is made apparent. Similarly, when running `nmap` on a system we can get extra information about which services are running behind which port, for example, `nginx` behind port `80`. By probing the system as a whole, the attacker is getting information about its subsystems.

### 5.2.2 ATTACKER HIJACKING (EXPLOITATION)

Now suppose that the attacker has a good grasp of the system behavior and architecture, and model the last step, in which the system is hijacked and exploited.

We distinguish between two main kinds of attacks:

1. *Rewriting attacks*, where the behavior of a (sub)system is changed. Practical examples of this are, for instance, exploiting a vulnerability in a WiFi card to rewrite its firmware, and using this change of behavior to progress towards obtaining administrative privileges over the whole machine;
2. *Rewiring attacks*, that are concerned with modifying the way subsystems communicate with each other.

**Definition 5.2.3** (Rewriting Attack). Given a $\mathbf{W}$-algebra $\mathcal{B}$, a *rewriting attack* for

$\mathcal{B}$ is a monoidal natural transformation

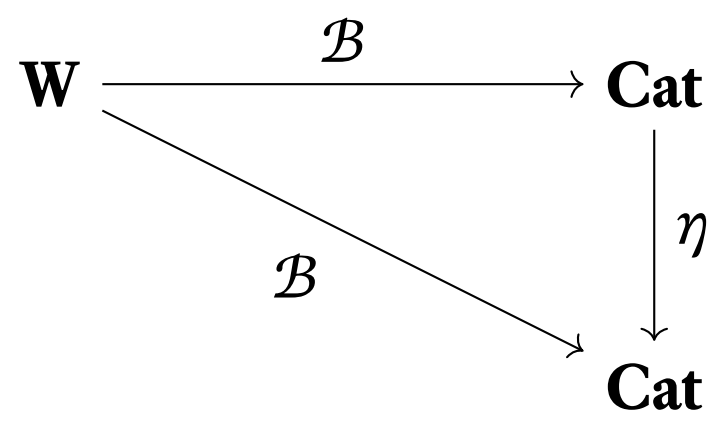

In a rewriting attack we do not change the possible behaviors that are assigned to wiring diagrams, but we reshuffle them. If $(X, S)$ was our system, then $S$ is an object of $\mathcal{B}X$, which is in turn an object of **Cat**. We then see that $\eta_X$ is a morphism $\mathcal{B}X \to \mathcal{B}X$ in **Cat**, that is, a functor $\mathcal{B}X \to \mathcal{B}X$. Applying it to $S$ we get that the behavior of our new system is $\eta_X S$.

The fact that $\eta$ is natural monoidal yields a compositional description of an attack. In modifying the behavior of a given subsystem, we can infer how the behavior of the whole system changes.

**Definition 5.2.4** (Rewiring Attack). Given a **W**-algebra $\mathcal{B}$, a *rewiring attack for* $\mathcal{B}$ is a functor

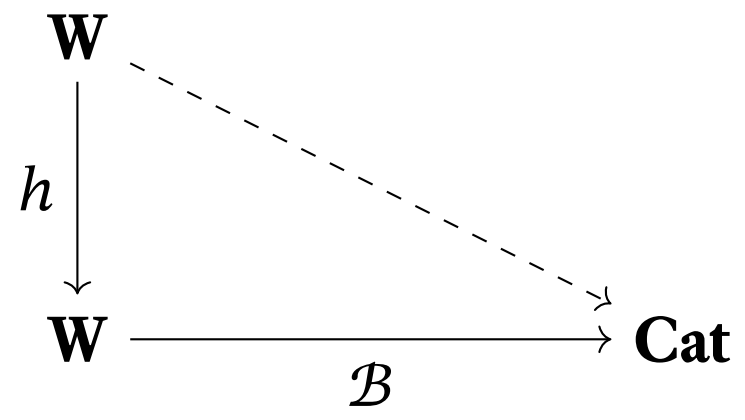

Here $h$ represents a reshuffling of the wirings and boxes. The new behavior can then be obtained by considering the composition $\mathcal{B} \circ h$. If we started with system $(X, S)$, now the behavior of the hijacked system is $(hX, (\mathcal{B} \circ h)X)$. Again, the nature of this attack is compositional: It describes how altering systematically a wiring in **W** resonates through all the systems modeled by $\mathcal{B}$.

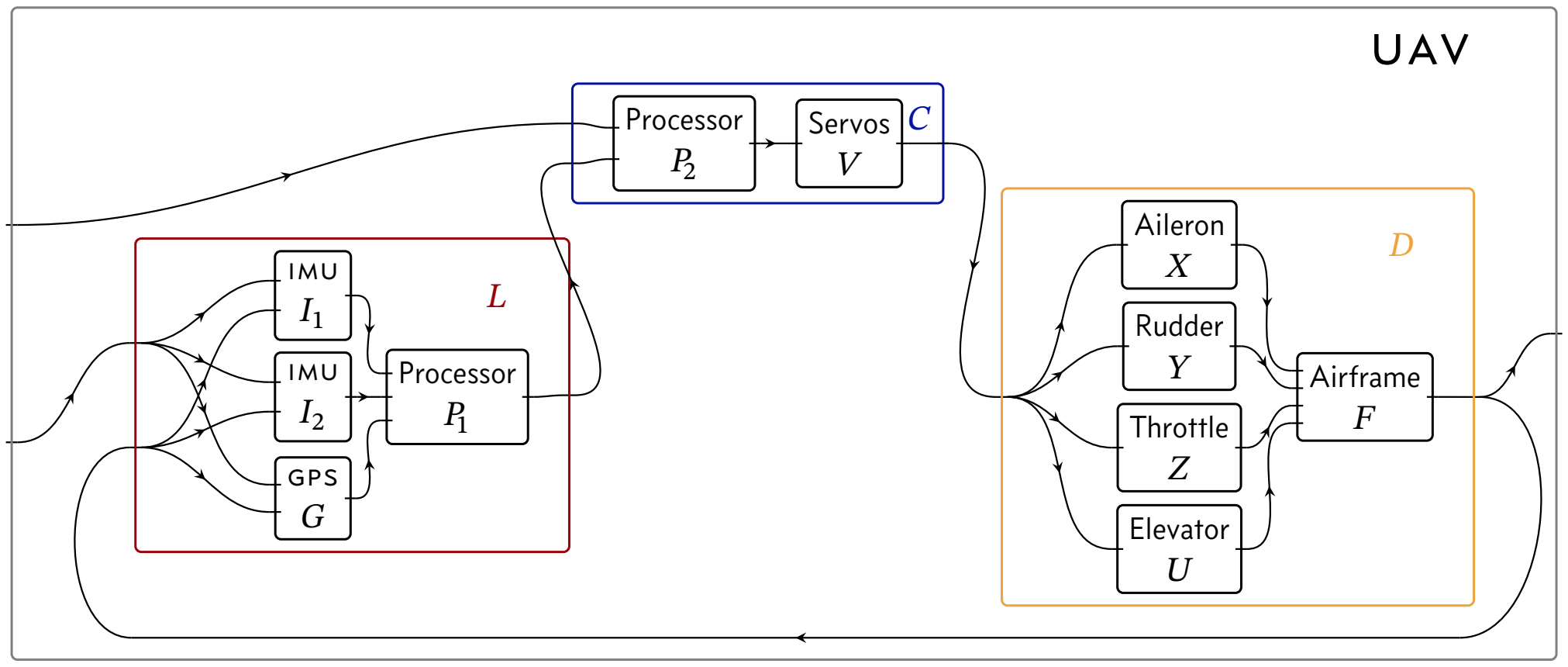

Figure 5.1: The hierarchical decomposition from behavior to candidate system architecture is formally contained within the slice category $\mathbf{C}/c$ in which there exist all possible design decisions that adhere to the behavioral model as defined in the abstraction above the system architecture. We segment here to subsystems following a behavior decomposition to sensors $L$, controller $C$, and dynamics $D$. Split wires indicate function duplication, $\Delta$.

Both rewriting and rewiring attacks form categories. This conforms with our intuition that attacks can be performed in batches, or stacked one on top of each other, which in categorical terms means that we can consider the nature of attacks as being compositional.

## 5.3 Yoneda reasoning models security posture

We demonstrate the use of the Yoneda lemma as a possible model for security tests and exploitation method over a system model of a UAV from the perspective of attacker actions (example 5.2.1).

The UAV is composed of a sensor unit (denoted $L$, in red), of a controller unit (denoted $C$, in blue), and and of a dynamics unit (denoted $D$, in yellow) (figure 5.1). Each of this unit is itself composed of various subsystems. There are multiple possible system architectures that can implement this higher level behavior. We assume that the attacker is familiar with the general class of vehicle cyber-physical systems. For illustrative purposes we focus on one possible but relatively simple

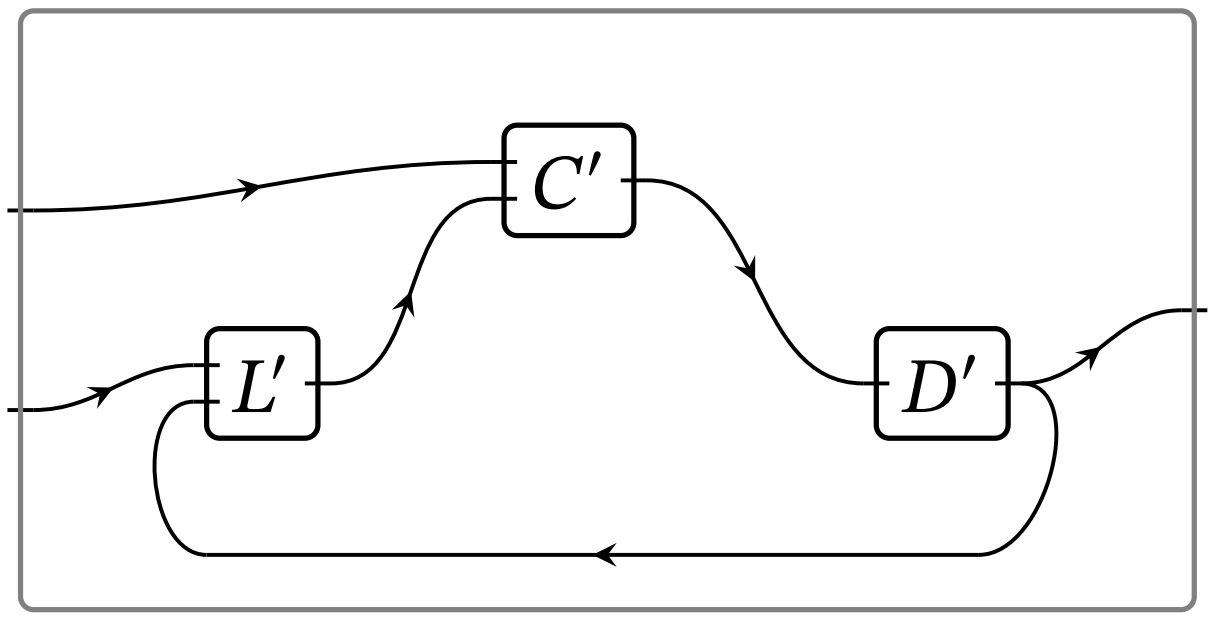

Figure 5.2: The attacker's understanding of the system after the first cycle of learning.

system architecture.

Let's start with the minimum observability possible—for the attacker the UAV is just a black box with three inputs and one output, which we denote by the box UAV. The first step of the attack is gathering information about its behavior.

Our systems-as-algebras model (section 3.2) represents the assignment of behaviors to wiring diagrams with a functor

$$\mathcal{B} : \mathbf{W} \to \mathbf{Cat}.$$

In our running example, $\mathcal{B}(\text{UAV})$ denotes the category of all the possible behaviors that we can assign to the UAV box. Thus, our UAV is a pair $(\text{UAV}, S)$ with $S$ an object of $\mathcal{B}(\text{UAV})$ representing the particular UAV model at hand.

In general information gathering, the attacker uses Yoneda reasoning to infer $S$. To do so, the attacker must be able to perform a set of tests on the system $S$. Any of such tests is a functor

$$\Theta : \mathcal{B}(\text{UAV}) \to \mathbf{Set}$$

and the result of the test $\Theta$ applied to $S$ is denoted $\Theta S$ (in category theory practice, it is customary to avoid parentheses whenever possible). In our particular case, $\Theta$ may be a test that analyzes the aerodynamics of the UAV during flight.

The more informative $\Theta$ is, and the bigger the number of $\Theta$'s the attacker can have access to, the more it will be likely to infer $S$. If the attacker finds that $S \simeq S'$ for some $S'$ in their knowledge database, $K_{\mathcal{B}(\text{UAV})}$, then the attacker will know how the UAV behaves.

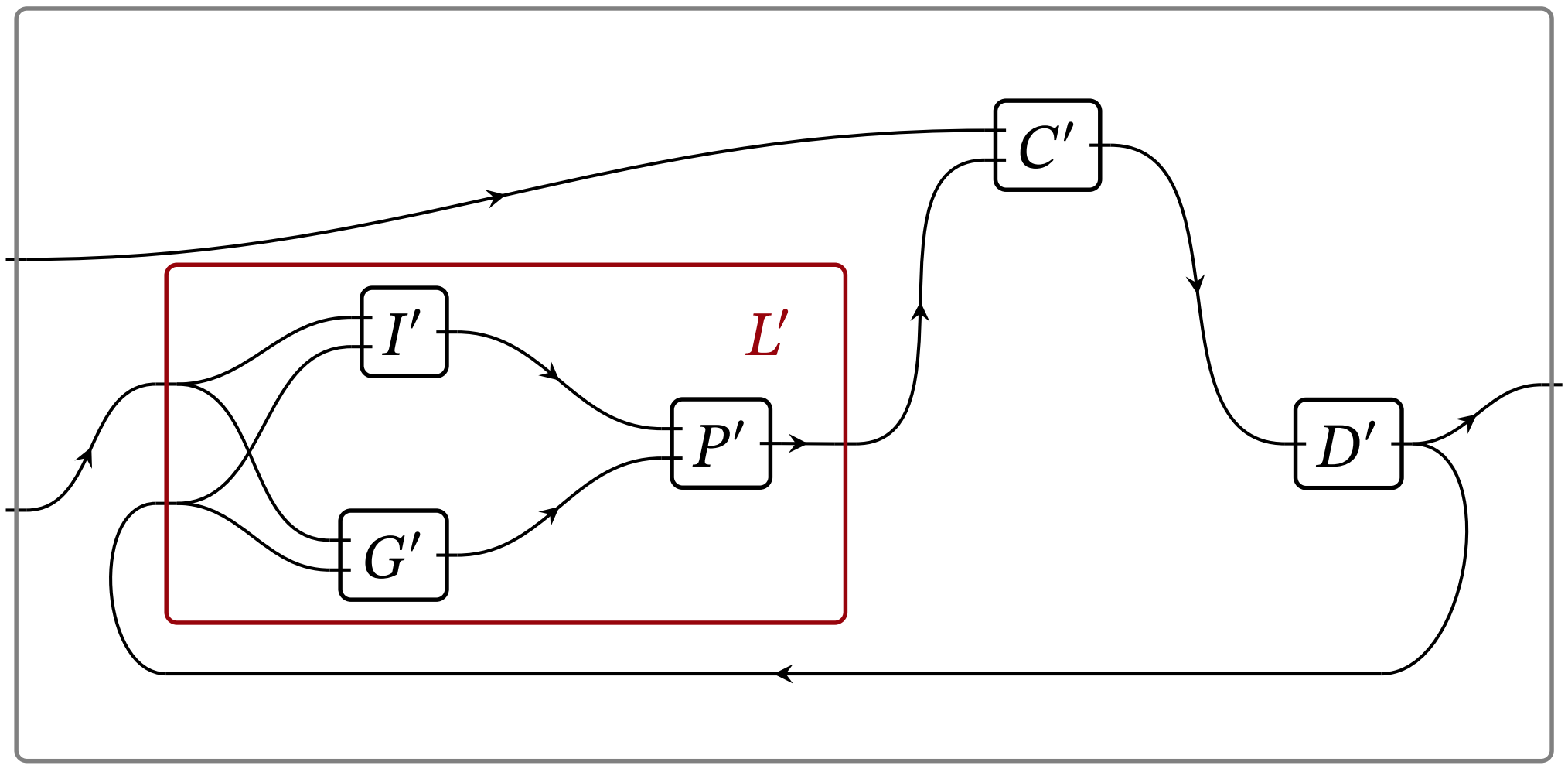

Figure 5.3: The architecture of the sensory system as understood by the attacker, which is in reality erroneous but behaviorally equivalent. The attacker found that there is one IMU (when in reality there are two) and a GPS.

In practice, it is very unlikely for the attacker to have access to *every* test $\Theta$. This entails that the attacker will not be able to infer with certainty that $S \simeq S'$: most likely, the attacker will be *prone to infer $S \simeq S'$ with a certain degree of confidence*. As such, the outcome of testing is probabilistic more than deterministic.

Assuming that the attacker inferred $S \simeq S'$ for some system $S'$ in their knowledge database, the particular design choices making up the UAV are still unknown to them. Applying Yoneda reasoning again to the category $\mathbf{W}/\text{UAV}$, they may be able to infer some of these design choices. For instance, it could be possible to infer an initial understanding of what the UAV is composed of (figure 5.2).

The attacker sees the system as decomposed into boxes $L'$, $C'$ and $D'$, which will be behaviorally equivalent to $L$, $C$ and $D$, respectively. The inner workings of such boxes are still unknown to the attacker, that we now assume focuses on understanding the sensory system, $L'$. This amounts to repeat the same cycle of Yoneda reasoning focusing on tests that target $L$ in particular. After running these test, the attacker sees a first approximation of the UAV (figure 5.3).

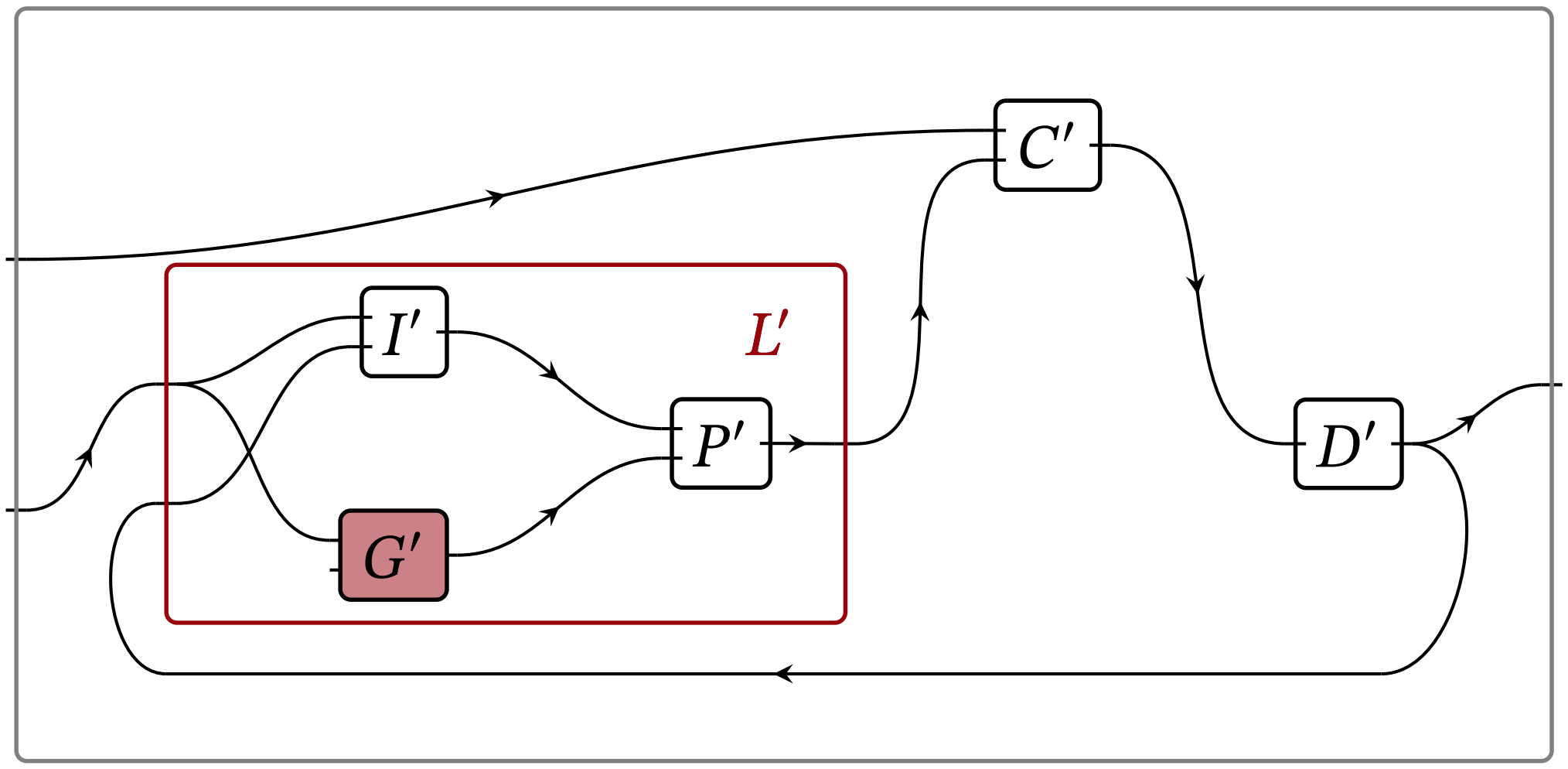

Figure 5.4: The compromised UAV, with GPS firmware hacked and input wires to the IMU units swapped.

The initial understanding of the attacker is slightly erroneous, because the two separated IMU units in UAV have been conflated into one. Still, the two wiring diagrams are behaviorally equivalent. This both reflects the fact that on one hand Yoneda reasoning is probabilistic in nature, and on the other that identification of system happens only up to behavioral equivalence.

Now, suppose the attacker decides to do two things: rewriting the firmware of the GPS module $G$, and cutting its feedback input. The first is a rewriting attack and the second is a rewiring attack.

The rewriting attack is represented by a monoidal natural transformation $\eta$ : $\mathcal{B} \to \mathcal{B}$. This means that for each box $X$, $\eta_X$ is a functor $\mathcal{B}X \xrightarrow{\eta_X} \mathcal{B}X$. The components on the wiring diagrams $I'$, $P'$, $C'$ and $D'$ are just the identity functors—these boxes are mapped to themselves unchanged. On $G'$, the functor $\eta_{G'} : \mathcal{B}G' \to \mathcal{B}G'$ is instead different from the identity. If $g$ is the object of $\mathcal{B}G'$ representing the particular GPS unit used in the system (UAV, $S$), then $\eta_{G'}g$ represents the GPS module with its firmware rewritten. The behavior of $\eta$ on the composite boxes $L', C', D'$ and their composition can be inferred from the fact that $\eta$ is a natural transformation, and hence the naturality squares have to commute.

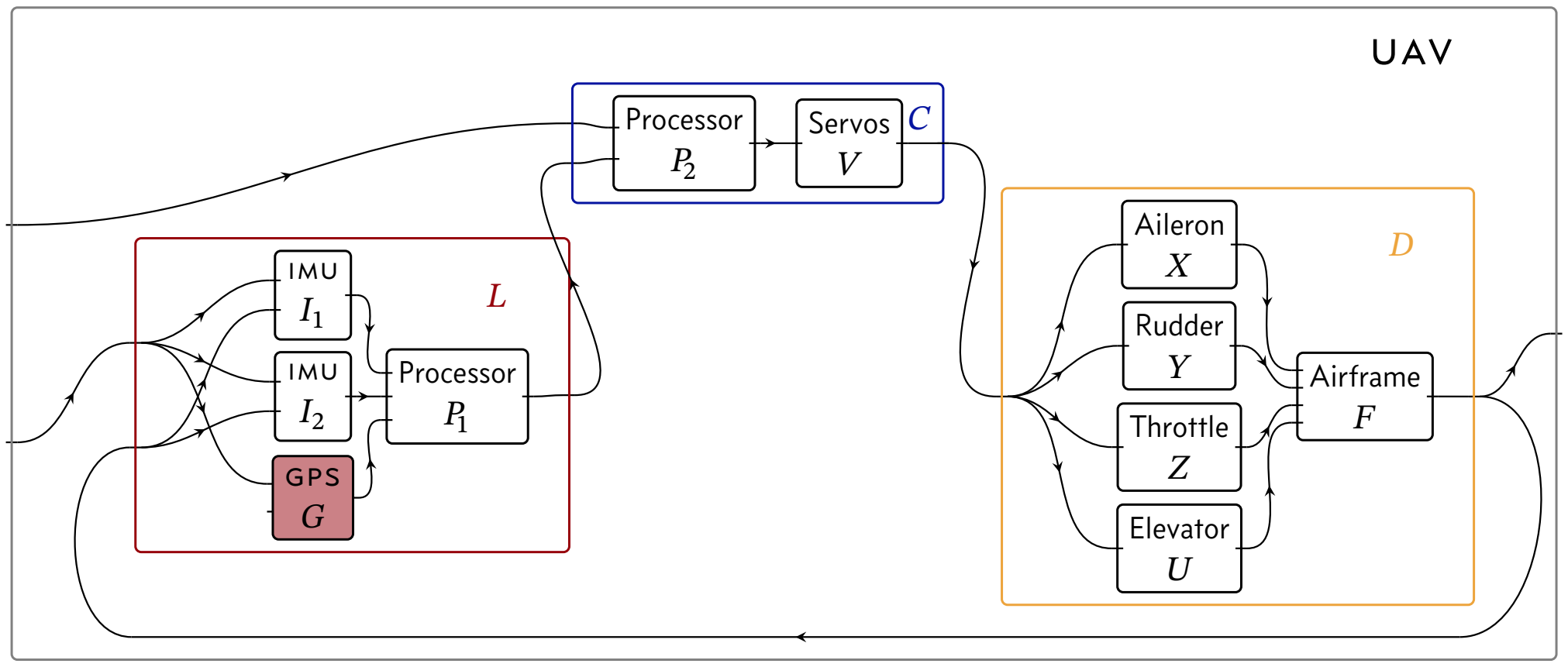

Figure 5.5: The UAV (figure 5.1), after the attack.

The rewiring attack, instead, is an endofunctor $\mathbf{W} \to \mathbf{W}$, which we set to be identity on objects. On morphisms, it remaps the wiring of $G'$ inside the box $L'$ (figure 5.3) such that there is no wire coming from the outside (figure 5.4). In sequence, we can represent the effect of both attacks as changes in boxes and wires within UAV, as perceived by the attacker. The resulting wiring diagram is completely determined from the initial attacker learning wiring diagram (figure 5.3) by the fact that behavior assignment is functorial; compositionally allows us to infer the behavior of the composite after having replaced $G'$ and having swapped the wirings of the IMU units.

The resulting wiring diagram (figure 5.4) represents how the attacker presumes the system will behave after the exploit is deployed, assuming the system was correctly profiled in phase 1. Since phase 1 has margin for error, the attacker can re-probe the system to assure that the perceived behavior is compatible with reality. This further round of testing is necessary to assert with confidence that the exploit has been deployed correctly.

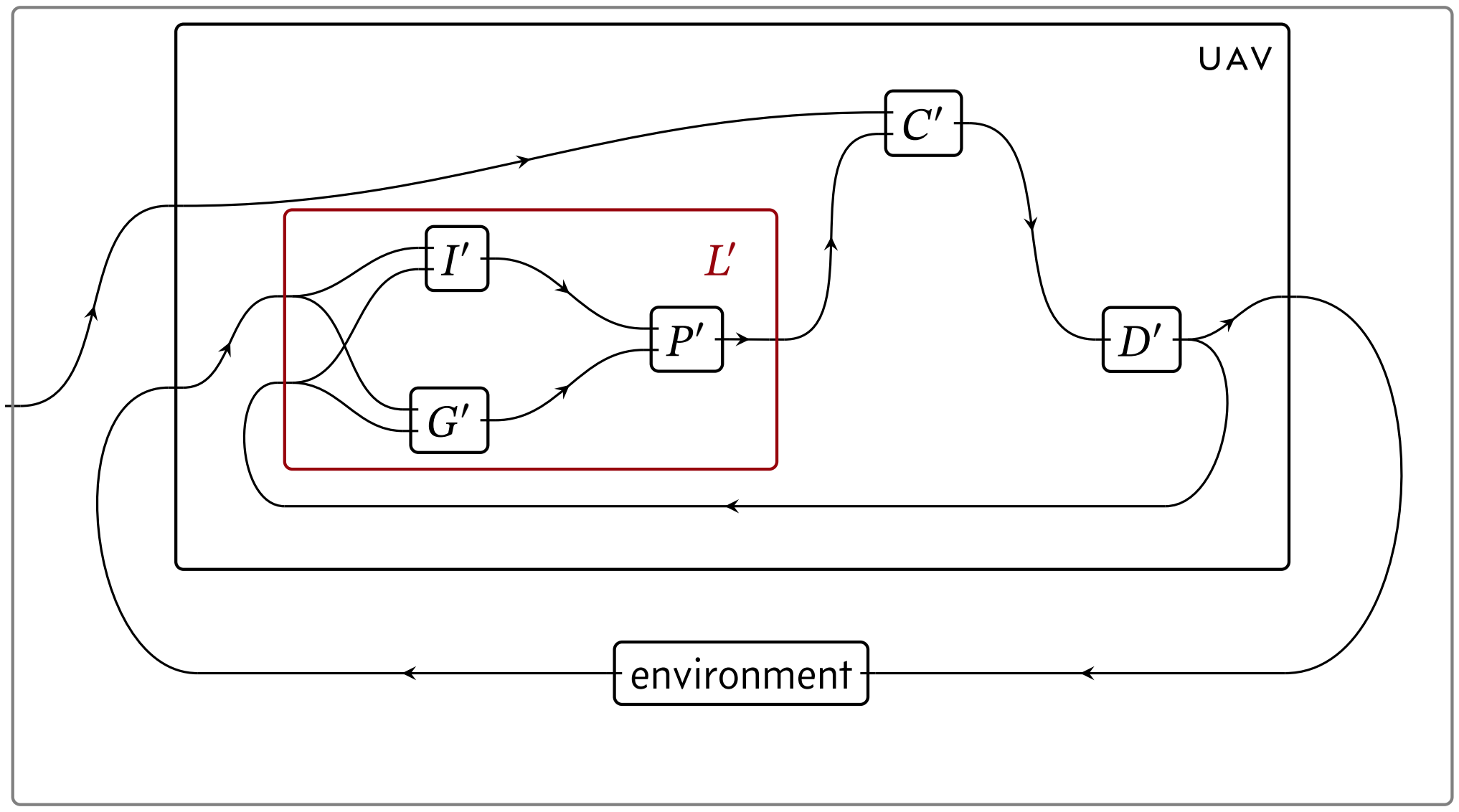

Figure 5.6: Feeding counterfeit GPS signals to the UAV hijacks the box "environment".

In this example, we postulated that the attacker was indeed able to correctly gather information about the UAV. Formally, we expressed this by stating that what we consider to be the actual UAV (figure 5.1) and the attacker's understanding of what the UAV is (figure 5.3) are behaviorally equivalent. Because functors preserve isomorphisms, we are able to describe how the attack impacts the real UAV (figure 5.5).

One other attack we can model in this framework is feeding a counterfeit GPS signal to the UAV to compromise it. This sort of attack is documented "CAPEC-627: Counterfeit GPS Signals" and is considered difficult to realize. The wiring diagrams formalism gives us an idea of why: feeding a counterfeit GPS signal does not involve modifying the GPS module in any way. What changes is the information traveling on the GPS wires which communicates with the outside world.

So, to understand this attack properly, one needs to model how the UAV interacts with the environment it is in (figure 5.6). Here, by the box "environment" we mean a process that given the UAV position in space and time returns the data sensed by the IMU and GPS units. In this sense, we see that spoofing a GPS sig-nal does not amount to intervening on the UAV, but on the environment itself. Attackers rarely have the capacity to control the environment within a region of space and time—radio waves from the GPS satellites in this particular case—that is big enough to influence the behavior of the single UAV.

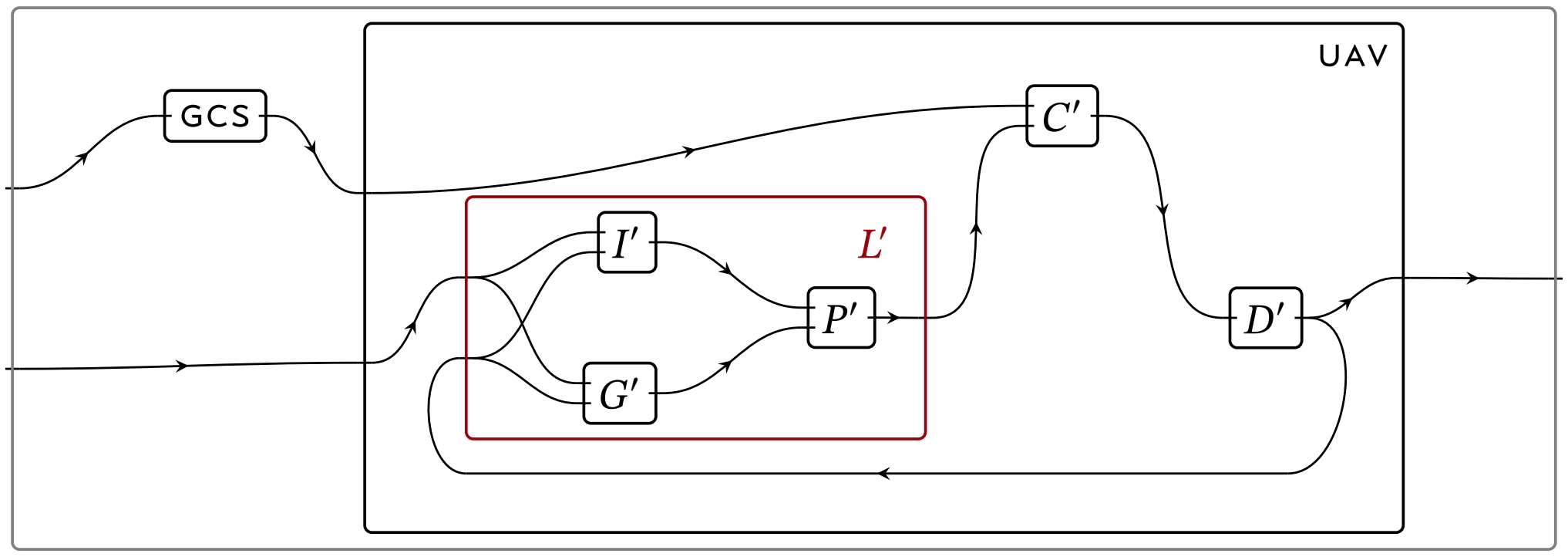

Figure 5.7: Social engineering attacks may hijack the ground control station (GCS).

To conclude, we present another possible attack, performed by means of *social engineering*. As with the previous example, social engineering does not exploit the UAV itself, but instead takes advantage of the human factor surrounding it. Examples of this may include bribing whoever programs the UAV goals, or making the control tower believe that a given order has been officially issued from whoever is in command, for example, as defined in "CAPEC-137: Parameter Injection."

As in the previous case, the behavior of the UAV as a wiring diagram is unchanged. Instead, what changes is the information traveling on the first input wire of the UAV box. From our perspective, this requires again to put the UAV into context (figure 5.7). An attack based on social engineering will consist in rewriting the box GCS, which abstracts away a possible ground control station.

We can implement the categorical semantics of cyber-physical security modeling for the UAV algorithmically (Listing 5.1 & 5.2).

Listing 5.1: Modeling attacker learning

```
-- Define category of wiring diagrams
W : Category
W  = Definition 2.2.1

-- Define functor for UAV modeling
𝓑 : Functor W → Cat
𝓑 = assignment of UAV behavior (LTIS) to boxes

-- Model UAV as a 2-input 1-output W-box
UAV  : W
UAV = (2,1)

-- Knowledge database
K_𝓑(UAV): List 𝓑(UAV)
K_𝓑(UAV) = attacker knowledge for systems of type UAV

-- Compare tests with target S
CompareTests : (Functor 𝓑(UAV) → Set) → 𝓑(UAV) → Bool
CompareTests ΘS′ = Θ(K_𝓑(UAV)(S′)) ≃ Θ(S)

-- Yoneda reasoning
for each Θ : Functor 𝓑(UAV) → Set
filter (CompareTests Θ) K

-- Running security tests reveals the following boxes
L′, C′ , D′  : W
L′, C′ = (2,1), D′ = (1,1)
diagram : Morphism W (L′ ⊗ C′ ⊗ D′) → UAV
diagram = (in, out)
 where
  in : Morphism Set UAV_in × (L′_out × C′_out × D′_out) → (L′_in × C′_in × D′_in)
  in u1 u2 l c d = (u2, d, u1, c)
  out : Morphism Set (L′_out × C′_out × D′_out) → UAV_out
  out l c d = d
```

Listing 5.2: Modeling hijacking

```
-- By iterating learning further decompose L′
I′, G′ , P′  : W
I′, G′ , P′ = (2,1)

Ldiagram : Morphism W (I′ ⊗ G′ ⊗ P′) → L′
Ldiagram = (in, out)
where
  in : Morphism Set (L′_in × (I′_out × G′_out × P′_out) → (I′_in × G′_in × P′_in))
  in l1 l2 i g p = (l1, l2, l1, l2, i, g)
  out : Morphism Set (I′_out × G′_out × P′_out) → L′_out
  out i g p = p

-- Rewriting attack
η : NatTrans (Functor W → Cat) → (Functor W → Cat)

-- η is the identity on everything but G'
η G′ : Morphism  ℬG′ → ℬG′
η G′ = firmware rewriting

-- Rewiring attack
Lattack : Morphism W (I′ ⊗ G′ ⊗ P′) → L′
Lattack = (in, out)
where
  in : Morphism Set (L′_in × (I′_out × G′_out × P′_out) → (I′_in × G′_in × P′_in)
  in l1 l2 i g p= (l1, l2, l1, 0, i, g)
  out : Morphism Set (I′_out × G′_out × P′_out) → L′_out
  out i g p = p

Rewiring : Functor W → W
Rewiring Ldiagram = Lattack

-- The modified behavior of the hijacked UAV
behavior : ℬ(UAV)
behavior = η UAV (ℬ(Rewiring (UAV))) (S′)
```

LIMITATIONS Based on how we defined Yoneda reasoning, we identify several limitations. These limitations can be overcome by enriching over metric spaces, which we will also discuss. The main point of this section is to set a strong theoretical footprint of category theory and the diagrammatic reasoning that emerges in the application of securing cyber-physical systems. Making the results probabilistically concrete is a potential future topic that can be based on the above formal methods.

One such limitation can be inspected from the resulting algorithm (Listing 5.1). As output we may have that Yoneda reasoning returns no result (the list being filtered from `K` is empty, meaning that the attacker does not have entries in the knowledge database that adequately model the target system), but we may also have that it returns more than one (the list being filtered from `K` having more than one element, meaning that the test performed were not fine-grained enough to pinpoint the target system with deterministic accuracy). This is mainly due to the nature of the tests performed. Indeed, some tests could be more informative than others; sending any system in $\mathcal{B}(\text{UAV})$ to the one-element set defines a functor to **Set** and hence a valid test, which is though maximally uninformative since the test outcome will be the same on all systems; on the other hand, any injective-on-objects functor $\Theta$ allows us to conclude with certainty that $\Theta S = \Theta S'$ implies $S = S'$, and is hence maximally granular. The formalism as we presented it has no way to express which subset of the tests we can perform allows to individuate the target system unambiguously.

Indeed, there are tests with different degrees of expressiveness, and the Yoneda lemma does not account for this; we are able to conclude $S \simeq S'$ using Yoneda lemma if and only if $S$ and $S'$ agree on *all* tests, including the maximally useless ones. This is why we speak of Yoneda reasoning as a heuristic and not as a deterministic procedure.

Looking at things more abstractly, the ultimate reason for this shortcoming lies in the fact that in our definition of category we considered homsets to be sets; that is, we speak of the *set* $\mathrm{Hom}_{\mathcal{C}}[A, B]$ of all possible morphisms from $A$ to $B$ in some category $\mathcal{C}$. Sets have very little structure, and in such an environment we cannot formulate the Yoneda lemma to be more expressive.

In a probabilistic setting, what we would like to have is a version of Yoneda reasoning that gives an *interval of confidence* relating $\Theta S \simeq \Theta S'$ and $S \simeq S'$ for any possible test $\Theta$. In other words, we want to attach to each $\Theta$ a measure of how informative $\Theta$ is in our context.

This is certainly doable by resorting to *enriched category theory*, which is a generalization of category theory where homsets can have more structure. In particular, we can reformulate our theory using categories enriched over metric spaces. This gives a natural way of talking about *distances* between sets and this can be used to ultimately define a measure on the tests we can perform. In the context of enriched category theory, the Yoneda lemma can be reformulated in what it is informally known as *ninja Yoneda lemma* [110], which takes into account this additional structure. We can use this to define a version of Yoneda lemma that has a notion of confidence in the tests we perform over the cyber-physical system model and, therefore, have some granularity of what it means for two systems to be behaviorally equivalent under *some* (informative) tests.

BENEFITS While we showed some limitations with regard to the flexibility of the model, it is important to point out that the same flexibility can be beneficial. Different formalisms can inhabit the boxes defining a systems behavior, from Petri nets, to transition systems, to ordinary differential equations. The development of our formalisms will allow us to speak about all these representations within one framework. For example, applications to security modeling using Petri nets [99] is currently congruent with research in category theory and Petri nets [13, 58–60] and could be leveraged to make the application of the preceding formalism more concrete as a model of (mis-)behavior. Similarly, both models of security violations in automata [166] and continuous controller behavior [137] can be represented within our framework and, therefore, allow for a plethora of analyses within one model.

However, in order to drill further in the possible directions of describing different types of continuous, for example false sensor data, or discrete, for example transitioning the system to a hazardous state, misbehaviors caused by exploitation we require a first formulation of security modeling categorically and algebraically. Relaxing some of the unrealistic assumptions we made is then a matter of incorporating developing work from category theory.

Additionally, this security framework is part of a alternate paradigm of systems modeling that has its foundations in categorical modeling. In this framework it is possible to provide formal traces requirements, behaviors, and architectures [20] but also describe a vast amount of dynamical systems with applications to robotics, event-based systems, and hybrid systems, to name a few [47, 55, 107, 172].

Finally, we addressed the theoretical underpinnings of security modeling in category theory. But, the recent surge of categorical modeling languages and software, such as Catlab [68] or idris-ct [61] or algebraic databases [152], can be used to create modeling tools and security assessment methods based on our work practical within compositional cyber-physical systems theory.

We developed a categorical semantics for cyber-physical systems security modeling that is able to determine that two system representations are behaviorally equivalent provided that they agree on every test. The implication of this statement is that it is possible to model attacker actions without necessarily needing to give the attacker full observability on the system under attack. Additionally, we model two types of attacks on the incomplete but erroneous view of the attacker and show its impact on (what we consider to be) the real system. These attacks can do either of two things: (1) rewriting some system component or (2) rewiring an input or output from or to a component. This model is particularly useful for cyber-physical systems, where in the future we would like to say how a particular attack can change system behavior and, therefore, potentially transition it to a hazardous state. Overall, we model how the attacker *learns* about a system and how an attacker then might attempt to *hijack* the system from the knowledge that they were able to gather in a formal, unified way. Finally, the categorical formalism can be considered foundational, in the sense that in addition to the contributions above it can subsume already developed formalisms for modeling attacker actions, for example attack graphs, or augment the information contained in the model by using security frameworks.

My last request: Everything I leave behind me … in the way of diaries, manuscripts, letters (my own and others's), sketches and so on, to be burned unread.

— Franz Kafka

# 6 On unification

Having manifested the wiring diagram formalism for behavior, architecture and requirements of a UAV, we now summarize and further discuss how this categorical interpretation of cyber-physical systems models leads to unification of these aspects of system design and analysis.

Starting with some cyber-physical process $Y$, we usually model its behavior, mathematically described for example via some equations, and also the requirements it satisfies or should satisfy. We earlier discussed Moore machines and linear time-invariant systems; there can be other algebras of system behavior,[1] so here we generically speak of the "behavior algebra" which is any one of them, using the notation $\mathcal{B}$. As we saw, categorically these are certain objects $B_Y \in \mathcal{B}(Y)$ of the category of all the possible behaviors (section 3.2), and similarly the requirements are objects $R_Y \in \mathcal{C}(Y)$ of the category of all contracts that could be associated to such a process, via lax monoidal functors

$$\mathcal{B}, \mathcal{C} : \mathbf{W} \to \mathbf{Cat}.$$

To formally discuss and capture the behavior and requirements in terms of subprocesses, the designer first chooses some valid architecture of $Y$ which is categorically expressed by choosing a morphism $f : X \to Y$ in the category $\mathbf{W}$, namely an element of the slice category $\mathbf{W}/Y$ (section 3.3). Then the behavior algebra and

[1] For example, *machines* serve as an all-inclusive general system notion that allows us to compose systems of different description [156, § 4].

requirements algebra, independently, produce assignments

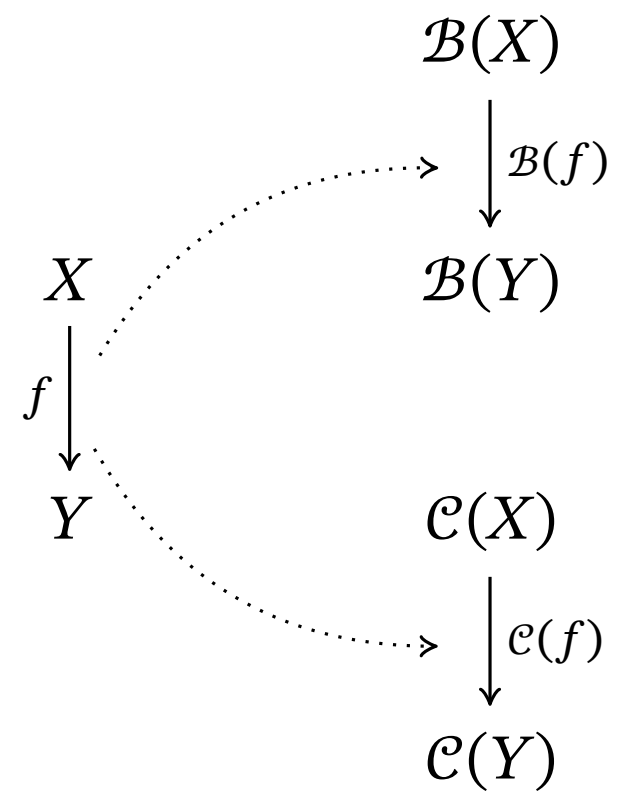

(6.1)

The designer then decides on "pre-image" objects $B_X \in \mathcal{B}(X)$ and $R_X \in \mathcal{C}(X)$ which, under these functors on the right-hand side, produce the original composite behavior and requirement on $Y$. As we saw, there could be multiple choices for $B_X$ and $R_X$. Also, the designer can decompose even further to subprocesses, on which the analysis carries on in the same formal way. Moreover, they may choose to go back and change the architecture to some alternative implementation $g : Z \to Y$, if that is physically sensible and allows to easier obtain the end results. Later on, using algorithms such tests could assist in deciding on the most optimal solutions.

The main contribution is we now sketch some additional connections between these two independent algebras of behavior and requirements, which further clarify their formal relation.

First of all, there is an *algebra map*[2] $\alpha : \mathcal{B} \Rightarrow \mathcal{C}$ which assigns to each specific physical behavior of a process $B_Y \in \mathcal{B}(Y)$, the *maximally satisfied* contract by it, $\alpha_Y(B_Y) \in \mathcal{C}(Y)$. Informally, if a box $\mathbb{R} \!-\!\boxed{X}\!-\! \mathbb{R}$ is inhabited by the function $f(x) = 6x$, its maximally satisfied contract is in effect $\{(a, 6a) \mid a \in \mathbb{R}\} \subseteq \mathbb{R}^2$. However, the system also satisfies the contracts $\mathbb{R} \times 6\mathbb{R}$ or $\mathbb{R} \times 3\mathbb{R}$, or even $\mathbb{R} \times \mathbb{R}$ as the maximum such. The fact that the assignment $\mathcal{B}(Y) \ni B_Y \mapsto \alpha_Y(B_Y) \in \mathcal{C}(Y)$ is an algebra map signifies in particular that the above mappings (6.1) are part of a commutative square relating system behavior and requirements for a specific wiring diagram $f : X \to Y$

$$\begin{array}{ccc} \mathcal{B}(X) & \xrightarrow{\alpha_X} & \mathcal{C}(X) \\ {\scriptstyle \mathcal{B}(f)}\downarrow & & \downarrow{\scriptstyle \mathcal{C}(f)} \\ \mathcal{B}(Y) & \xrightarrow[\alpha_Y]{} & \mathcal{C}(Y) \end{array}$$

[2] Formally, this is a monoidal natural transformation between the two lax monoidal functors [112].

Intuitively, this says that for a given system decomposition into subcomponents, first composing the behaviors of the internal boxes using the behavior algebra and then talking about the contract that composite satisfies is the *same* as first computing the maximal contracts the components satisfy individually and then composing using the contract algebra. This provides extra flexibility for passing between different models, not only for this specific algebra map example but also for other maps relating different algebras that may be established.

Another way to combine the behavior $\mathcal{B}$ and requirements $\mathcal{C}$ algebra is to construct a new algebra of *contracted behaviors* that, to each process placeholder $Y$ assigns a pair of a physical behavior along with some contract it satisfies. This allows us to compose using both algebras simultaneously and choosing which information to look at; this abstract algebra is already defined [156, Prop. 4.5.5] for a specific behavior algebra and provides a tool that allows us to essentially relate two algebras via some desired condition inside their product.

We have addressed a unification between requirements and system behaviors. Another unification present in this dissertation is that of system behavior and architecture in the form of traceable decompositions between a specific behavioral formalism and a candidate architecture that implements this formalism in hardware and software. This is implemented categorically through the notion of slice category, where all design choices reside. This means that we still need the expertise of systems designers to construct a *correct* system architecture. The traces between system behavior and system architecture (de)compose properly as long as the candidate system architecture constitutes a viable solution to the particular design problem of controlling a process.

Finally, there is composition between analyses techniques. Using the scaffolding of verified composition between requirements, system behaviors, and systems architectures it is possible to assure a subset of safety properties via contracts. Then, using the system architecture it is possible to do a security analysis via tests using the system architecture. This can take the form of finding attack patterns, weaknesses, and vulnerabilities in a hypothetical scenario and then using that information to mitigate possible hazardous states by knowing which parts of the system are most vulnerable to attackers.

The result of this theoretical foundation is precisely this unification between distinct models that can then be used to better assure the safety and security of the system as a whole only by examining the relationships between its parts and interfaces. This does not mean we eradicate emergent behavior but that we have a strategy of controlling the effects of emergent behavior.

Concretely, the contributions of this thesis are as follows:

contribution 1 — A type-oriented algebraic defition of contract theory and a compositional model of linear time-invariant systems using the ideas of functorial semantics.

contribution 2 — A unification between safety requirements and system behaviors within contract theory and control using the fundamental notion of the natural transformation.

contribution 3 — A hierarchical way of decomposing to architectures using the notion of the slice category that can then be used to do component-based security analysis.

Category theory is often called *abstract nonsense*. Yet, we show that category theory can be used to solve a significant and practical problem in cyber-physical systems. By studying how different notions of a system relate with each other instead of what each individual notion is, it is possible to construct semantics that could lead to unification of modeling paradigms, under one formalism, which could then allow metrics to propagate throughout the requirements generation, behavioral description, and architecture implementation stages of a systems lifecycle. In the future this could be streamlined by the creation of new tools based on categorical notions of system models.

*Telos.*

# Endnotes

Chapter 0: The general idea for this chapter is adapted from Paruchuri [138] and partially includes content from an article I wrote for Increment [15].

Chapter 1: The chapter partially contains material from an article written with Subrahmanian and Fleming [23].

Chapter 2: The title of the chapter is inspired by a similar one by Spivak [158]. Part of this chapter was written collaboratively with Fleming, Genovese and Vasilakopoulou [20, 21].

Chapter 3 & 4: The chapter is edited from work conducted and coauthored with Fleming and Vasilakopoulou [20].

Chapter 5: The chapter is edited from work conducted and coauthored with Fleming and Genovese [21].

Chapter 6: The chapter is edited from work conducted and coauthored with Fleming and Vasilakopoulou [20].